\documentclass[]{aastex631}

\shorttitle{Gravitationally decoupled strange star model...}
\shortauthors{Maurya et al.}

\graphicspath{{./}{figures/}}

\begin{document}

\title{Gravitationally decoupled strange star model beyond standard maximum mass limit in Einstein-Gauss-Bonnet gravity}

\author[0000-0003-4089-3651]{S. K. Maurya}
\affiliation{Department of Mathematics and Physical Science, College of Arts and Science,University of Nizwa,\\ Sultanate of Oman, Email: sunil@unizwa.edu.om}

\author[0000-0001-9778-4101]{Ksh. Newton Singh}
\affiliation{Department of Physics, National Defence Academy, Khadakwasla, Pune 411023, India, Email: ntnphy@gmail.com}

\author[0000-0001-6110-9526]{M. Govender}
\affiliation{Department of Mathematics, Durban University of Technology, Durban 4000, South Africa, Email: megandhreng@dut.ac.za}

\author[0000-0000-0000-0000]{Sudan Hansraj}
\affiliation{Astrophysics and Cosmology Research Unit, University of KwaZulu-Natal, Private Bag X54001, \\Durban 4000, South Africa, Email: hansrajs@ukzn.ac.za}

\begin{abstract}
The recent theoretical advance known as the Minimal Geometric Deformation (MGD) method has initiated renewed interest in investigating higher curvature gravitational effects in relativistic astrophysics. In this work we model a strange star within the context of Einstein-Gauss-Bonnet gravity with the help of the MGD technique. Starting off with the Tolman metric ansatz together with the MIT Bag model equation of state applicable to hadronic matter, anisotropy is introduced via the superposition of the seed source and the decoupled energy-momentum tensor. The solution of the governing systems of equations bifurcates into two distinct models, namely the mimicking of the $\theta$ sector to the seed radial pressure and energy density and a regular fluid model. Each of these models can be interpreted as self-gravitating static, compact objects with the exterior described by the vacuum  Boulware-Deser solution. Utilizing observational data for three stellar candidates, viz., PSR J1614-2230, PSR J1903+317, and LMC X-4 we subject our solutions to rigorous viability tests based on regularity and stability. We find that the Einstein-Gauss-Bonnet parameter and the decoupling constant compete against each other for ensuring physically realizable stellar structures. The novel feature of work is the demonstration of stable compact objects with stellar masses in excess of $M= 2 M_{\odot}$ without appealing to exotic matter. The analysis contributes new insights and physical consequences concerning the development of ultra-compact astrophysical entities.
\end{abstract}

\keywords{Neutron stars (1108); Compact objects (288); Theoretical models (2107)}

\section{Introduction}\label{sec1}

Recently \cite{drake} conjectured that the observed star  RX J1856.5–3754 could be a quark star. While subsequently, this suggestion has drawn many criticisms largely to do with the speculated radius, the question of the existence of quark stars, which are composed of constituents of subatomic particles,  remains open. Such stars are believed to be denser than neutron stars. Evidence for quark-matter cores in massive neutron stars was considered by \cite{annala} who found that the properties of matter in the cores of neutron stars with mass corresponding to 1.4 solar masses ($M_{\bigodot}$) compare favourably with nuclear model calculations. Moreover in the case of  neutron stars with mass twice the solar mass,  quark matter is related to the   sound speed and in a certain bound  massive neutron stars are predicted to have  quark-matter cores of significant size. It was \cite{bodmer} and  \cite{witten} that speculated on the existence of compact stars consisting of hadronic matter of deconfined quarks made up of the three lightest quark flavor states.  However, much of the analysis has been conducted in the context of the MIT bag model \citep{chodos1,chodos2,peshier} which is deemed simplistic.  Recently \cite{ayann1} constructed a quark star model using numerical methods and with an interacting quark equation of state in the framework of 4D EGB theory introduced by  \cite{glavan}. Although the dimensional regularization process remains under debate, useful insights were derived.   It was found that the fourth order correction parameter of the Quantum Chromodynamics (QCD) perturbation and the EGB coupling constant $\alpha$ contribute significantly to the mass-radius ratio as well as the stability of quark stars. It is interesting to analyse quark star behaviour in the full blown EGB theory whose  contributions become dynamic only in dimensions 5 and higher. This is the purpose of the present investigation. 

At least at the solar system level, Einstein's general relativity (GR) has proved its reliability in multiple settings. Besides the celebrated classical tests, the successes include the detection of gravitational waves and the recent exploits of the Event Horizon Telescope in capturing a black hole shadow. Nevertheless, there is a growing viewpoint that a successor to GR  is required. The motivation lies largely in the inability of GR to adequately deal with the observed accelerated expansion of the universe without invoking the existence of exotic forms of matter such as dark energy, dark matter, quintessence and phantom fields.  Additionally GR appears to experience challenges in high gravitational fields such as those near black holes and neutron stars. In view of the foregoing, extensions and modifications of GR have been pursued so that anomalies may be successfully resolved but without sacrificing all the gains of GR.

In the market place of ideas on gravitational field theory, there exist a number of contenders with GR. At the very least it is expected of a theory of gravity that it  satisfies diffeomorphism invariance and  generates up to second order equations of motion. Rastall gravity \citep{rastall1,rastall2,hans1}, $f(R, T)$ theory \citep{Harko,hans2} and unimodular (also known as trace-free) gravity \citep{ellis1,ellis2,hans3} have all been extensively investigated for their cosmological and astrophysical implications. Geometrically all three of these propositions are equivalent \citep{visser,darabi}.  The first two of these have the problem of non-conservation of energy momentum while the last mentioned permits the addition of energy conservation by hand as a constraint on the system of equations. The idea known as $f(R)$ gravity \citep{staro} succeeds in explaining the cosmic accelerated expansion however it has the drawback of generating ghosts in the form of derivatives of order higher than 2. Moreover, the theory has been shown to be conformally equivalent to scalar tensor theory such as Brans-Dicke. 

Lovelock gravity \citep{love1,love2} involves polynomial invariants of the Riemann tensor, Ricci tensor and Ricci scalar in constructing the Lagrangian. Remarkably the theory yields second order equations of motion despite the higher curvature effects and higher dimensions. These properties prompt many to refer to Lovelock gravity as a natural extension of GR in higher dimensions. Note that investigations into higher dimensional spacetimes is not novel. It began with the works of Kaluza \citep{kaluza} and Klein \citep{klein} who endeavoured to explain the Maxwell field in the context of 5 dimensional gravity. More recently,  brane-world cosmology and astrophysics  which relies on five dimensional gravity \citep{maartens} evoked considerable interest.  A particular case of Lovelock order two is the Einstein-Gauss--Bonnet (EGB) theory which is intimately connected with string theory. It appears in the Lagrangian of low energy heterotic string theory \citep{gross} and for this reason has been studied in gravitation as well since a link with quantum field theory is also a goal of gravity research seeking a grand unified field theory. EGB gravity, which has critical dimensions 5 and 6, has been thoroughly studied in the context of relativistic astrophysics with a few exact solutions having been found in five dimensions  \citep{hans4,hans5,hans6,kang} and six dimensions \citep{hans7,hans8}  after the seminal works of Boulware and Deser \citep{boul} who found the vacuum solution and Wiltshire \citep{wilt} who electrified the solution of Boulware and Deser. Davis \citep{davis} recently devised the junction conditions in general form for an interior spacetime to match the exterior Boulware-Deser or Wiltshire metric.  Some anisotropic stellar models have also been proposed recently and their properties studied. For example see \citep{sharif,tangphati,abbas,bhar,pani,Rincon12}. 

In Einstein gravity, finding exact solutions is often nontrivial on account of the nonlinearity of the field equations. The situation in EGB is exacerbated further by the inclusion of higher curvature terms. Nevertheless it is possible to profit from a recent insight known as the Minimal Geometric Deformation (MGD) approach which originated within the context of the Randall-Sundrum \citep{randall1,randall2} brane-world proposition \citep{Ovalle2013}. The MGD approach has been exploited in several areas in GR including black holes \citep{casadio,ovalle1, Contreras2018,Contreras2019,Rincon2020}, brane-world stars \citep{ovalle,casadio1}, gravitational lensing \citep{cavalcanti} and other stellar configurations \citep{casadio2,rocha1,rocha2}. The basic idea of the MGD is that one can start from a simple spherically symmetric source with a known solution and use that to couple on more complicated sources thus setting up an auxiliary system of equations that must be solved. Superposition of all the independent solutions solves the entire system. Essentially a function $f(r)$ of the radial component is added to the inverse of the radial metric potential and then the system is solved to find this unknown function $f(r)$.  The important requirement for the scheme to work is that there is no exchange of energy momentum between the sources. An advance of the MGD approach is the Extended Geometric Deformation (EGD) proposal where exchanges in energy between the sources is mandatory. Detailed treatments may be found in  \cite{ovalle3,sharif1,sharif2,Maurya2019,Maurya2020,Maurya11,Maurya12}.  In addition, it must be noted that the MGD technique extends the seed matter distributions to anisotropic domains. Historically isotropic stars were investigated thoroughly as equilibrium stages of stellar evolution, however, compact objects with unequal tangential and radial stresses have also been considered as representing realistic objects. Anisotropy may arise in various circumstances including viscosity effects and through intense magnetic fields. However, the anisotropy can be also introduced in the system using gravitational decoupling via MGD technique. Several authors have considered the MGD process for different configurations such as \cite{Maurya2020a,Maurya2020b,Maurya2021,Ortiz2020,Contreras2020,Abell2020,Rincon11,Zubair2020,Zubair2021,Sharif2020}. Futhermore, some pioneering works on solutions for Yang-Mills-Einstein-Dirac field equations, axial symmetry, Hairy black holes have been done in the context of MGD \citep{Ovalleprd1,Ovalleprd2,Ovalleprd3}. 
From a survey of the literature, it is evident that this very promising MGD technique has not been applied to investigate strange star models under EGB gravity which is more formidable than the Einstein counterpart. Our intention in this work is to explore this option in the scenario of an anisotropic five dimensional fluid sphere which is prospectively a higher dimensional stellar model. Note that while some skeptism may be encountered concerning the  existence of extra dimensions beyond the physically accessible four, it should be borne in mind that these extra dimensions are taken to be of Planck scale and to be topologically curled microscopic circles. Despite their small magnitude, our investigation reveals that they exert significant effects on stellar structure. While the Large Hadron Collider experiment failed to detect the presence of large extra dimensions, perhaps due to the capability of the instrument, small higher dimensions were not ruled out (\cite{choudhury}).  Hence this presents a strong inducement to analyse the structure of stars in such a context as EGB which has all the necessary ingredients for a viable theory of gravitation beyond general relativity  but which contains GR as a special case.

The article is organized as follows:  Section \ref{sec2} contains a brief  review of EGB gravity theory together with the salients aspects of the gravitational decoupling via MGD method. Here, we also discuss the energy-momentum tensor for different sources with the MIT Bag equation of state (EoS). In Sec. \ref{sec3} we obtain a gravitationally decoupled solution by adopting the well-behaved Tolman IV metric potential for the seed spacetime geometry
to guarantee a well defined horizon free spacetime.  The main purpose here is to obtain the deformation function $\Phi(r)$. To find this, we propose two different procedures in subsections \ref{sec3.1} and \ref{sec3.2} in order to close the system of equations for the extra  source which is introduced by gravitational decoupling. The exterior spacetime and matching conditions have been discussed in Sec.\ref{sec4} where we match the decoupled interior solution governed by the anisotropic matter distribution to the  exterior Boulware--Deser vacuum  solution at a suitable boundary. In Sec.\ref{sec5}, we discuss the physical properties of gravitationally decoupled strange stars for the  solutions  obtained in subsections \ref{sec3.1} and \ref{sec3.2}. The stability analysis of the model is considered in subsection\ref{sec5.2}. The most important physical features of the mass ($M$) and radius $(R$) measurements of the model via the $M-R$ curves is presented in Sec. \ref{sec5.4}, while the  mass and Bag constant measurements of the strange star models via equi-plane diagrams is depicted in Sec. \ref{sec5.4}. In the last Sec. \ref{sec6}, we present the conclusions with some astrophysical implications of the models.   Some relevant lengthy expressions of physical quantities have been relegated to  the Appendix.  

 \section{Basic field equations of EGB gravity for gravitational decoupling} \label{sec2}
The modified $D$-dimensional action for Einstein--Gauss--Bonnet (EGB) gravity with matter field by introducing an extra source may be written in the form: 
\begin{eqnarray}
	\mathcal{I}_{G}=\frac{1}{16 \pi }\int d^{D}x\sqrt{-g}\left[ \mathcal{R}-2\Lambda +\alpha \mathcal{L}_{\text{GB}} \right]+
\mathcal{S}_{\text{matter}}+\beta\,\int \mathcal{S}_{\theta}\sqrt{-g}~d^{4}x, \label{eq1}
\end{eqnarray}
where $\mathcal{R}$ and $\Lambda$ denote the $D$--dimensional Ricci scalar and the cosmological constant, respectively. Here, 
$\mathcal{S}_{\text{matter}}$ and $\mathcal{S}_{\theta}$ define the Lagrangian of the matter field and extra source, respectively. The dimensionless constant $\beta$ is called the decoupling constant while the EGB coupling constant $\alpha$ is taken to be positive definite quantity   since it associates with the inverse string tension with dimension of [$\text{length}$]$^2$.  At this point it is worth noting that there are no experimental values of $\alpha$ known to date however various orders of magnitude have been speculated in some instances.  \citep{amendola} have argued for an $\alpha$ value  as high as of the order of $10^{23}$ in the context of solar system tests of EGB theory.  \citep{dehghani} used a negative $\alpha$ to explain the accelerated cosmic expansion if higher curvature Gauss-Bonnet invariants are present. Moreover, Doneva and Yazadjiev \citep{doneva}  considered stars in the 4D EGB theory propounded by Glavan and Lin \citep{glavan} and employed positive and negative $\alpha$ values of unit order to generate physically plausible models. They also demonstrated that negative $\alpha$ resulted in stable black holes.  

The Gauss--Bonnet 
Lagrangian $\mathcal{L}_{\text{GB}}$ is the combination of Riemann curvature tensor ($\mathcal{R}_{\mu\nu kl}$), Ricci tensor ($\mathcal{R}_{\mu\nu}$), and Ricci scalar $(\mathcal{R})$ which can be given by
\begin{eqnarray}
&& \mathcal{L}_{\text{GB}}=\mathcal{R}^{\mu\nu kl} \mathcal{R}_{\mu\nu kl}- 4 \mathcal{R}^{\mu\nu}\mathcal{R}_{\mu\nu}+ \mathcal{R}^2\label{GB}.  \label{eq2}
\end{eqnarray}
The decoupled equation of motion can be directly obtained by the variation of the action  (\ref{eq1}) with respect to $g^{\mu\nu}$ in the form 
\begin{equation}\label{eq3}
G_{\mu\nu}+\alpha H_{\mu\nu} = \frac{8 \pi G }{c^4} T_{\mu\nu}~,~~\mbox{where}~~~T_{\mu\nu}= \hat{T}_{\mu\nu}-\beta\,\theta_{\mu\nu},
\end{equation}
with
\begin{eqnarray}
\hat{T}_{\mu\nu}=\frac{-2}{\sqrt{-g}}\frac{\delta\left(\sqrt{-g}\,\mathcal{S}_{matter}\right)}{\delta g^{\mu\nu}},~\text{and}~\theta_{\mu\nu}=\frac{2}{\sqrt{-g}}\frac{\delta\left(\sqrt{-g}\,\mathcal{S}_\theta\right)}{\delta g^{\mu\nu}},  \label{eq4}
\end{eqnarray}
where $G_{\mu\nu}$ denotes the Einstein tensor and $H_{\mu\nu}$ is the contribution of Gauss-Bonnet (GB) term that can given by the following expressions
\begin{eqnarray}
 && G_{\mu\nu} = \mathcal{R}_{\mu\nu}-\frac{1}{2}\mathcal{R}~ g_{\mu\nu} ~~~~~~~~\text{and}~~ \nonumber\\
&& H_{\mu\nu} =  2\Big( \mathcal{R} \mathcal{R}_{\mu\nu}-2\mathcal{R}_{\mu k} \mathcal{R}^k_\nu -2 \mathcal{R}_{\mu\nu kl}\mathcal{R}^{kl} - \mathcal{R}_{\mu kl\delta} \mathcal{R}^{kl\delta}_\nu\Big)- \frac{1}{2}~g_{\mu\nu}~\mathcal{L}_{\text{GB}}.  \label{eq5}
\end{eqnarray}
Here, it is mentioned that the GB term has no effect on the gravitational dynamics in $D$-dimensional spacetime when $D\le4$, since the Gauss-Bonnet invariants become a total derivative. Now in order to describe the strange star compact object, we assume a $D$ dimensional static and spherically symmetric line element of form,
\begin{eqnarray}\label{eq6}
ds^2_{D}= - W(r)\, c^2dt^2 + H (r) dr^2 + r^{2} d\Omega^2_{D-2},
\end{eqnarray}
where  $W\equiv W(r)$ and $H\equiv H(r)$ are the metric functions which depend on radial coordinate $r$ only, while  $d\Omega^2_{D-2}$ is the metric on the unit $D-2$- dimensional sphere. Then  Eq. (\ref{eq3}) together with the $D$-dimensional space-time (\ref{eq6}) provide the following non-vanishing components of the energy momentum tensor in  EGB gravity,
\begin{small}
\begin{eqnarray}
&&\hspace{-0.5cm} \frac{8\pi G }{c^4} T^0_0 = \frac{(D-2) (H-1) [\alpha (D-5) (H-1) + (D-3)\, H r^2]}{2 H^2 r^4} - \,\frac{(D-2) H^{\prime} [4 \alpha (H-1) + H r^2]}{2 H^2 r^3 W},  \label{eq7} \\
&& \hspace{-0.5cm} \frac{8\pi G }{c^4} T^1_1= \frac{(D-2) (H-1) [\alpha (D-5) (H-1) + (D-3)\, H r^2]}{2 H^2 r^4} + \,\frac{(D-2) W^{\prime} [4 \alpha (H-1) + H r^2]}{2 H^2 r^3 W},  \label{eq8}\\
&&\hspace{-0.5cm} \frac{8\pi G }{c^4} T^2_2 =\frac{(2-D)}{12 H^3 r^4 W^2}\, \Big[H r^2 \big\{W^{\prime 2} H r^2 + r (-4 W^{\prime} H + H^{\prime} W^{\prime} r  - 2 W^{\prime \prime} H r) W + 
    2 (D-3) [(H-1) H + H^{\prime} r] W^2\big\} \nonumber\\
&& \hspace{0.8cm} + 
 2 \alpha \big\{2 W^{\prime 2} (H-1) H r^2 + 
    2 W^{\prime} r [12 H + H^{\prime} (H-3) r] W  - (H-1) W \big(4 W^{\prime \prime} H r^2 + (D-5) [(H-1) H  - 2 H^{\prime} r] W\big) \big\} \Big].  \label{eq9}
\end{eqnarray}
\end{small}
 Quintessentially EGB theory is a higher dimensional and higher curvature proposal generating up to second equations of motion. It is acknowledged that efforts have been made to examine the effects of the GB terms in 4 dimensional gravity through a process of  dimensional regularisation \citep{glavan,tomozawa}. Essentially the GB term undergoes a rescaling. The method however has generated considerable criticisms and is still not free of controversy \citep{gurses1,gurses2}. For these reasons and the motivations to study higher dimensional stellar structures provided earlier, we direct our attention to 5 dimensional hyperspheres. Then the static  spherically symmetric line element (\ref{eq6}) in 5--dimensional spacetime may be written as, 
\begin{eqnarray}\label{eq10}
ds^2_{5}= - W(r) \, c^2dt^2 + H (r) dr^2 + r^{2} d\Omega^2_3, 
\end{eqnarray} 
where $d\Omega^2_3=\big(d \theta^2+\sin^2 \theta~ d\phi^2 +\sin^2 \theta \sin^2 \phi ~d\psi^2 \big)$.  Now we suppose that the compact stellar object is filled with anisotropic fluid which can be described by the following energy momentum tensor as
\begin{eqnarray}
T_{\mu\nu} = (c^2\,\epsilon + P_{t})\,u_{\mu} u_{\nu}+ P_{t}\, g_{\mu\nu}+(P_r-P_{t})\chi_{\mu}\,\chi_{\nu}, \label{eq11}
\end{eqnarray}
where $P_r$ and $P_{t}$ are called radial and tangential pressures, respectively and $\epsilon$ describe the energy density for the  decoupled energy tensor ($T_{\mu\nu}$). Moreover, $u^{\nu}$ is the contravariant $5$-velocity satisfying $u^{\nu} u_{\nu}=-1$,  while $\chi^\mu=\sqrt{1/H(r)}\,\delta^\mu_1$ is the unit space-like vector in the radial direction. 
Now, using the Eqs. (\ref{eq6}) and (\ref{eq11}) with (\ref{eq3}) one could obtain the non-vanishing components of the gravitational field equations as,
\begin{small}
\begin{eqnarray}
&&\hspace{-0.5cm} -\frac{8\pi G }{c^2} T^0_0  = \frac{12\, \alpha H^\prime (H-1) +3r\, H (H^\prime r + 2 H^2-2H )}{2 H^3 r^3},  \label{eq12} \\
&& \hspace{-0.5cm}  \frac{8\pi G }{c^4}  T^1_1 =\frac{12 \alpha\, W^\prime (H-1) +3 H r (W^\prime r - 2 (H-1) W)}{2 H^2 r^3 W},~~~~~ \label{eq13}\\
&&\hspace{-0.5cm}  \frac{8\pi G }{c^4}  T^2_2  = \frac{1}{4 H^3 r^2 W^2}\,\Big[4 \alpha \{W^{\prime 2} (H-1) H + H^\prime W^\prime (H-3) W   - 2\, W^{\prime\prime} (H-1)\, H\, W\}  + H \big\{W^{\prime 2} H r^2 + W^\prime r (H^\prime r  -4 H ) W  \nonumber\\ && \hspace{0.8 cm} - 2 W [W^{\prime\prime} H r^2 + 2 (H - H^2 - H^\prime r) W]\big\} \Big].  \label{eq14} ~~~~~~
\end{eqnarray}
\end{small}
where $\prime$ denotes the derivative with respect to the radial coordinate $r$, only. It is important to note that the Einstein tensor $G_{\mu\nu}$ and the 
Gauss--Bonnet tensor $H_{\mu\nu}$ are individually conserved \citep{love1,love2}. Then from this fact, the decoupled  
energy-momentum tensor, $T_{\mu\nu}$, in Eq. (\ref{eq3})  is also divergence--free i.e. $\nabla^\mu\,T_{\mu\nu}=0$, which yields the equation,
\begin{eqnarray}
-\frac{W^\prime}{2W}\,({T}^1_1-{T}^0_0)-({T}^1_1)^\prime+\frac{3}{r}({T}^2_2-{T}^1_1)=0,  \label{eq15}
\end{eqnarray}
Then using Eq.(\ref{eq11}), the above equation (\ref{eq15}) leads to a general Tolman-Oppenheimer-Volkoff (TOV) equation for the decoupled system in EGB gravity as,  
\begin{eqnarray}
-\frac{W^\prime}{2W}(\epsilon+P_r)-P_r^{\prime}+\frac{3}{r}( P_{t}-P_r)=0.\label{eq16}
\end{eqnarray}
The above equation is known as a general hydrostatic equation for 5D Einstein-Gauss-Bonnet gravity under the spacetime (\ref{eq6}). \\
Following the work of Wright \citep{Wright}, we define an arbitrary function $F(r)$ in the form 
\begin{eqnarray}
F(r)=\frac{H(r)-1}{H(r)}, \label{eq17}
\end{eqnarray}
to find the mass function. 
 Then from Eq.(\ref{eq12}) and (\ref{eq17}), we can write:
\begin{eqnarray}
\frac{16\,\pi G}{3\,c^2}\,\epsilon\,r^3 &=& 4\alpha\,F^\prime\,F+ r^2\,F^\prime+2\,r\,F = ( 2\,\alpha\,F^2+r^2\,F)^\prime ~. \label{eq18}
\end{eqnarray}
By integrating above equation (\ref{eq18}) with the limit $0$ to $r$, we get
\begin{eqnarray}
&&( 2\,\alpha\,F^2+r^2\,F)=\frac{16\,\pi G}{3\,c^2}\,\int^r_0 \epsilon(x)\,x^3 dx ~. \label{eq19}
\end{eqnarray}
It is now convenient to determine the mass function in the form  
\begin{eqnarray}
m(r)=\frac{8\,\pi G}{3\,c^2}\,\int^r_0 \epsilon(x)\,x^3 dx~, \label{eq20}
\end{eqnarray}
where the factor of $1/3$ has been introduced for the higher dimensional mass function since $D$-dimensional spacetime possesses an extra factor of $1/(D -2)$ \citep{Ponce}. 
Then from Eqs.(\ref{eq19}) and (\ref{eq20}), we find the function $F(r)$
\begin{eqnarray}
2\alpha F^2+r^2 F-2\,m=0 \Longrightarrow F=\frac{r^2}{4 \alpha}
\left(-1\pm\sqrt{1+\frac{16\alpha m}{r^4}}\right).\label{eq21}
\end{eqnarray}
The above equation (\ref{eq21}) with positive signature and the relation (\ref{eq17}) provides the metric function $H(r)$ as,
\begin{eqnarray}
\frac{1}{H(r)}=1+\frac{r^2}{4 \alpha}
\left(1-\sqrt{1+\frac{16\,\alpha\, m}{r^4}}\right).  \label{eq22}
\end{eqnarray}
which resembles the Boulware-Deser spatial potential. The temporal potential will of course bear no relationship to the Boulware-Deser metric since we are working in the interior of an anisotropic star. The mass function (\ref{eq20}) at the boundary can be evaluated by matching of the interior metric (\ref{eq6}) with the appropriate components of the  exterior Boulware--Deser metric  \citep{boul} (vacuum) solution at the boundary of the star which will be the total mass of the compact star. 
Our next aim is to solve the EGB field equations for strange star models. For this purpose, we apply the  gravitational decoupling via MGD under the specific transformation along the gravitational potential, 
\begin{eqnarray}
&& W(r) \longrightarrow A(r)+\beta\, \chi(r), \label{eq23}\\
&& H(r) \longrightarrow \frac{1}{X(r)+\beta\, \Phi\,(r)}, \label{eq24}
\end{eqnarray}
where $\Phi(r)$ and $\chi(r)$ are the geometric deformation functions associated with  the spatial and temporal metric components. The parameter $\beta$ allows us to manipulate the  deformation suitably. In  the special case   $\beta =0$, the standard EGB scenario is regained.\\
While there is considerable latitude in nominating the deformation contribution, we elect to analyse the simplest case of the minimal deformation of the metric. This entails setting either $ \chi(r) = 0$ with  $\phi(r)\ne 0$ or $\phi(r)\ne 0$ with $\chi(r) = 0$.  The first case generates a deformation of the radial component only while the temporal evolution is unaffected.  Observe that the anisotropy in the system is introduced through  the  deformation of the radial component (\ref{eq24})  through the  anisotropic tensor $\theta_{\mu\nu}$. 
\begin{figure}
    \centering    
     \includegraphics[width=8.5cm]{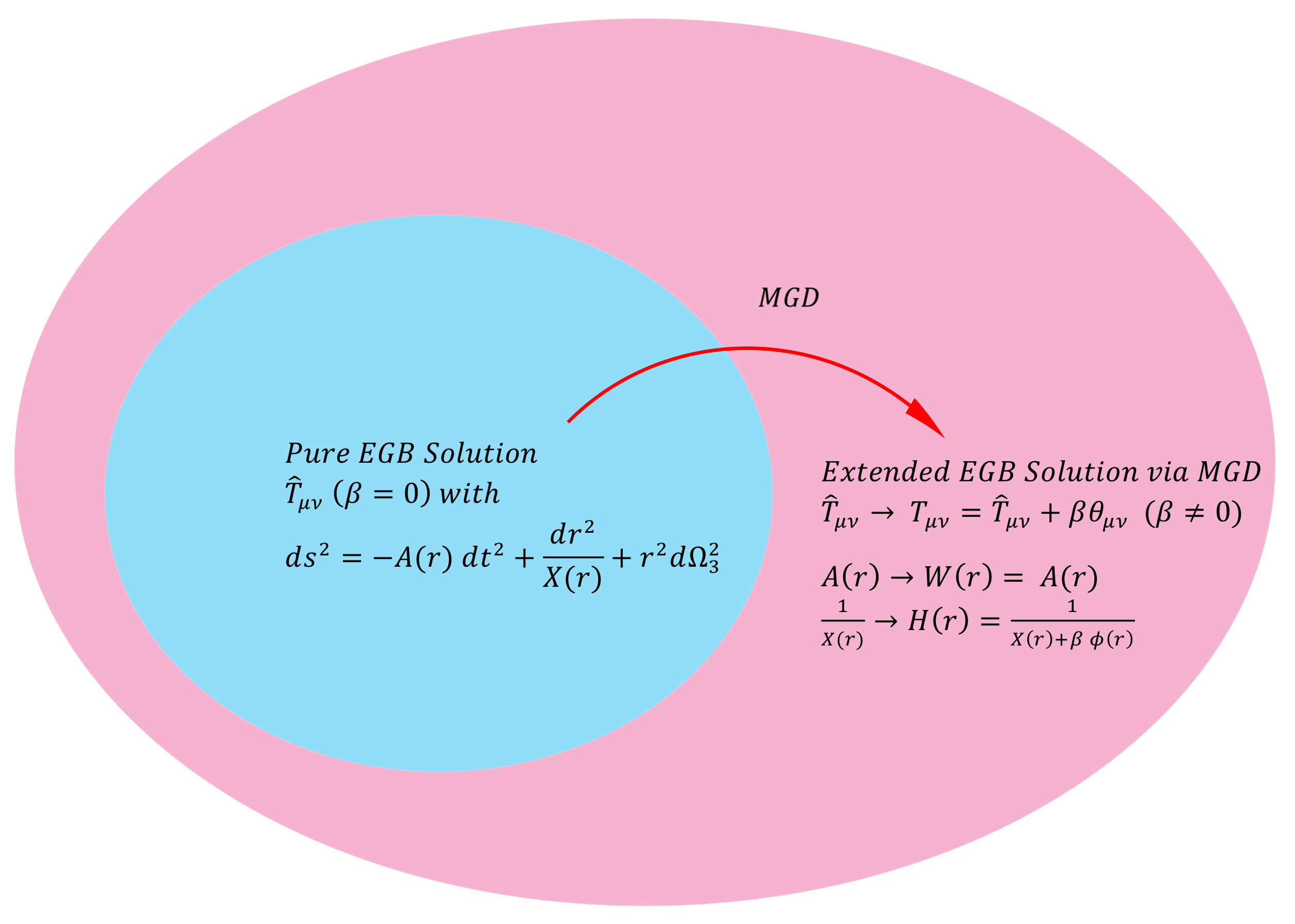}
    \caption{The above diagram describes that how pure EGB solutions can be extended via MGD to anisotropic domains.}
    \label{fig1}
\end{figure}  
The consequence of the transformation of (\ref{eq24}) is that the field equations (\ref{eq12})-(\ref{eq14}) with Eq.(\ref{eq3}) bifurcate into two categories.  First we consider the standard EGB field equations that correspond to the anisotropic case when $\beta = 0$. Depending on the gravitational potentials $X$ and $W$ the system  assumes the form,
\begin{eqnarray}
&&\hspace{-0.5cm} -\frac{8\pi G }{c^4} \hat{T}^0_0 =\frac{12\alpha X^\prime (X-1)-3 r (X^\prime r+2 X-2)}{2r^3} ,\label{eq25}\\
&&\hspace{-0.5cm} \frac{8\pi G }{c^4} \hat{T}^1_1=\frac{12 \alpha W^\prime X (1- X)+3\,r [W^\prime X r+2(X-1) W]}{2 r^3 W} ,\label{eq26}\\
&&\hspace{-0.5cm}  \frac{8\pi G }{c^4} \hat{T}^2_2=\frac{1}{4r^2W^2}\big[W^\prime r (X^\prime r + 4 X) W -W^{\prime 2} r^2 T + 4 \alpha \{W^{\prime 2} (X-1) X + X^\prime W^\prime (1 - 3 X) W  - 2 W^{\prime\prime} (X-1) X W\}  \nonumber\\
 && \hspace{0.9cm} + 2 W (W^{\prime\prime} r^2 X + 2 (X^\prime r + X-1) W)\big],~~~~\label{eq27} 
\end{eqnarray}
with the  assumption that the following  conservation quantity (\ref{eq15}) reduces to
\begin{eqnarray}
-\frac{W^\prime}{2W}\,(\hat{T}^1_1-\hat{T}^0_0)-(\hat{T}^1_1)^\prime+\frac{3}{r}(\hat{T}^2_2-\hat{T}^1_1)=0.  \label{eq28}
\end{eqnarray}
Then the seed solution can be described by following spacetime,
\begin{eqnarray}
ds^2=- W(r) dt^2 + \frac{ dr^2 }{X(r)}+ r^{2}\big[d \theta^2+\sin^2 \theta ~(d\phi^2+\sin^2 \phi ~d\psi^2) \big], \label{eq29} 
\end{eqnarray}
and corresponding seed mass function ($m_s$) can be determined by the formula (\ref{eq20}) as,
\begin{eqnarray}
m_s(r)=\frac{8\pi G}{3c^2} \int^r_0 \hat{\epsilon}(x)\, x^3 dx.  \label{eq30}
\end{eqnarray}
where the metric function $X(r)=1+\frac{r^2}{4 \alpha}
\left(1-\sqrt{1+\frac{16\,\alpha\, m_s}{r^4}}\right)$. Now when we turn on the decoupling constant $\beta$, that is $\beta \neq 0$, we obtain another set of equations corresponding to an extra source $\theta_{\mu\nu}$ satisfying the equations, 
\begin{eqnarray}
&&\hspace{-0.7cm}\frac{8\pi G }{c^2} \theta_0^0 = \frac{3\, \beta\, [4 \alpha\,X^\prime\,\Phi-r \,(2\, \Phi + \Phi^\prime\, r) + 4 \,\alpha\, \Phi^\prime\, (\beta\, \Phi + X-1)]}{2\,r^3}, ~~~~~~\label{eq31}\\
 &&\hspace{-0.7cm}\frac{8\pi G }{c^4}\theta_1^1 = -\frac{3 \beta\,\Phi [4 \,\alpha\, W^\prime\, (1 - \beta\, \Phi -2 X) + r (W^\prime\, r + 2 W)]}{2\,r^3 W},~~~~~~~\label{eq32}\\
&&\hspace{-0.7cm} \frac{8\pi G }{c^4} \theta_2^2 = \frac{-\beta}{4r^2W^2}\,\big[W^\prime r (4 \Phi + \Phi^\prime r) W -W^{\prime 2} \Phi r^2 + 
   2 W (W^{\prime\prime} \Phi r^2 + 2 (\Phi + \Phi^\prime r) W) + 
   4 \alpha \{W^{\prime\prime} \Phi ( \beta\, \Phi -1 \nonumber\\&& \hspace{0.8cm} + 2\, X) - 2\, W^{\prime\prime}\, \Phi\, (\beta\, \Phi + 2\, X-1)\, W + 
      W^\prime (\Phi^\prime - 3 \beta \Phi^\prime \Phi - 3 X^\prime \Phi - 3 \Phi^\prime X) W\}\big]. \label{eq33}
\end{eqnarray}

The conservation equation in this case  $\nabla^\mu\,\theta_{\mu\nu}=0$ explicitly reads as 
\begin{eqnarray}
-\frac{W^\prime}{2W}\,(\theta^0_0-\theta^1_1)+(\theta^1_1)^\prime+\frac{3}{r}(\theta^1_1-\theta^2_2)=0. \label{eq34}
\end{eqnarray}
Thus it can be seen that the two sources $\hat{T}_{\mu\nu}$ and $\theta_{\mu\nu}$ have been  successfully decoupled by means of the MGD process.   Under these circumstance, it may be noted that  a decoupling without exchange of energy between the sources occurs (see Ref. \cite{Ovalle2018} for more). Also the mass function $m_{\Phi}$ contribution due to this extra source $\theta_{\mu\nu}$ is given as,
\begin{eqnarray}
&& m_{\Phi}= \frac{8\pi G}{3\,c^2}\,\int_0^r \theta^0_0 (x)\, x^3 dx,  \label{eq35}
\end{eqnarray}
Then equations (\ref{eq31}) and (\ref{eq35}) give the relation,
\begin{eqnarray}
2\,m_{\Phi}= 2\,\alpha\,[\beta\,\Phi^2+2\,(X-1)\,\Phi]-r^2\,\Phi,  \label{eq36}
\end{eqnarray}
 which may be solved to yield the  deformation function $\Phi$ in the form,
\begin{eqnarray}
\Phi(r)= \frac{4 \alpha + r^2 \pm \sqrt{16 \alpha \beta m_{\Phi} + [r^2 - 4 \alpha (X-1)]^2} - 4 \alpha X}{4\,\alpha\,\beta}.  \label{eq37}
\end{eqnarray}
\section{Gravitationally decoupled solution} \label{sec3}
Now we need to solve both system of equations (\ref{eq25})-(\ref{eq27}) and (\ref{eq31})-(\ref{eq33}) corresponding to sources $\hat{T}_{\mu\nu}$ and $\theta_{\mu\nu}$. Here the energy momentum tensor $\hat{T}_{\mu\nu}$ can describe either a perfect fluid or an anisotropic fluid matter distribution. We  mention that if $\hat{T}_{\mu\nu}$ represent a perfect fluid matter distribution then the anisotropy will be appeared in system through this extra source $\theta_{\mu\nu}$ but if $\hat{T}_{\mu\nu}$ already describes the anisotropic fluid then this $\theta_{\mu\nu}$ may enhance the anisotropy of the system. \\          
But here we consider the anisotropic matter distribution corresponding to energy tensor $\hat{T}_{\mu\nu}$ that can be defined as,
\begin{eqnarray}
\hat{T}_{\mu\nu} = (\hat{\epsilon} + \hat{P}_{t})\,u_{\nu} u_{\nu}+ \hat{P}_{t}\, g_{\mu\nu}+(\hat{P}_r-\hat{P}_{t})\chi_{\mu}\,\chi_{\nu}, \label{38}
\end{eqnarray}
where, $\hat{\epsilon}$, $\hat{P}_{r}$, and $\hat{P}_{t}$ denote the seed energy density, seed radial and seed tangential pressure for seed matter distribution described by energy tensor $\hat{T}_{\mu\nu}$. Then the components of the seed matter distribution are as follows:
\begin{eqnarray}
\hat{T}^0_0=-\hat{\epsilon},~~~~\hat{T}^1_1=\hat{P}_r,~~~\hat{T}^2_2=\hat{P}_t. \label{eq39}
\end{eqnarray}
Then using Eqs.(\ref{eq3}) and (\ref{eq11}) together with Eq.(\ref{eq39}) we get,
\begin{eqnarray}
\epsilon=\hat{\epsilon}+\beta\,\theta^0_0,~~~~P_r=\hat{P}_r-\beta\,\theta^1_1,~~~P_t=\hat{P}_t-\beta\,\theta^2_2. \label{eq40}
\end{eqnarray}
Since we are interested in strange quark star models corresponding to seed spacee-time (\ref{eq29}) whose internal structure is defined by the MIT bag equation of state (EOS) \citep{Chodos:1974}. Then the MIT bag model that represents a degenerated Fermi gas of quarks up ($u$), down ($d$) and strange ($s$) \citep{Chodos:1974,Farhi:1984} can  describe a  non-interacting EoS for the matter variables $\hat{\epsilon}$ and $P_r$ as,
\begin{eqnarray} \label{eq41}
\hat{P}_r=\sum_{f} P^{f}-\mathcal{B}_g,~~~~f=u,~d,~s
\end{eqnarray}
where ${p^f}$ denotes the individual pressures for the $(u)$, $(d)$ and $(s)$ quark flavor which is neutralized by the total external Bag pressure or Bag constant
$\mathcal{B}_g$. In this situation, the energy density  ($\hat{\epsilon}$) of deconfined quarks interior corresponding to the MIT Bag model can be given as, 
\begin{eqnarray}
\hat{\epsilon}=\sum_{f} \epsilon^{f}+\mathcal{B}_g.\label{eq42}
\end{eqnarray}
where equation $\epsilon^f=3P^f$ leads the matter density ($\epsilon^f$) in terms of pressure ($p^f$) for each flavor. Then by combining of Eqs. (\ref{eq41}) and (\ref{eq42}) together with the relation $\epsilon^f=3P^f$, the final form of MIT bag equation of state (EOS) for strange quark stars can be given as
\begin{eqnarray} \label{eq43}
\hat{P}_r=\frac{1}{3}(\hat{\epsilon}-4\mathcal{B}_g),  
\end{eqnarray} 
Now from the Eqs.(\ref{eq25}), (\ref{eq26}), and (\ref{eq43}) together with Eq.(\ref{eq39}), we obtain the  following equation in $X$ and $W$,
\begin{eqnarray}
&&\hspace{-0.6cm} 12 \alpha\, (X-1) (3 W^\prime\, X + X^\prime\, W) -
 r \big[ 9 W^\prime\, r\, X + (-24 + 3\,X^\prime\, r  + 8\, \mathcal{B}\, r^2 + 24\,X) W \big]=0. \label{eq44}
\end{eqnarray}
where henceforth we shall use $\mathcal{B}=8\pi\mathcal{B}_g$.  In order to integrate the above equation, we nominate the potential ansatz corresponding to the Tolman IV metric for  $X(r)$ as,
\begin{eqnarray}
X(r)=1/(1+Lr^2+Nr^4),  \label{eq45}
\end{eqnarray}
where $L$ and $N$ are constants with dimensions $\text{length}^{-2}$ and $\text{length}^{-4}$, respectively.
Now by plugging $X(r)$ into Eq.(\ref{eq44}) and integrating, we obtain,
\begin{eqnarray}
&&\hspace{-0.6cm} W(r)=C\,\text{exp}\Big[\frac{2}{9} \Big\{2\, [\, 3 (L - 4 \alpha N) + \mathcal{B} (-1 + 4 \alpha L - 16 \alpha^2 N)\,] r^2+ (3N-\mathcal{B} L+ 4 \alpha \mathcal{B} N) r^4 - 2/3 \mathcal{B} N r^6\nonumber\\
&&\hspace{0.4cm} -\frac{W_1(r)\,W_2(r)}{W_4(r)}  + \frac{W_2(r)\,W_3(r)}{W_4(r)} + \frac{3}{2} \ln(1 + L r^2 + N r^4)\Big\}\Big].  \label{eq46}
\end{eqnarray}
where, $C$ is an arbitrary constant of integration and,  
\begin{eqnarray}
&&\hspace{-0.5cm} W_1(r)=4 \alpha (3 + 4 \alpha \mathcal{B}) \big[L^2 + L \{-8 \alpha N + W_4(r)\} + 2 N \{-1 + 8 \alpha^2 N - 2 \alpha W_4(r)\}\big],\nonumber\\
&&\hspace{-0.5cm} W_2(r)=\ln[L + 4 \alpha N - W_4(r)+2 N r^2],\nonumber\\
&&\hspace{-0.5cm} W_3(r)=4 \alpha (3 + 4 \alpha \mathcal{B}) \big[L^2 -L \{8 \alpha N + W_4(r)\}  + 2 N \{-1 + 8 \alpha^2 N + 2 \alpha W_4(r)\}\big],\nonumber\\
&&\hspace{-0.5cm} W_4(r)=\sqrt{L^2 - 8 \alpha L N + 4 N (-1 + 4 \alpha^2 N)}.\nonumber
\end{eqnarray}
On inserting of $X(r)$ and $W(r)$ into Eqs.(\ref{eq25})-(\ref{eq27}) together with (\ref{eq39}) and (\ref{eq43}) we find the expressions for $\hat{\epsilon}$, $\hat{P}_r$ and $\hat{P}_t$ (taking $G=c=1$ from now) as, 
\begin{eqnarray}
&& \hspace{-0.6cm} 8\pi\,\hat{\epsilon}= \frac{3}{(1 + L r^2 + N r^4)^3} \big[L^3 r^4 + N r^2 (3 + 8 \alpha N r^2 + 4 N r^4 + N^2 r^8) + L (2 + 12 \alpha N r^2 + 7 N r^4 + 3 N^2 r^8) \nonumber\\
&& \hspace{0.4cm} + L^2 (4 \alpha  + 3 (r^2 + N r^6))\big],  \label{eq47}\\
&& \hspace{-0.6cm} 8\pi \hat{P}_r= \frac{1}{3 (1 + L r^2 + N r^4)^3} \big[-4 \mathcal{B} (1 + L r^2 + N r^4)^3 + 
 3 \big(L^3 r^4 + L^2 (4 \alpha + 3\, r^2 + 3\, N\, r^6) + N r^2 (3 + 8 \, \alpha \,N \,r^2 \nonumber\\
 && \hspace{0.4cm} + 4\, N\, r^4  + N^2 r^8) +L (2 + 12 \alpha N r^2 + 7 N r^4 + 3 N^2 r^8)\big) \big],  \label{eq48}\\
&& \hspace{-0.6cm}  8\pi \hat{P}_t=\frac{\hat{P}_{t1}(r)+\hat{P}_{t2}(r)}{81 (1 + L r^2 + N r^4)^4 [1 + L r^2 + N r^4 + 4 \alpha (L + N r^2)]}.~~~~~~~~ \label{eq49}
\end{eqnarray}
where $\hat{P}_{t1}(r)$ and $\hat{P}_{t2}(r)$ are given in the Appendix. Now we focus on the solution of the  $\theta$-sector in order to find the $\theta^0_0$, $\theta^1_1$, and $\theta^2_2$ components. Since these $\theta$-sector components depends on the deformation function $\Phi(r)$. For this purpose we apply the two different well-known mimic approach procedures: (i) The mimicking of $\theta$-component to seed energy density: $\theta^0_0=\hat{\epsilon}$, and (ii) The mimicking of $\theta$-component to seed radial pressure: $\theta^1_1=\hat{P}_r$, to find the deformation function, which are discussed in the next section as: 

\subsection{The mimicking of $\theta$-component to seed energy density: $\theta^0_0=\hat{\epsilon}$~:} \label{sec3.1}  
Now using the Eqs.(\ref{eq25}) and (\ref{eq31}) together with relation,  $\theta^0_0=\hat{\epsilon}$, yields following differential equation, 
\begin{eqnarray}
\frac{d\Phi}{dr}+\frac{2 (6\, \alpha X^{\prime}\,\Phi - 3 \Phi r - \hat{\epsilon}\, r^3)}{
 3 (-4 \alpha + 4 \alpha\, \beta\, \Phi - r^2 + 4 \alpha\,X)}=0,\label{eq50}
\end{eqnarray}
After integration of Eq.(\ref{eq50}) we get the deformation function $\Phi(r)$ 
\begin{eqnarray}
&&\hspace{-0.6cm} \Phi(r)=\frac{1}{4 \,\alpha\,\beta\, (1 + L r^2 + N r^4)}\big[r^2 + 4 \alpha L r^2 + L r^4 + 4 \alpha N r^4 + N r^6-\sqrt{r^4 (1 + L r^2 + M r^4)^2 +\Phi_1(r)}\,\big].~~~~~~~~  \label{eq51}
\end{eqnarray}
where,
\begin{eqnarray}
&&\hspace{-0.6cm}\Phi_1(r)=8\,\alpha\, (1 + \beta)\, r^4 \big[L + L^2 r^2 + 2 L N r^4 + N r^2 (1 + N r^4)\big]  + 16 \alpha^2 \big[r^4 (L + N r^2)^2 - \beta (1 + 2 L r^2 + 2 N r^4)\nonumber\\&&\hspace{0.6cm} +  \beta^2  (1 + L r^2 + N r^4)^2 C\big].\nonumber
\end{eqnarray}
Here $C$ is a constant of integration. As we know that the deformation function $\Phi(r)$ must vanish at the centre in order to have $H(0)=1$. Then $\Phi(0)=0$ yields $C=\frac{1}{\beta}$.  
The components of $\theta$-sector are given by,
\begin{eqnarray}
&&\hspace{-0.6cm} 8\pi\theta^0_0=\frac{3}{(1 + L r^2 + N r^4)^3} \big[L^3 r^4 + N r^2 (3 + 8 \alpha N r^2 + 4 N r^4 + N^2 r^8) + L (2 + 12 \alpha N r^2 + 7 N r^4 + 3 N^2 r^8) \nonumber\\&& \hspace{0.4cm} + L^2 (4 \alpha  + 3 (r^2 + N r^6))\big],  \label{eq52}\\
&&\hspace{-0.6cm} 8\pi\theta^1_1=\frac{-\Phi\,\big[\theta_{11}(r)+\theta_{22}(r)\big]}{3 (r + L r^3 + N r^5)^2 [1 + L r^2 + N r^4 + 4 \alpha (L + M r^2)]}.~~~~~~  \label{eq53}
\end{eqnarray}
Due to long expressions for $\theta^2_2$, the expression is not written here. 
\subsection{The mimicking of $\theta$-component to seed radial pressure: $\theta^1_1=\hat{P}_r$~:} \label{sec3.2}
Now using the Eqs.(\ref{eq26}) and (\ref{eq32}) together with relation $\theta^1_1=\hat{P}_r$ provides the deformation function, 
\begin{eqnarray}
\Phi(r)=\frac{\big[4 \alpha W^\prime + W^\prime\, r^2 - 8\, \alpha\, W^\prime X + 2\, r\, W \pm \Phi_2(r)\big]}{8\,\alpha\, \beta\, W^\prime} ,  \label{eq54}
\end{eqnarray}
with
\begin{eqnarray}
&&\hspace{-0.6cm} \Phi_2(r)=\sqrt{[\alpha  W^\prime (4 - 8 X) + r ( W^\prime r + 2  W)]^2 +16\, \alpha \,\beta \, \Phi_3(r)},~~ \Phi_3(r)=  W^\prime [4 \alpha W^\prime (X-1)\,X - r ( W^\prime\,X\, r + 2\,W\, (X-1)].\nonumber
\end{eqnarray}
Now plugging the values of $X(r)$ and $W(r)$ from Eqs. (\ref{eq45}) and (\ref{eq46}), we obtain,  
\begin{eqnarray}
&&\hspace{-0.6cm} \Phi(r)=\frac{(1 + L r^2 + N r^4)}{8 \alpha\, \beta\, \Phi_4(r)\, (1 + L r^2 + N r^4)^2} \Big[ \big\{4 \alpha (L r^2 + N r^4-1)\Phi_4(r) + r \,[r\, \Phi_4(r)+2] (1 + L r^2 + N r^4)\big\} - \sqrt{\Phi_5(r)}\Big],~~~~ \label{eq55}
\end{eqnarray}
where,
\begin{eqnarray}
&&\hspace{-0.6cm}\Phi_4(r)=-{\Phi_6(r)}/{\Phi_7(r)},\nonumber\\
&&\hspace{-0.6cm} \Phi_5(r)=[4 \alpha \Phi_4 (L r^2 + N r^4-1) + r (2 + \Phi_4 r) (1 + L r^2 + N r^4)]^2  - 16\, \alpha\, \beta\, \Phi_4\,r^2\, \big[\Phi_4 (1 + L r^2 + N r^4 + 4 \alpha (L + N r^2)) \nonumber\\
&& \hspace{0.4cm} - 2 r (L + L^2 r^2 + 2 L N r^4 + N r^2 (1 + M r^4))\big],\nonumber\\
&&\hspace{-0.6cm} \Phi_6(r) = 8 N r \big[4 \mathcal{B} (1 + L r^2 + N r^4)^3 -  3 (4 L^3 r^4 + L^2 (4 \alpha + 9 r^2 + 12 N r^6) + 2 N r^2 (3 + 4 \alpha N r^2 + 5 N r^4 + 2 N^2 r^8) \nonumber\\
&& \hspace{0.4cm} + 
 L (5 + 12 \alpha N r^2 + 19 N r^4 + 12 N^2 r^8))\big],\nonumber \\
&&\hspace{-0.6cm} \Phi_7(r) =36 N [1 + L r^2 + N r^4 + 4 \alpha (L + N r^2)]\, (1 + L r^2 + N r^4).\nonumber
\end{eqnarray}

We avoid writing expressions for the $\theta$-components due to their length. Now the thermodynamical quantities for energy momentum tensor ($T_{\mu\nu}$) such as $\epsilon$, $P_r$, and $P_t$ are,
\begin{eqnarray}
\epsilon=\hat{\epsilon}+\beta\,\theta^0_0,~~~P_r=\hat{P}_r-\beta\,\theta^1_1,~~~~\text{and}~~~~ P_t=\hat{P}_t-\beta\,\theta^2_2 .  \label{eq56}
\end{eqnarray}
Then the conservation equation and mass $m(r)$ can be cast as, 
\begin{eqnarray}
&&\hspace{-0.4cm} -\frac{W^\prime}{2W}\,(\hat{\epsilon}+\hat{P}_r)-\hat{P}_r^\prime+\frac{3}{r}(\hat{P}_t-\hat{P}_r)+\beta\,\Big[-\frac{W^\prime}{2W}\,(\theta^0_0-\theta^1_1) +(\theta^1_1)^\prime+\frac{3}{r}(\theta^1_1-\theta^2_2)\Big]=0,  \label{eq57}\\
&&\hspace{-0.4cm} m(r)= \frac{8\,\pi}{3} \int_0^r \epsilon(x)\,x^3 dx=\underbrace{\frac{8\,\pi}{3} \int_0^r \hat{\epsilon}(x)\,x^3 dx}_{m_s}  +\underbrace{\frac{8\,\pi}{3} \beta \int_0^r \theta^0_0(x)\,x^3 dx}_{m_{\Phi}}.  \label{eq58}
\end{eqnarray}
where $m_s$ and $m_{\phi}$ are mass functions in pure EGB gravity and extra source given, respectively (see equations (\ref{eq30}) and (\ref{eq35})). 
\section{Exterior space--time and matching conditions} \label{sec4}
For defining the boundary conditions of the gravitationally decoupled solution, it is necessary to find a suitable  exterior space--time for matching of the internal manifold $\mathcal{M}^{-}$ with the exterior manifold $\mathcal{M}^{+}$ at the boundary $r=R$. The suitable exterior spacetime can be given by Boulware--Deser \citep{boul} exterior (vacuum) solution as, 
\begin{eqnarray}
&& \hspace{-0.8cm}ds^2_{5} = -\Bigg[1+\frac{r^2}{4 \alpha}
\left(1-\sqrt{1+\frac{16\alpha M}{r^4}}\right)\Bigg] dt^2 + \Bigg[1+\frac{r^2}{4 \alpha}
\Big(1-\sqrt{1+\frac{16\alpha M}{r^4}}\Big)\Bigg]^{-1}{dr^2} + r^{2}d\Omega^2_3, \label{eq59}
\end{eqnarray}
where $M$ is associated with the total gravitational mass of the object  and $d\Omega^2_3=\big(d \theta^2 +\sin^2 \theta ~d\phi^2+\sin^2 \theta \sin^2 \phi ~d\psi^2 \big)$. 
It can be observed that limit $\alpha \to 0$ reduces the above metric to the $5D$ Schwarzschild solution. Moreover, the inclusion of the new source $\theta_{\mu\nu}$ into the matter distribution could in principle change matter content and it geometry for the exterior space--time and then the stellar compact object will be not immersed in vacuum space--time anymore. In this situation, if we assume that the contributions for the new source $\theta_{\mu\nu}$ are confined within the stellar interior only \citep{Ovalle2018}, then the compact object will remain embedded into a vacuum space--time (\ref{eq59}). On the other hand, the deformed interior spacetime is given by the following line element,
\begin{eqnarray}\label{exterior}
&& \hspace{-0.8cm}ds^2_{5} = -W(R) ~dt^2 + \frac{dr^2}{X(r)+\beta\,\Phi(r)} + r^{2}d\Omega^2_3, \label{eq60}
\end{eqnarray}
Now by matching of the exterior (\ref{eq59}) and interior (\ref{eq60}) spacetimes across the boundary, one can find the suitably arbitrary constant parameters. The subsequent manifolds have boundaries given by the time--like hyper--surfaces 
\begin{equation}
ds^{2}_{\Sigma} = - d\tau^2+R^{2}\big(d \theta^2+\sin^2 \theta~ d\phi^2+\sin^2 \theta \sin^2 \phi ~d\psi^2 \big),  \label{eq61}
\end{equation}
with the intrinsic coordinates of $\Sigma$ being
$\xi^{\mu}$ = $(\tau, \theta, \phi, \psi)$ in $\Sigma$, and $\tau$ is the proper time on the boundary.
Now, take the field equations projected on the shell $\Sigma$ 
[generalized Darmois--Israel \citep{darmois,israel} formalism for Einstein--Gauss--Bonnet theory] are  (see Refs. \cite{davis,gravanis} for more)
\begingroup
\small
\begin{eqnarray}  
2\langle K_{\mu\nu}-K h_{\mu\nu}\rangle + 4\alpha \langle 3J_{\mu\nu}-Jh_{\mu\nu}+2P_{iklj}K^{kl}\rangle =-\kappa^{2}S_{\mu\nu}, \label{eq62}
\end{eqnarray}
\endgroup
where the $\langle \cdot\rangle$ is the jump of a given quantity across the hyper--surface $\Sigma$. Here $h_{\mu\nu} = g_{\mu\nu} - n_{\nu}n_{\mu}$ is the induced metric on $\Sigma$ with the divergence free part of the Riemann tensor is defined by
\begin{eqnarray}
&& \hspace{-0.7cm} \mathcal{R}_{\mu\nu kl} = \mathcal{R}_{\mu\nu kl}+(\mathcal{R}_{\nu k}h_{l\mu}- \mathcal{R}_{\nu l}h_{k\mu})-(\mathcal{R}_{\mu k}h_{l\nu}- \mathcal{R}_{\mu l}h_{k\nu})  +\frac{1}{2}\mathcal{R}(h_{\mu k}h_{l\nu}-h_{\mu l}h_{k\nu}), \label{eq63}
\end{eqnarray}
and $J$ is the trace of 
\begin{small}
\begin{eqnarray}
&& \hspace{-0.9cm} J_{\mu\nu}=\frac{1}{3}\left[2KK_{\mu k}K^{k}_{\nu}+K_{kl}K^{kl}K_{\mu\nu}-2K_{\mu k}K^{kl}K_{l\nu}-K^{2}K_{\mu\nu}\right].~~~~~~ \label{eq64}
\end{eqnarray}
\end{small}

Hence, the form of the extrinsic curvature in the present case is given as, 
\begin{equation}
{K}^{\pm}_{\mu\nu}=-n^{\pm}_{\mu}\left(\frac{\partial^{2}X^{\mu}}{\partial\xi^{\mu}\partial\xi^{\nu}}+\Gamma^{\mu}_{\alpha \beta}\frac{\partial X^{\alpha}}{\partial\xi^{\mu}}\frac{\partial X^{\beta}}{\partial\xi^{\nu}}\right)_{r = R}, \label{eq65}
\end{equation}
where $\xi^{\mu}$ denote the intrinsic coordinates at the surface and the sign $\pm$ depends on the signature of the junction hyper--surface. \\
Now we investigate the junction by matching of the exterior and inner metrics at the surface $r=R$. From smooth matching, we require the continuity of the first and second fundamental forms at the boundary. Then from the first fundamental form at the boundary implies that $g_{tt}^{-} = g_{tt}^{+}$ and $g_{rr}^{-} = g_{rr}^{+}$, which gives
\begin{small}
\begin{eqnarray}
W^{-}(r)|_{r=R}=W^{+}(r)|_{r=R} \quad \mbox{and} \quad   H^{-}(r)|_{r=R}=H^{+}(r)|_{r=R}, \label{eq66}
\end{eqnarray}
\end{small}
which yields
\begin{eqnarray}
&&\hspace{-0.4cm} \frac{1}{H(R)} = X(R)+\beta \Phi(R) =\bigg[1 + \frac{R^2}{4 \alpha}\bigg(1 - \sqrt{1 + \frac{16\, \alpha M}{R^4}}\bigg)\bigg],~~\text{and}~~~
W(R)=\bigg[1 + \frac{R^2}{4\,\alpha}\bigg(1 - \sqrt{1 + \frac{16\, \alpha M}{R^4}}\bigg)\bigg].\label{eq67}
\end{eqnarray}
where $X(R) = \big[1 + \frac{R^2}{4\,\alpha}\big(1 - \sqrt{1 + \frac{16\, \alpha\, M_{EGB}}{R^4}}\big)\big]$ with $M_{EGB} = m_{s}(R)$ is the mass of the compact object with radius $R$ in pure EGB gravity for the metric (\ref{eq29}) i.e., for the non--deformed space--time. Then, using (\ref{eq58}), we have
\begin{eqnarray}\label{effectivemass}
M=M_{EGB}+ \frac{1}{2}\beta  \Phi(R) \left[2\alpha\beta \Phi(R)-\sqrt{R^4 +16 \alpha M_{EGB}}\right].~ \label{eq68}
\end{eqnarray}
Now the continuity of second fundamental form or the extrinsic curvature at the boundary leads to, 
\begin{eqnarray}
\big[(G_{\mu\nu}+\alpha\,H_{\mu\nu})\,r^{\nu}\big]_{\Sigma}=0 , \,\label{eq69}
\end{eqnarray}
where $r^{\nu}$ is a unit radial vector. Now, the following condition (\ref{eq69}) together with (\ref{eq3}) leads 
\begin{eqnarray}
\big[T_{\mu\nu}\,r^{\nu}\big]_{\Sigma}=0\, ,\label{eq70}
\end{eqnarray}
which gives,
\begin{eqnarray}
\big[P_r\big]_{\Sigma}=0~~\Longrightarrow~~\,\big[\hat{P}_r-\beta\,\theta^1_1\big]_{\Sigma}=0,\label{eq71}
\end{eqnarray}
where the $\Sigma$ denotes the surface which is defined at $r=R$. Condition (\ref{eq71}) determines the size of the compact objects i.e. radius ($R$). 
The reason for this behaviour is that it is expected for closed compact objects that the pressure decreases radially outwards and vanishes across a suitable boundary. At this junction we seek to match the interior and exterior spacetimes. Essentially this follows from the continuity of the second fundamental form as discussed by \cite{davis} and \cite{gravanis} in the context of EGB theory. Expressed otherwise, it may be said that the  matter distribution is confined in a finite space--time region. Consequently the star does not expand indefinitely beyond the boundary  $\Sigma$. This matching condition then may be expressed as
\begin{eqnarray}
\hat{P}_r(R)-\beta\,(\theta^1_1)^{-}(R)=-\beta\,(\theta^1_1)^{+}(R), \label{eq72}
\end{eqnarray}
where $(\theta^1_1)^{-}(R)$ and $(\theta^1_1)^{+}(R)$ represent the $\theta$--components for interior and exterior space--times at $r=R$, respectively. The condition (\ref{eq72}) is called a general expression for the second fundamental form for deformed spacetime  associated with the equation of motion for EGB gravity that is given by Eq. (\ref{eq3}).\\
Now, we determine $\theta^1_1$ from (\ref{eq32}) and plugged into the Eq. (\ref{eq72}), we get a modified form the second fundamental form as, 
\begin{eqnarray}
&&\hspace{-0.8cm} \hat{P}_r(R)+\frac{3 \beta\,\Phi_{\Sigma} [4 \,\alpha\, W_{\Sigma}^\prime\, (1 - \beta\, \Phi_{\Sigma} -2 X_{\Sigma}) + R\, (W_{\Sigma}^\prime\, R + 2 W_{\Sigma})]}{16 \pi\,R^3 W_{\Sigma}}=-\beta\,(\theta^1_1)^{+}(R),~~ \label{eq73}
\end{eqnarray}
where the notations are $\Phi_{\Sigma} = \Phi(R)$, $X_{\Sigma}=X(R)$, and $W^\prime_{\Sigma}={\partial_r W}\big|_{r=R}$, respectively. 
Furthermore, using the Eq. (\ref{eq32}) for
the exterior geometry in Eq. (\ref{eq73}), which  yield
\begin{eqnarray}
&&\hspace{-0.8cm} \hat{P}_r(R)+\frac{3 \beta\,\Phi_{\Sigma} [4 \,\alpha\, W_{\Sigma}^\prime\, (1 - \beta\, \Phi_{\Sigma} -2 X_{\Sigma}) + R\, (W_{\Sigma}^\prime\, R + 2 W_{\Sigma})]}{16 \pi\,R^3 W_{\Sigma}} =\frac{3\,\beta\,\Phi^{\ast}_{\Sigma}}{8\,\pi\,R^3} \Bigg[ \frac{R\, \bigg(\sqrt{1 + \frac{16\, \alpha\, \mathcal{M}}{R^4}}-1\bigg)}{\sqrt{1 + \frac{16 \,\alpha\, \mathcal{M}}{R^4}} \bigg[4 \,\alpha +R^2\, \bigg(1 - \sqrt{1 + \frac{16\,\alpha\, \mathcal{M}}{R^4}}\bigg) \bigg]}\nonumber\\
&&\hspace{0.2cm} \times \bigg\{-4\,\alpha\,\beta\, \Phi^{\ast}_{\Sigma} -8\,\alpha\,\bigg[1 + \frac{R^2}{4\,\alpha}\bigg(1 - \sqrt{1 + \frac{16\, \alpha\, \mathcal{M}}{R^4}}\bigg)\bigg]+4\,\alpha+R^2\bigg\}+R\Bigg], \label{eq74}
\end{eqnarray}
where $\Phi^{\ast}_{\Sigma}=\Phi^{\ast}(R)$ is a decoupling function for the exterior space--time at $r=R$ in the presence of the source $\theta_{\mu\nu}$, can be determined by the
following $5D$ spacetime as
\begin{eqnarray}
ds^2_{5} &=& - \bigg[1 + \frac{r^2}{4\,\alpha}\bigg(1 - \sqrt{1 + \frac{16\, \alpha\, \mathcal{M}}{r^4}}\bigg)\bigg] dt^2 + \bigg[1 + \frac{r^2}{4\,\alpha}\bigg(1 - \sqrt{1 + \frac{16\, \alpha\, \mathcal{M}}{r^4}}\bigg)+\beta\,\Phi^{\ast}(r)\bigg]^{-1} dr^2 \nonumber\\
&&+ r^{2}\left(d \theta^2+\sin^2 \theta~ d\phi^2+\sin^2 \theta~ \sin^2 \phi ~d\psi^2 \right), \label{eq75}
\end{eqnarray}
The conditions (\ref{eq67}) and (\ref{eq74}) are the necessary and sufficient conditions for matching the interior deformed metric (\ref{eq60}) to the exterior ``vacuum" static and spherically symmetric space--times given in (\ref{eq59}). Since here we considered that the contributions for the new source $\theta_{\mu\nu}$ are confined within the stellar interior only and the exterior geometry is given by the exact Boulware--Deser solution. Then we must substitute $\Phi^{\ast}_{\Sigma}=0$ in Eq. (\ref{eq75}). Hence, we find the final condition by substituting $\Phi^{\ast}_{\Sigma}=0$ in equation (\ref{eq74}) as,
\begin{eqnarray}
&&\hspace{-0.6cm} P_r(R) =  \hat{P}_r(R)+ \frac{1}{16 \pi\,r^3 W} \big[3 \beta\,\Phi\, \big\{4 \,\alpha\, W^\prime\, (1 - \beta\, \Phi -2 X)  + r (W^\prime\, r + 2 W)\big\}\big] =0,~ \label{eq76}\\
&&\hspace{-0.6cm} \text{which~can~be~written~as,}\nonumber\\
 &&\hspace{-0.6cm} P_r(R)=\hat{P}_r(R)-\beta\,\theta^1_1(R) = 0.~~ \label{eq77} 
\end{eqnarray}
We conclude that the star will be in equilibrium in a true  (Boulware--Deser) vacuum only if the radial pressure ($P_r$) vanishes at the surface of the star.
\section{Physical Properties of Strange Stars} \label{sec5}
\subsection{Regular behavior of the Models} \label{sec5.1}
\begin{figure*}
    \centering
    \includegraphics[width=8cm,height=6.2cm]{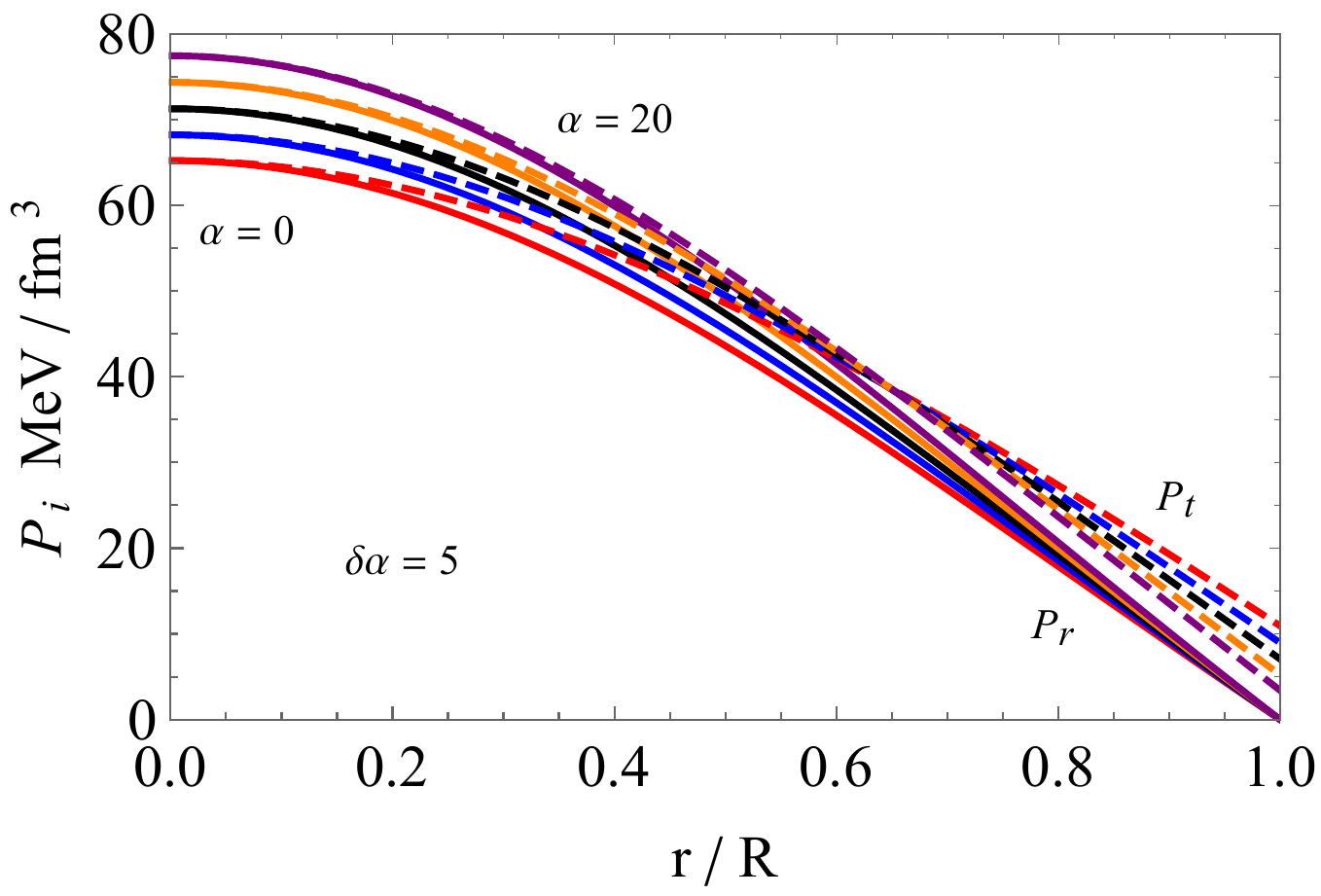}~~~ \includegraphics[width=8cm,height=6.2cm]{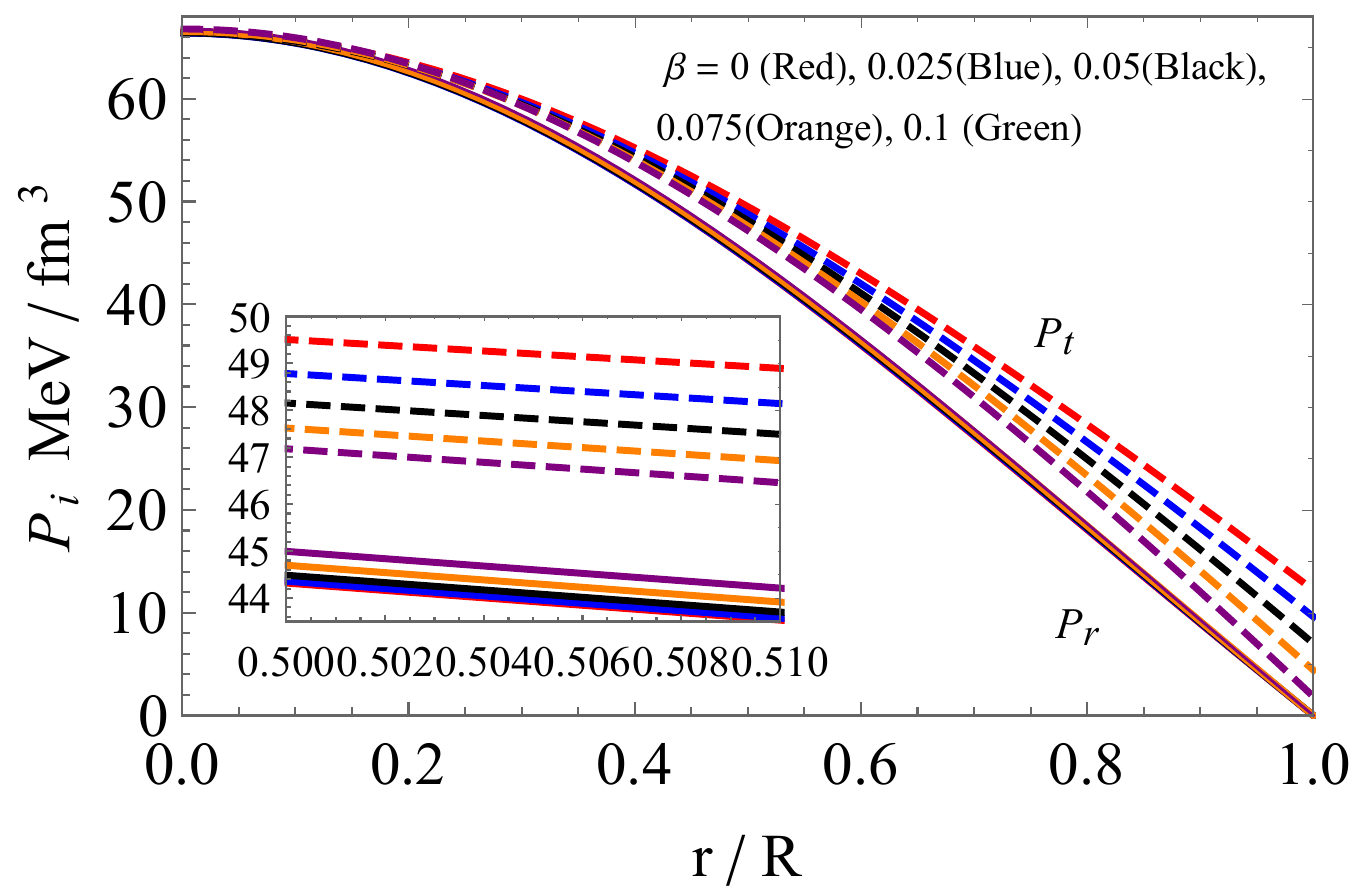}\\
     \includegraphics[width=8cm,height=6.2cm]{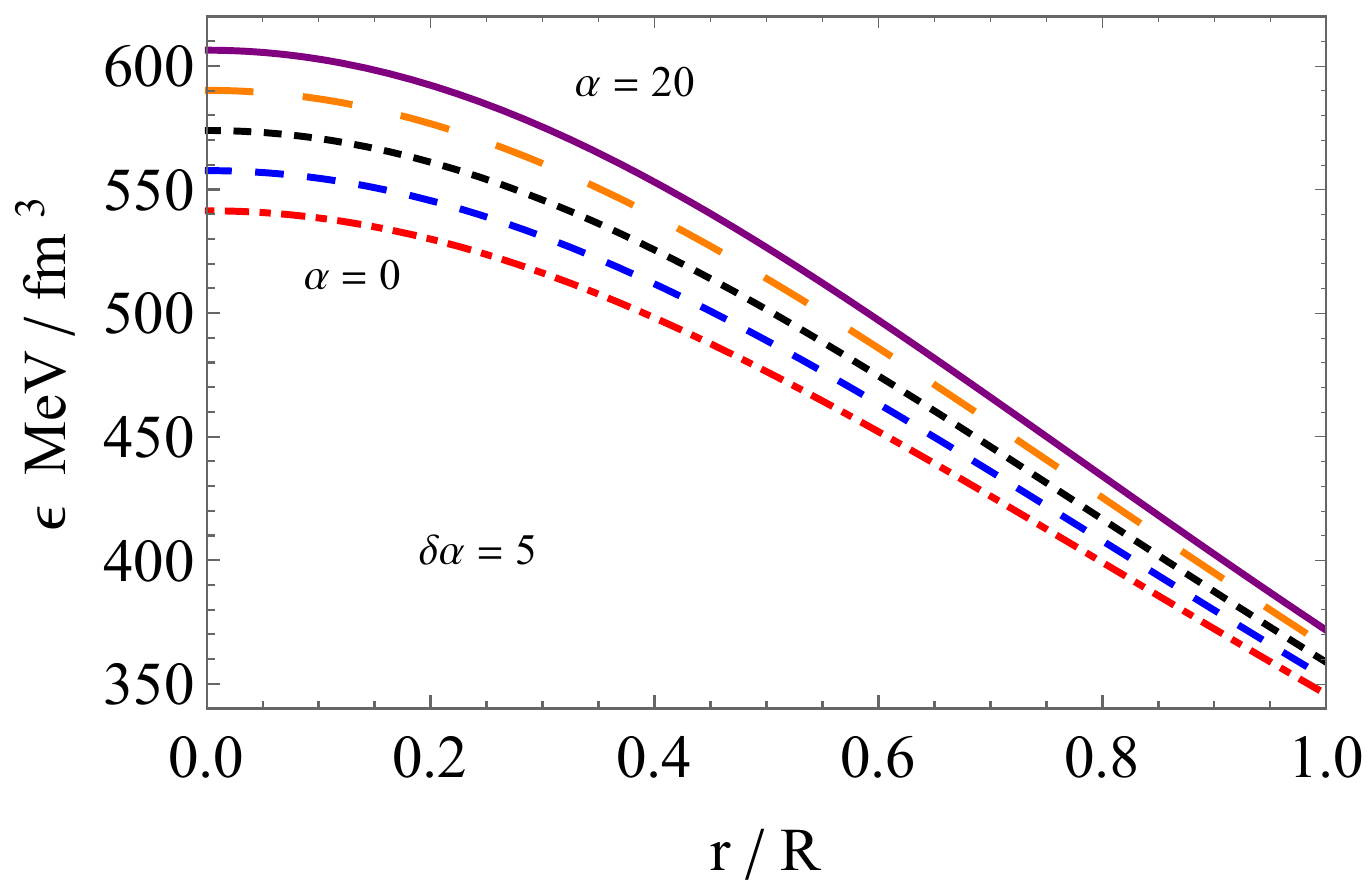}~~~  \includegraphics[width=8cm,height=6.2cm]{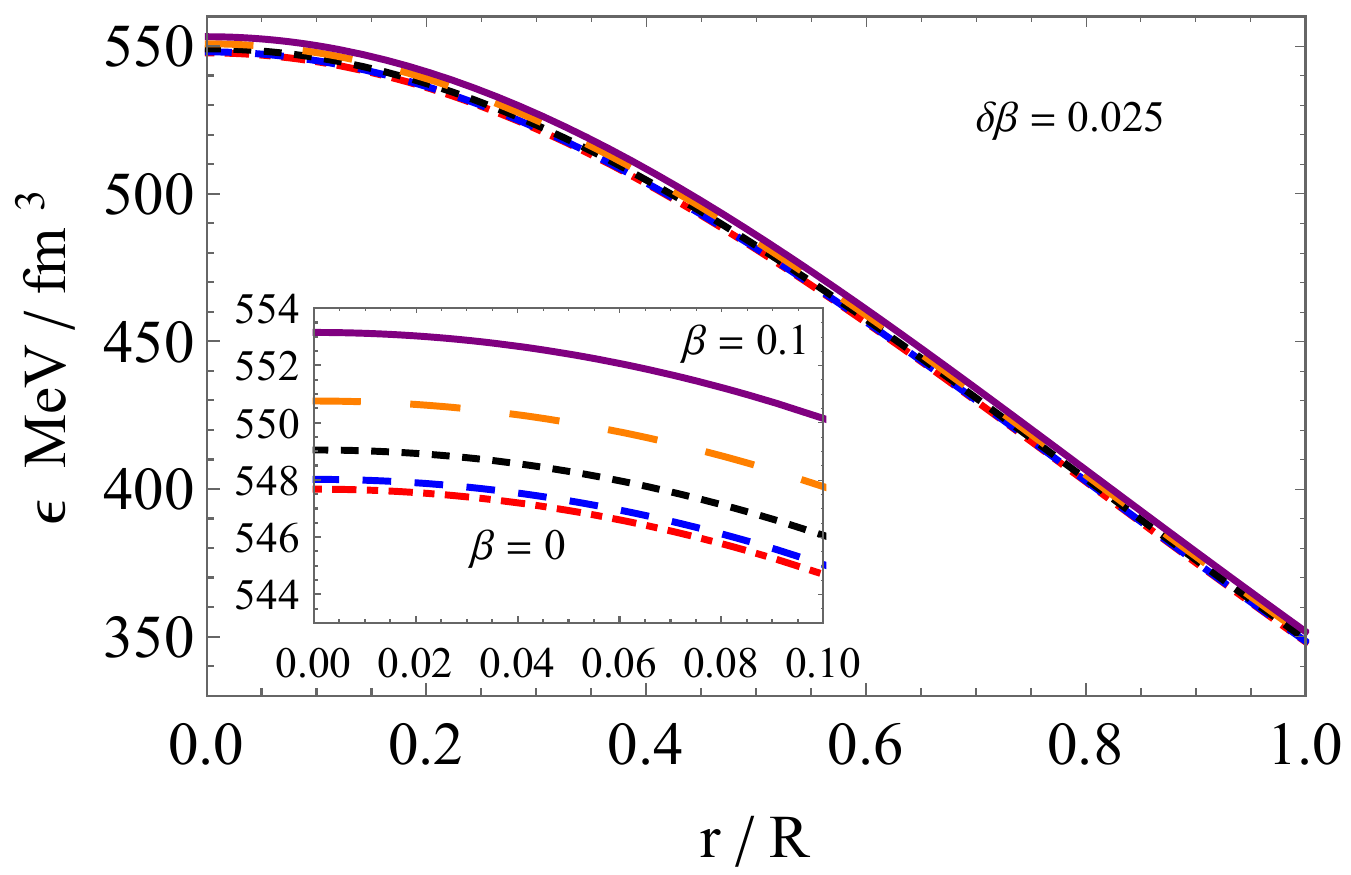}
    \caption{\textit{Top panels} and \textit{Bottom panels} show the  pressures [ radial ($P_r$) and tangential ($P_{t}$)] energy density with respect to $r/R$ for different $\alpha$ and  $\beta$, respectively for the $\theta^0_0=\hat{\epsilon}$ solution.
 We set the numerical values $~L = 0.003/km^2,~N = 10^{-7}/km^4,~
R = 11\,km$ for plotting of left panels and right panel when $\beta=0.02$ and $\alpha=2\, km^2$, respectively.}
    \label{fig2}
\end{figure*}

In this section we provide a physical analysis of our model with the emphasis on regularity and stability  connected to EGB contributions and anisotropisation via decoupling. The behaviour of the radial and transverse pressures are displayed in Fig. \ref{fig2}.
In the top left panel, we observe the effect of varying the EGB coupling constant in increments of 5 units $(0 \leq \alpha \leq 20)$ while the decoupling parameter, $\beta$ is fixed. We observe that both the radial and tangential pressures are monotonically decreasing functions of the scaled radial coordinate, $\frac{r}{R}$. The radial pressure vanishes for some finite radius which demarcates the boundary between the interior spacetime and the vacuum exterior described by the Boulware-Deser solution. An increase in the EGB coupling constant is accompanied by an increase in both the radial and tangential pressures respectively. It is interesting to note that for some interior point within the fluid configuration, the effect due to EGB contributions to the tangential pressure changes. We observe that from the center $(r = 0)$ to some finite radius $(r = r_0)$, an increase in $\alpha$ results in an increase in the tangential pressure. However, for $r > r_0$ an increase in $\alpha$ results in a decrease in the tangential pressure, with $P_t > P_r$ in this region. In Fig.\ref{fig2}, top right panel, we fix the EGB coupling constant and vary the decoupling parameter, $\beta$. The effect of varying $\beta$ on the radial pressure is minimal. There is certainly a noticeable impact of the variation of $\beta$ on the tangential pressure, particularly towards the surface layers. An increase in the decoupling parameter suppresses the tangential pressure, with this effect being amplified as one moves closer to the boundary layers of the star. \\
\begin{figure*}
    \centering
    \includegraphics[width=8cm,height=6.3cm]{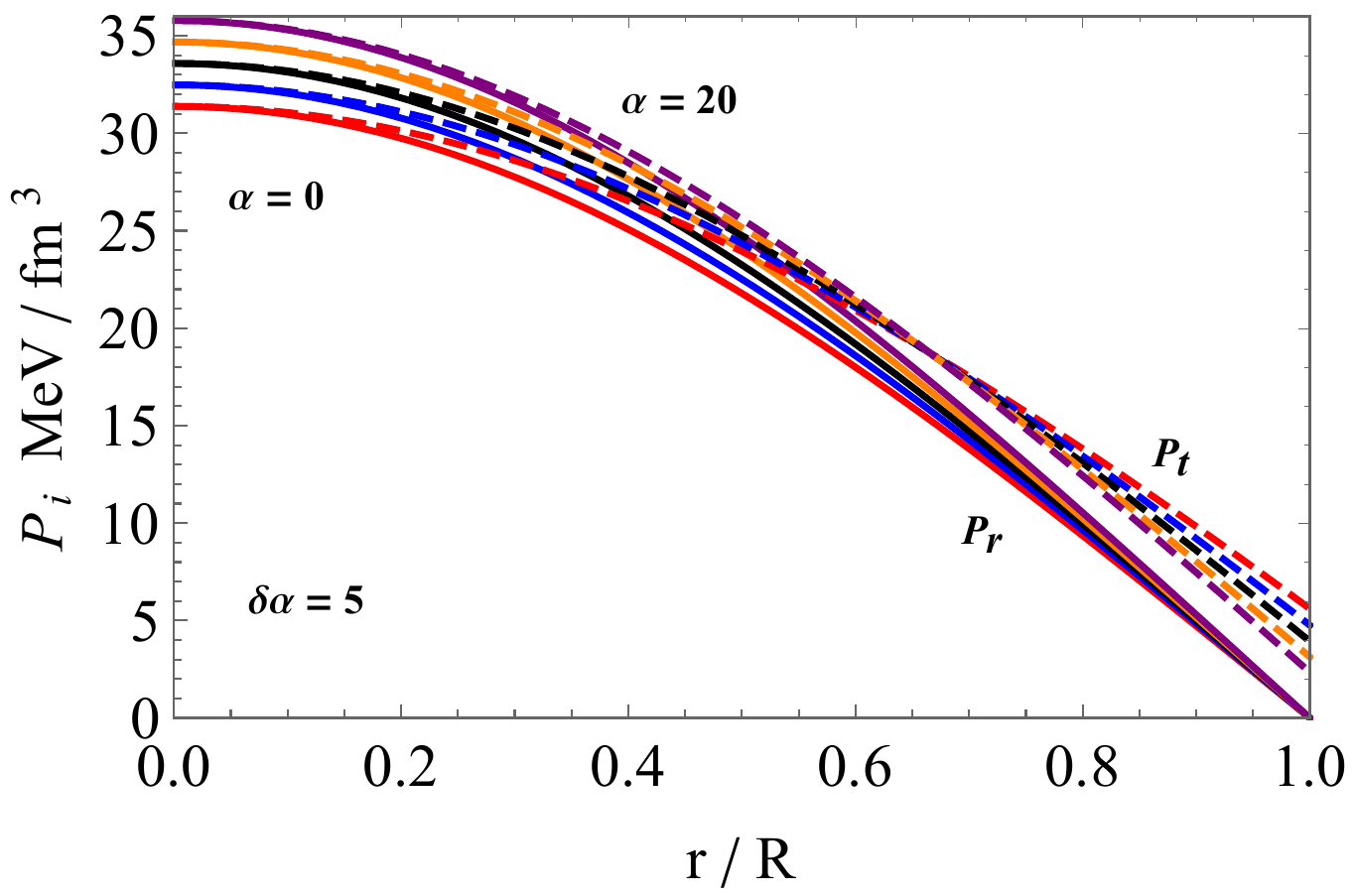}~~~ \includegraphics[width=8cm,height=6.3cm]{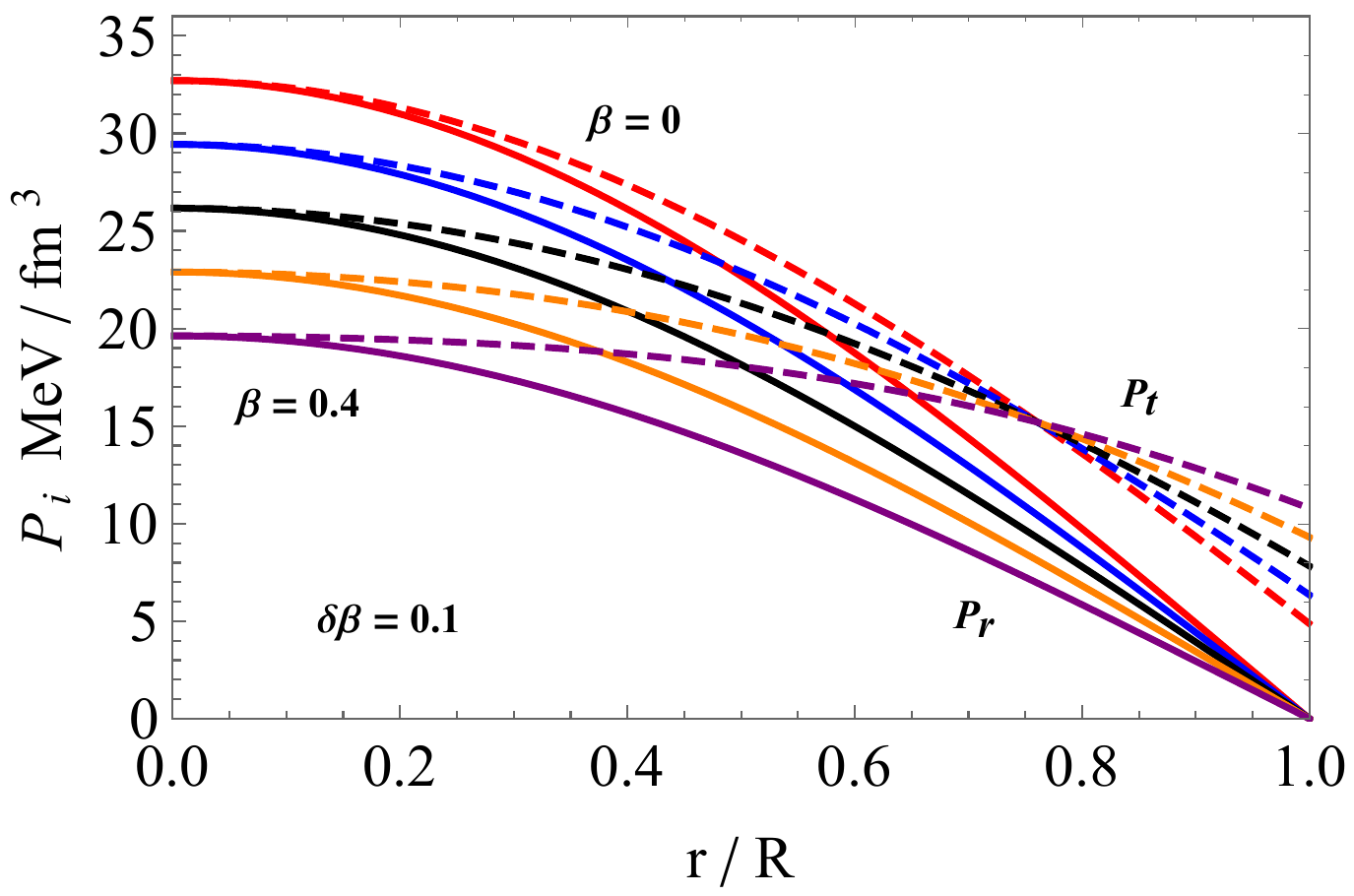}\\
     \includegraphics[width=8cm,height=6.3cm]{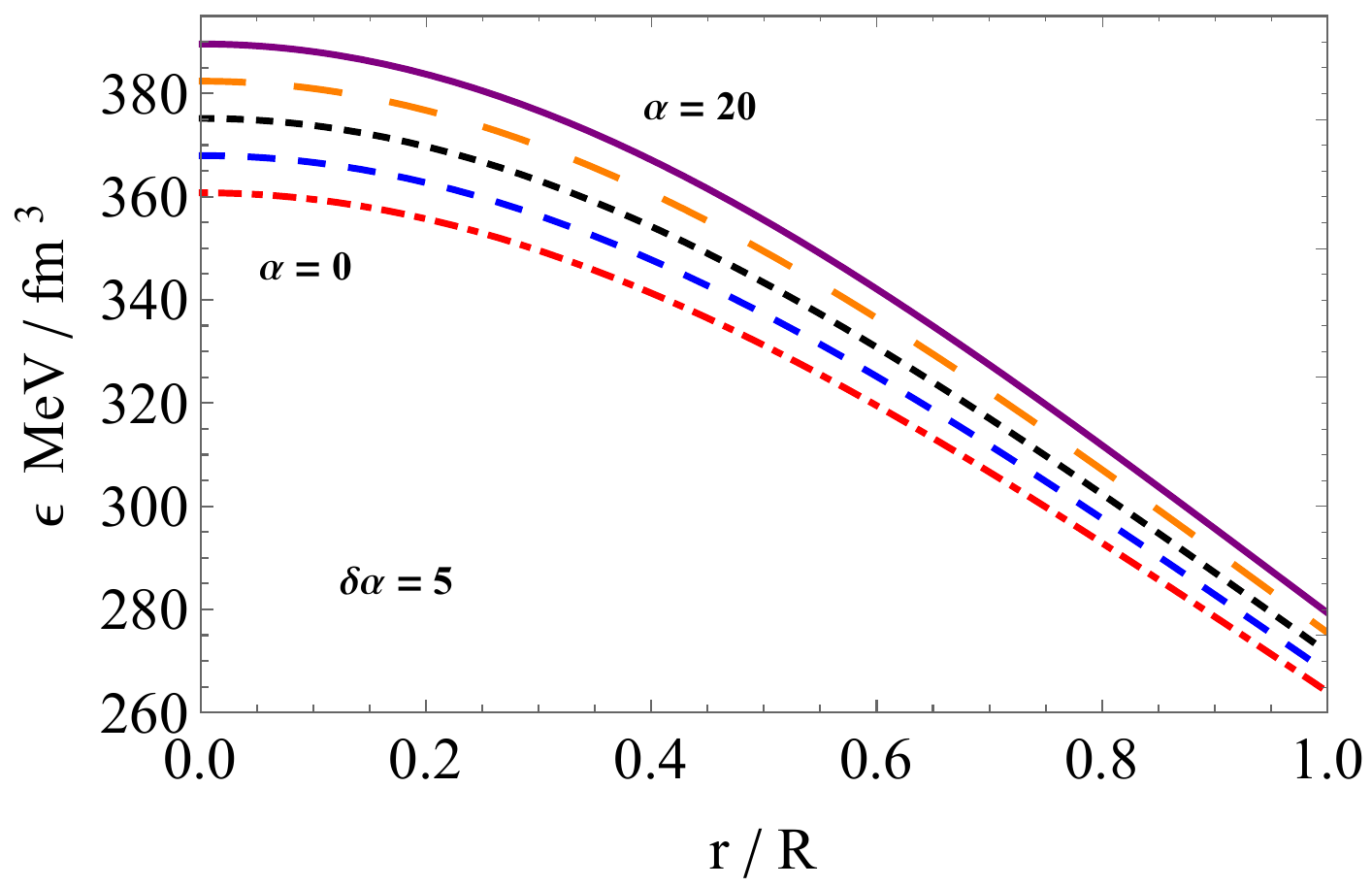}~~~  \includegraphics[width=8cm,height=6.3cm]{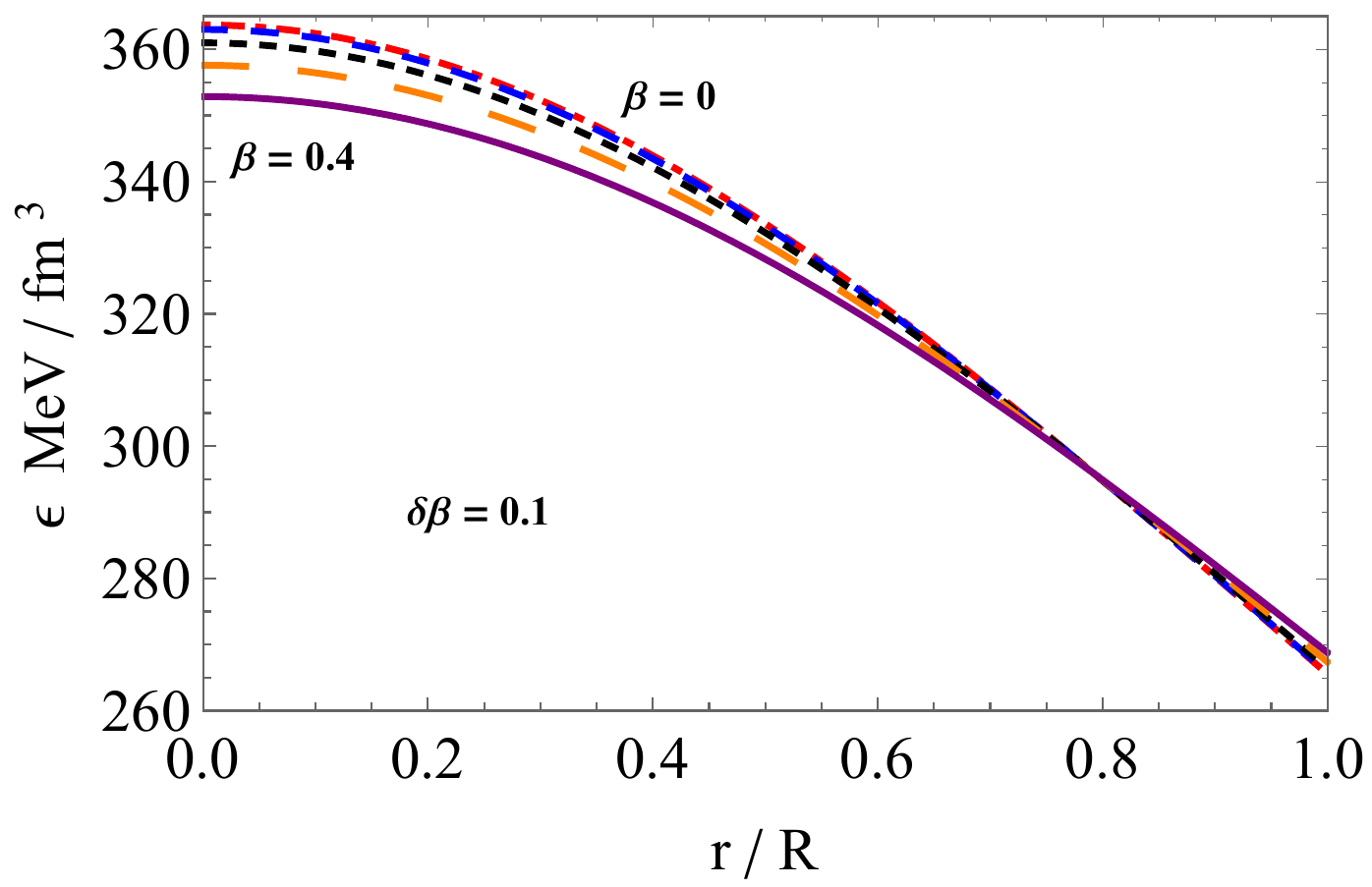}
    \caption{\textit{Top panels} and \textit{Bottom panels} show the effective pressures [ radial ($P_r$) and tangential ($P_{t}$)] effective energy density with respect to $r/R$ for different $\alpha$ and  $\beta$, respectively for the $\theta^1_1=\hat{P}_r$ solution.  We set the numerical values $~L = 0.003/km^2,~N = 10^{-7}/km^4,~
R = 11\,km$ for plotting of left panels and right panels when $\beta=0.02$ and $\alpha=2\, km^2$, respectively.}
    \label{fig3}
\end{figure*}
The energy density as a function of the scaled radial coordinate is plotted in the lower panels of Fig.\ref{fig2}. On the left panel, we observe the variation of the density profile with an increase in the EGB coupling constant. As $\alpha$ is increased, the density increases. This is expected as the radial pressure and density are connected via the linear EoS. The bottom left panel shows that an increase in the decoupling parameter decreases the density throughout the stellar configuration. \\
In Fig. \ref{fig3}, we present the behaviour of the pressure and density profiles for the mimicking of $\theta$-component to seed radial pressure. The top left panel reveals the behaviour of the radial and tangential pressures for varying $\alpha$ and fixed $\beta$. Both the radial and tangential pressures drop off smoothly towards the boundary. We observe that the magnitudes of $P_r$ and $P_t$ are lower than those in the  $\theta^0_0 =\hat{\epsilon}$ model. An increase in $\alpha$ is accompanied by an increase in the corresponding pressure. We again observe the switch over of the tangential pressure at an interior point within the fluid configuration. The top right panel depicts the behaviour of the pressure profiles for varying $\beta$ and constant EGB coupling constant. We again observe the suppressive nature of the decoupling constant. The magnitude of the radial pressure decreases as $\beta$ is increased. A similar trend can be seen in the tangential pressure up to some fixed radius. In the lower left panel, the density profile is a decreasing function of the scaled radial coordinate. An increase in the magnitude of the EGB coupling constant leads to higher energy densities within the core. As with the radial pressure, the magnitude of the energy density is lower than their counterparts in the $\theta^0_0 =\hat{\epsilon}$ model.\\
\begin{figure*}
    \centering
    \includegraphics[width=8cm,height=6.3cm]{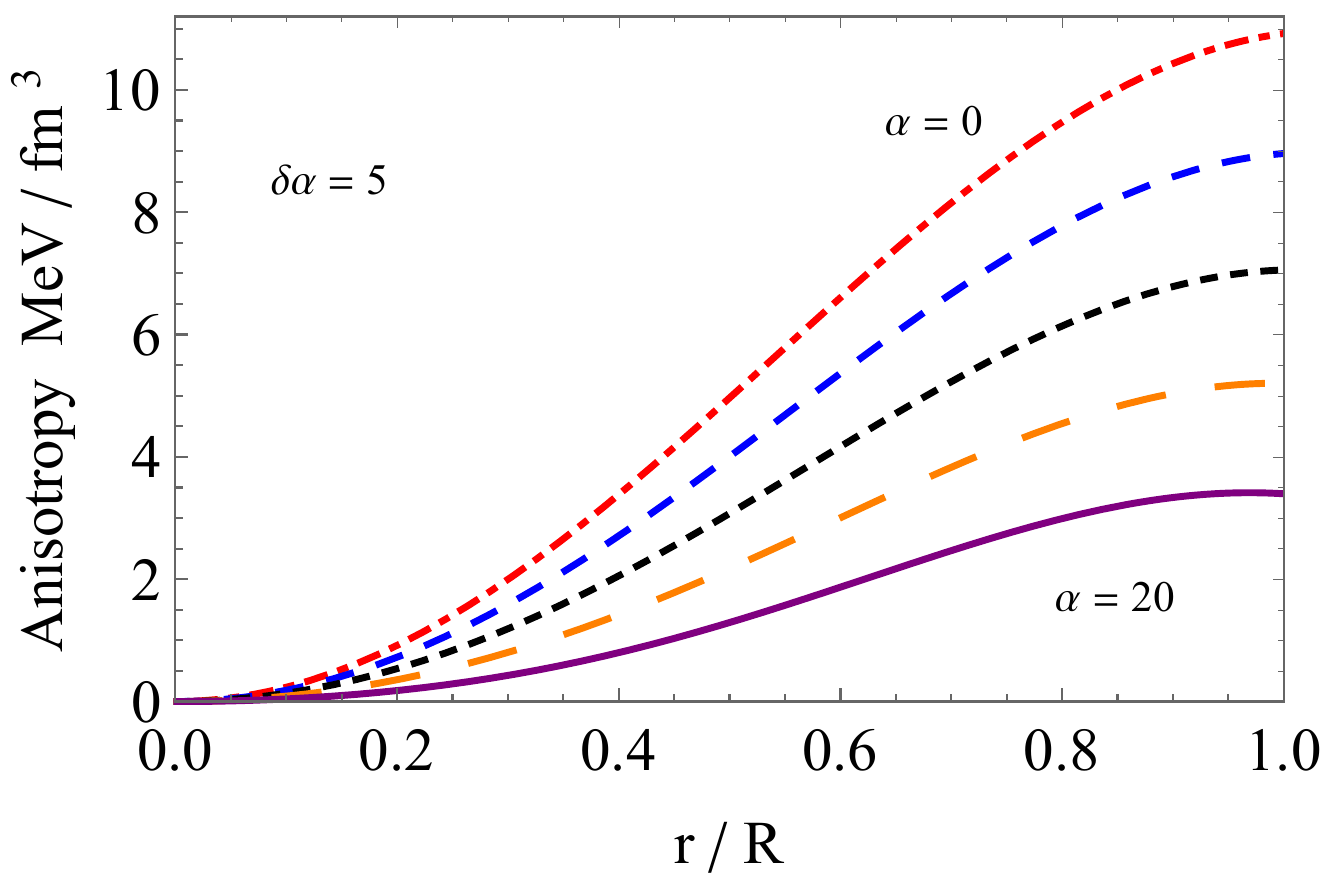} ~~~ \includegraphics[width=8cm,height=6.3cm]{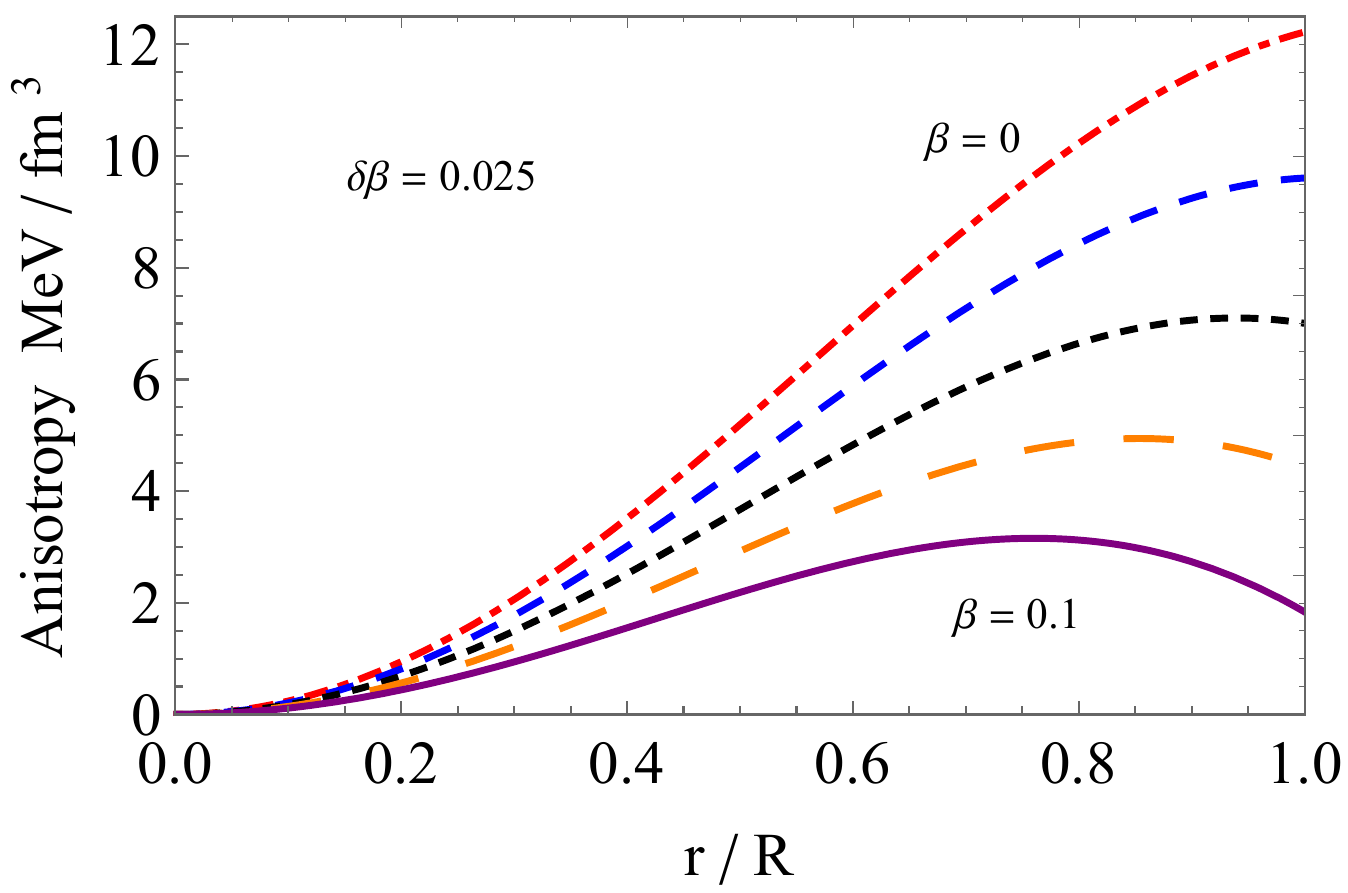}
     \includegraphics[width=8cm,height=6.3cm]{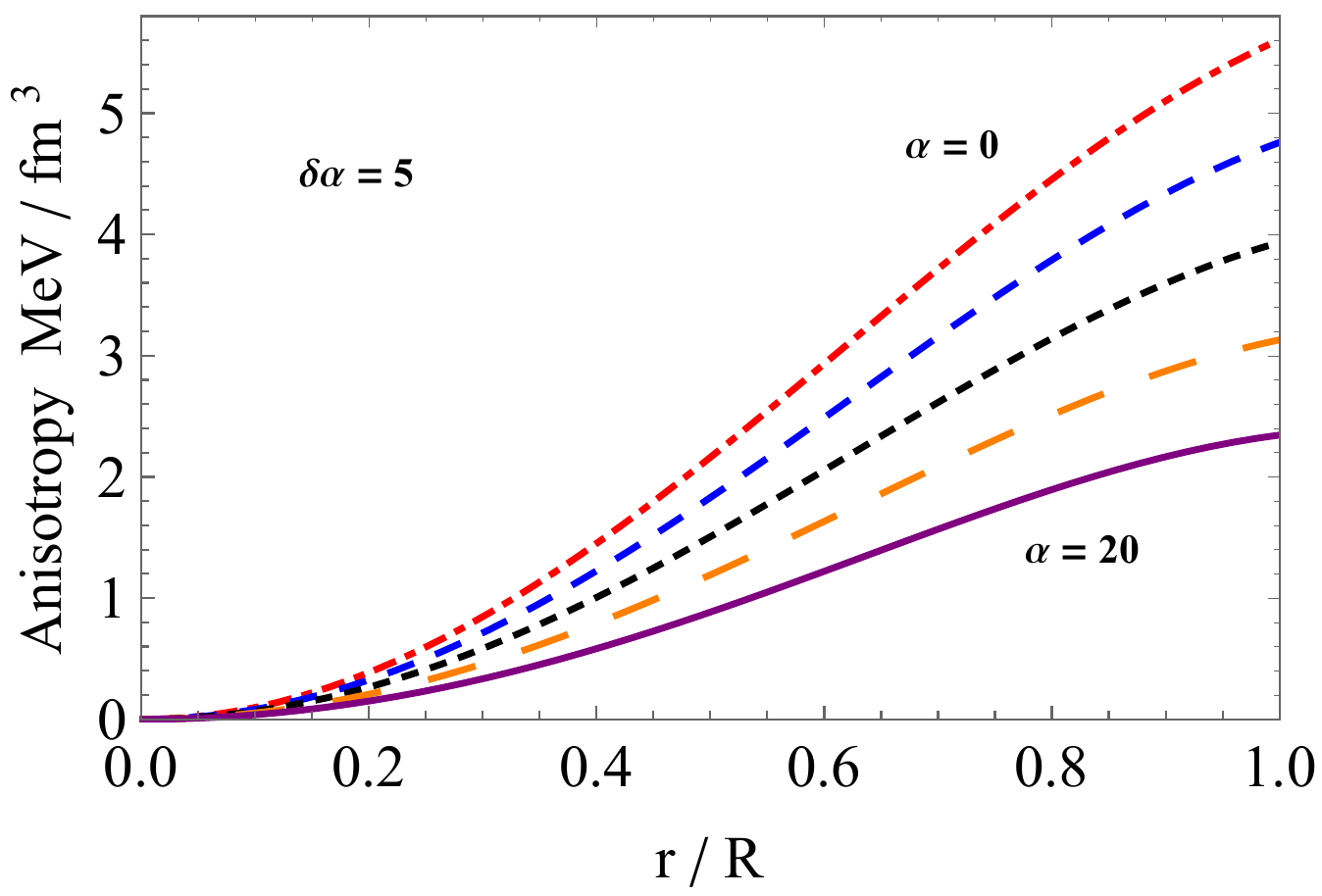} ~~~ \includegraphics[width=8cm,height=6.3cm]{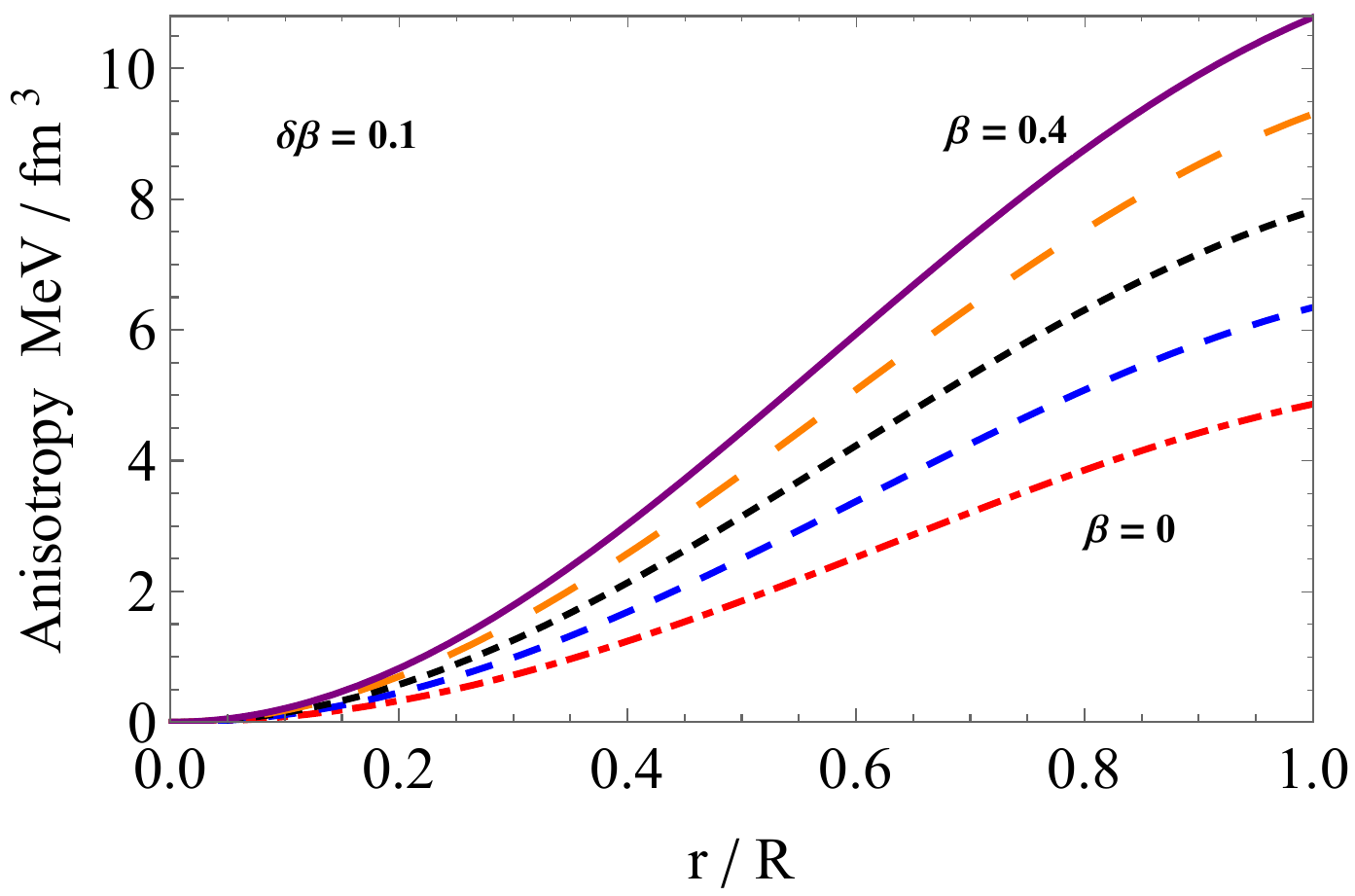}
    \caption{\textit{Top panels} (for solution $\theta^0_0=\hat{\epsilon}$) and \textit{Bottom panels}  (for solution $\theta^1_1=\hat{P}_r$) show the effective anisotropy ($\Delta$) with respect to $r/R$ for different $\alpha$ and  $\beta$. We set same numerical values as used in Fig.\ref{fig1} }
    \label{fig4}
\end{figure*}
Fig. \ref{fig4} shows the variation of the anisotropy parameter for the two sectors. We observe that the anisotropy, $\Delta = P_t > P_r$ is positive at each interior point of the stellar configuration. The domination of the tangential pressure  over the radial stress gives rise to a repulsive force due to anisotropy. This repulsive force helps stabilise the star against the inwardly driven gravitational force. We also note that the degree of anisotropy can be controlled by the EGB coupling constant and the decoupling constant. Increasing one while keeping the other fixed results in a decrease in the amount of anisotropy present within the gravitating body, one exception being in the $\theta^0_0 =\hat{\epsilon}$ model. In this case, fixing $\alpha$ and increasing $\beta$ is accompanied by an increase in anisotropy. What we do observe overall is that the anisotropy is greatest as one approach the surface layers of the compact object.
\begin{figure*}
    \centering
    \includegraphics[width=8.5cm,height=6.3cm]{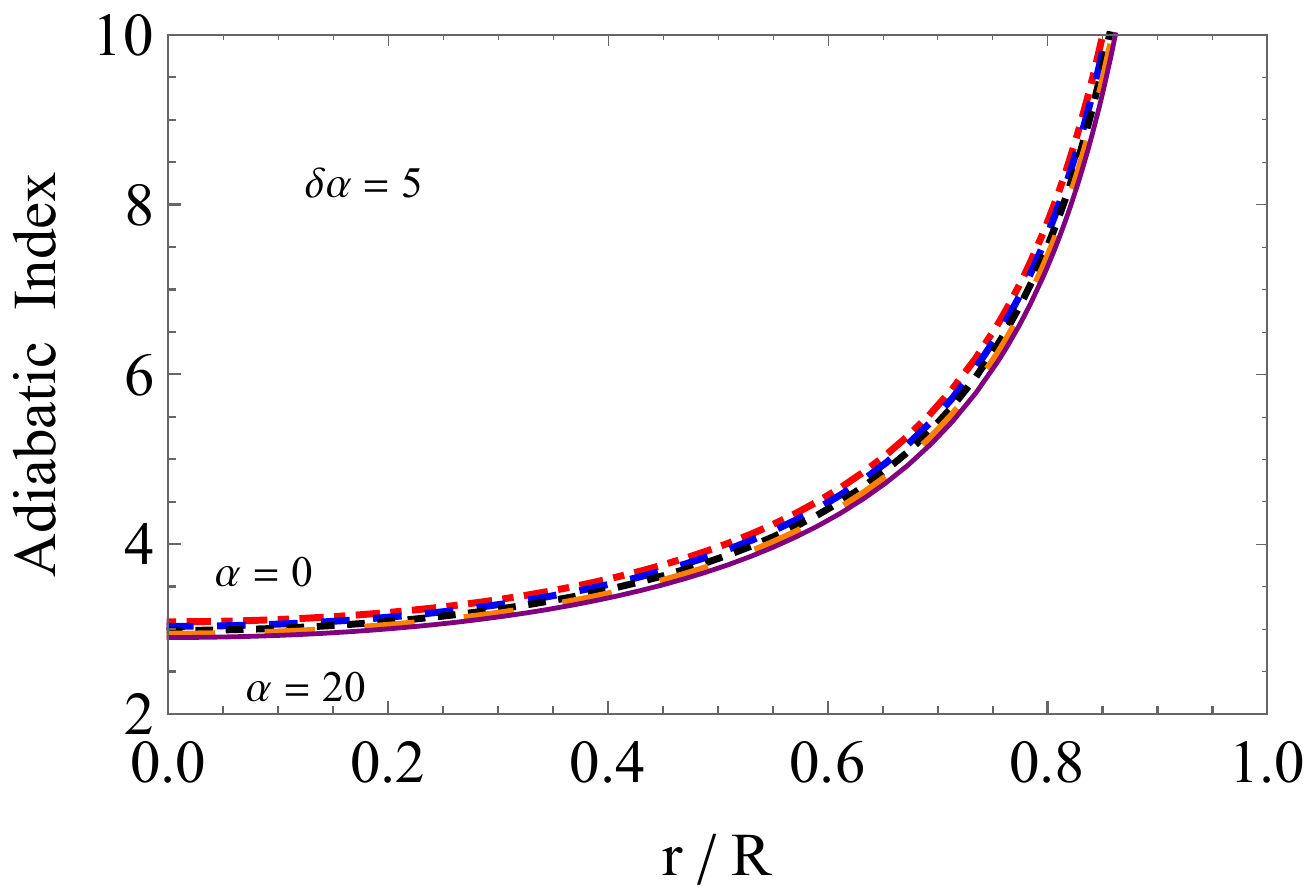}~~~  \includegraphics[width=8.5cm,height=6.3cm]{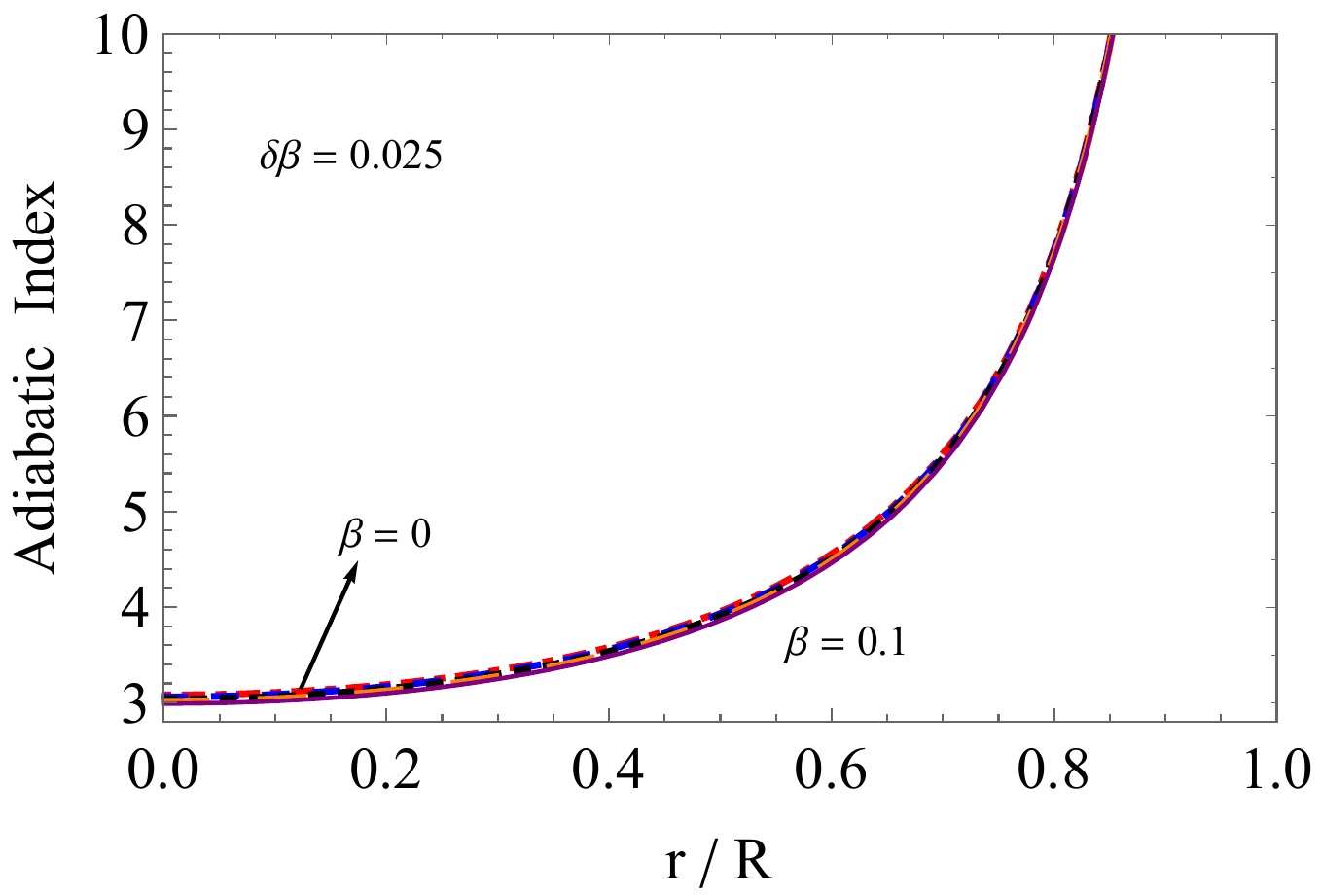}
    \includegraphics[width=8.5cm,height=6.3cm]{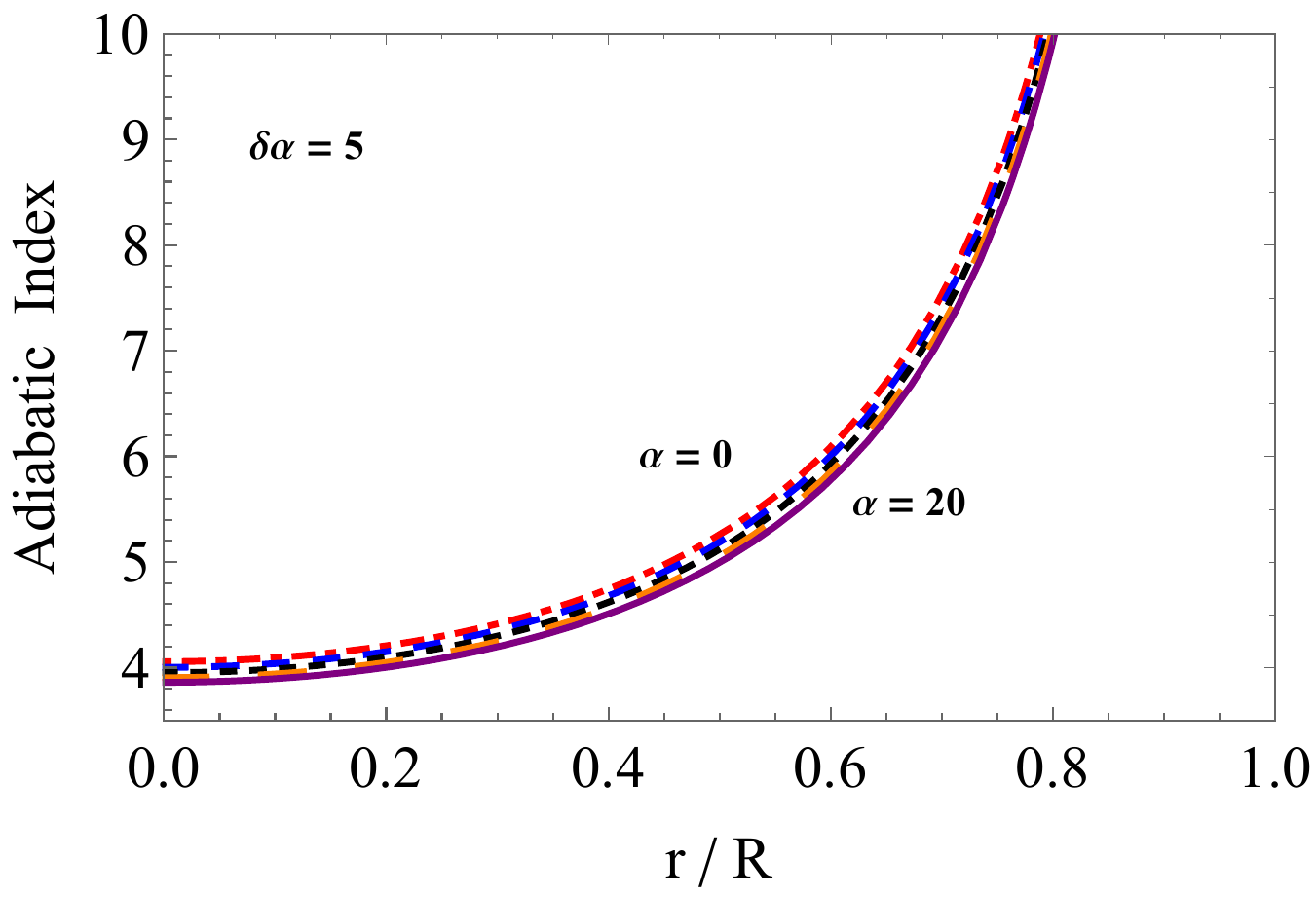}~~~  \includegraphics[width=8.5cm,height=6.3cm]{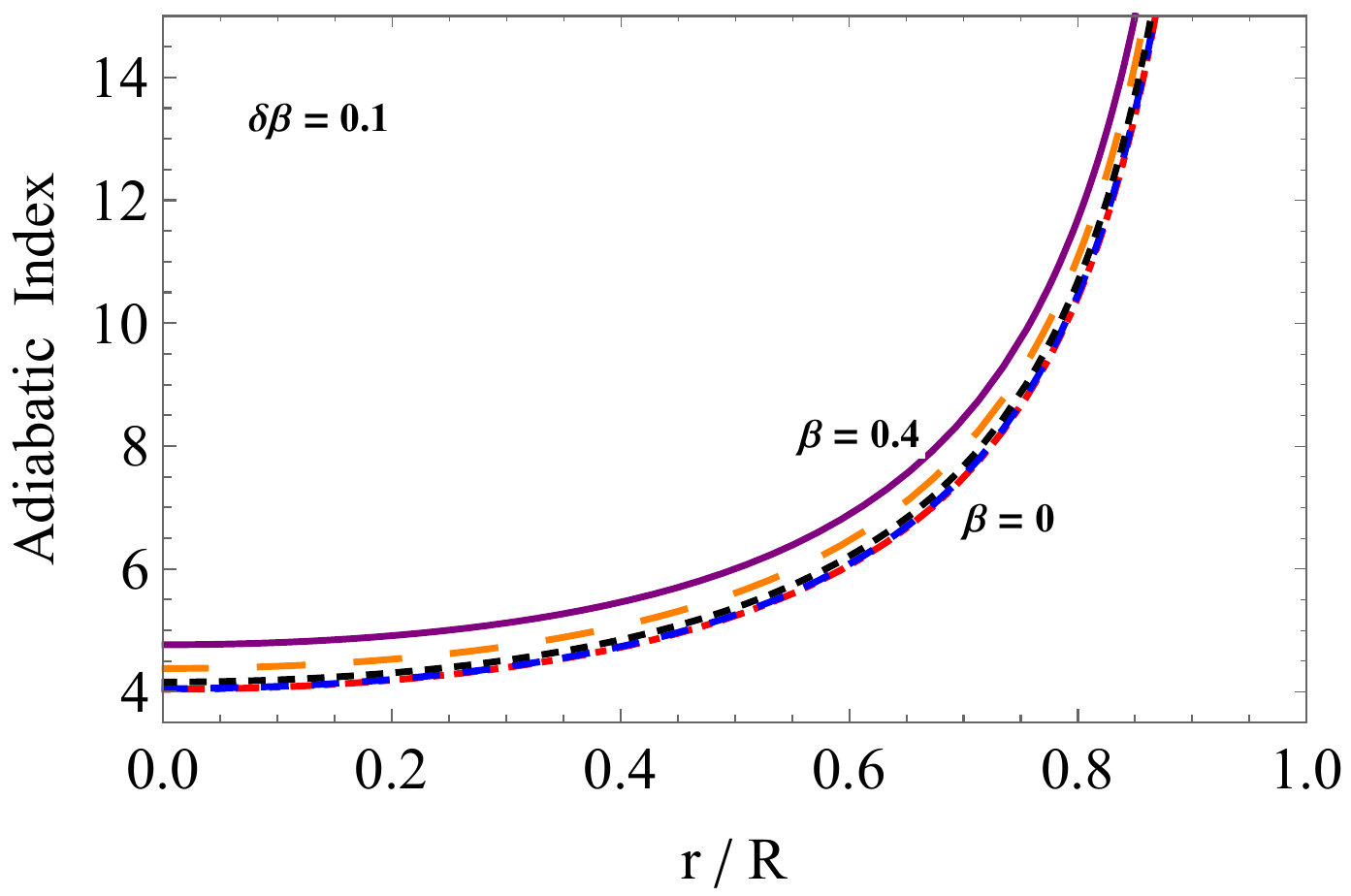}
    \caption{\textit{Top panels} (for solution $\theta^0_0=\hat{\epsilon}$) and \textit{Bottom panels}  (for solution $\theta^1_1=\hat{P}_r$) show the adiabatic index ($\Gamma$) with respect to $r/R$ for different $\alpha$ and  $\beta$. We set same numerical values as used in Fig.\ref{fig1}. }
    \label{fig5}
\end{figure*}
\begin{figure*}
    \centering
    \includegraphics[width=8.5cm,height=6.3cm]{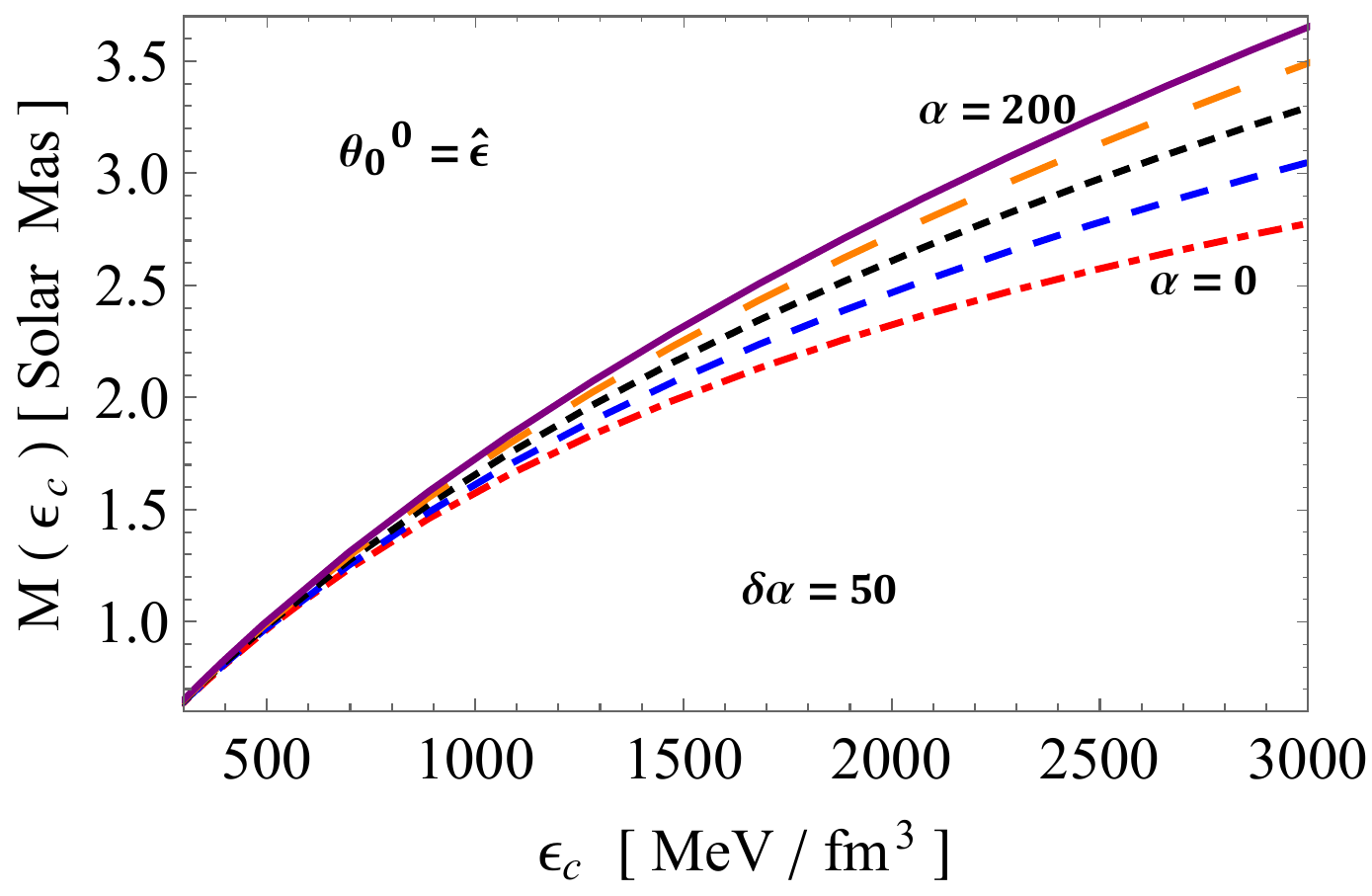}~~~ \includegraphics[width=8.5cm,height=6.3cm]{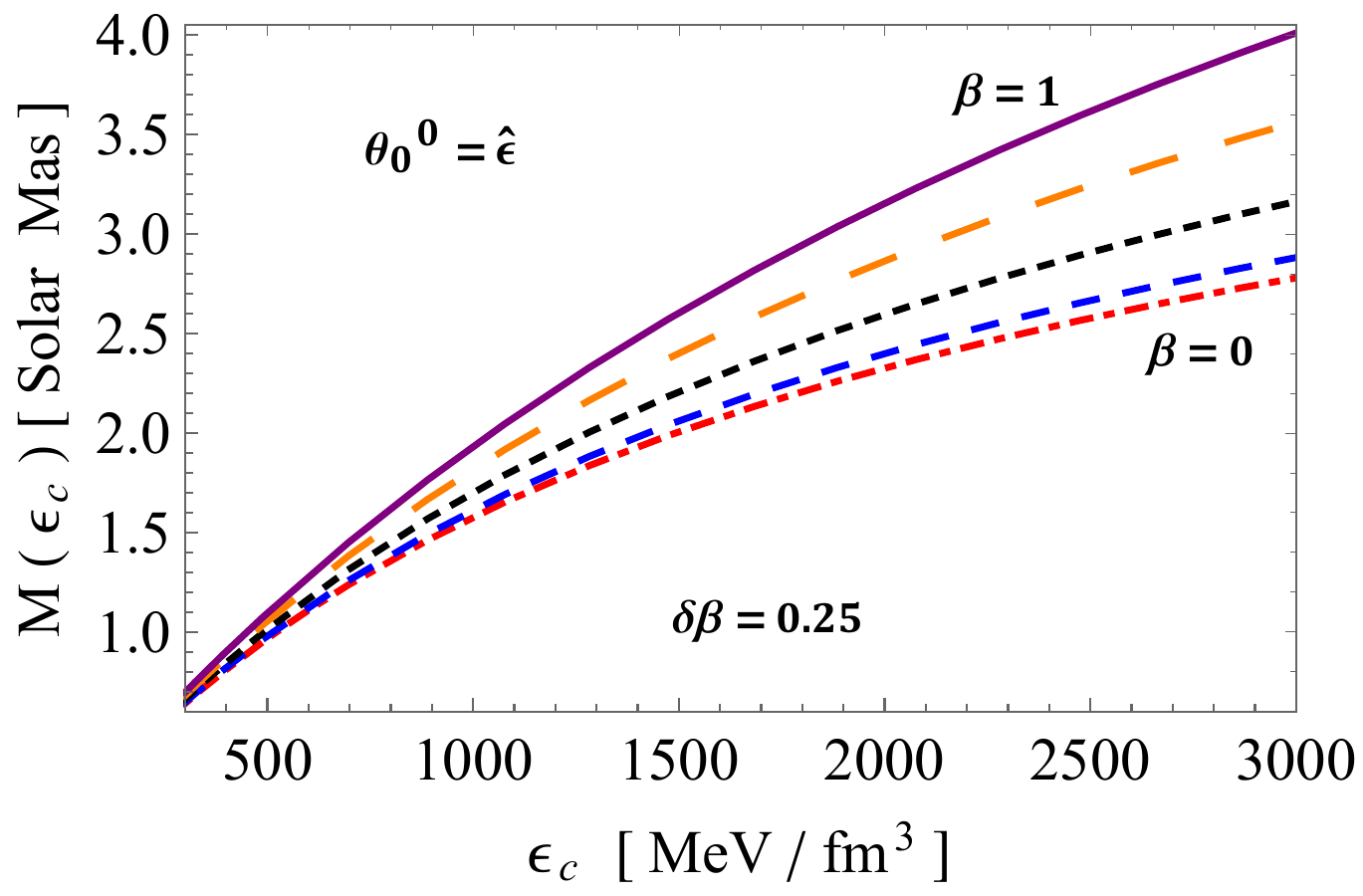}
    \caption{Mass versus central density for different $\alpha$ and  $\beta$ for the solution $\theta_0^0=\hat{\epsilon}$. We set same numerical values as used in Fig.\ref{fig1}.}
    \label{fig6}
\end{figure*}
\subsection{Stability Analysis} \label{sec5.2}
Here, our main interest is to discuss the stability of the compact strange star model. In this connection, Chandrasekhar \citep{chandra1,chandra2}  and Merafina \& Ruffini \citep{Merafina} derived the adiabatic stability criterion $\Gamma = \left(1+\frac{\epsilon}{P_r}\right) \left(\frac{dP_r}{d\epsilon}\right)_S$ showing that $\Gamma > 4/3$ for stars exhibiting isotropic pressures, where $\frac{dP_r}{d\epsilon}$ is the velocity of sound
and the subscript $S$ denotes a constant specific entropy. Herrera and co-workers demonstrated that this condition gets modified in the presence of anisotropy and dissipation.  The modified adiabatic index in the presence of anisotropy for a collapse scenario is given by 
\begin{equation} \label{modgamma}
\Gamma < \frac{4}{3} + \left[-\frac{4}{3}\frac{(P_r - P_t)}{|P'_r|r}\right]
\end{equation}
where the prime denotes differentiation with respect to the radial coordinate. We observe from (\ref{modgamma}) that in the absence of relativistic effects, we regain the Newtonian limit, $\Gamma < \frac{4}{3}$. Heintzmann and Hillebrandt \citep{hill} concluded that a positive, increasing anisotropy factor, $\Delta = P_t - P_r > 0$ ensures stability of a relativistic fluid configuration if $\Gamma > \frac{4}{3}$. This is due to a positive anisotropic factor giving rise to a repulsive force at each interior point within the compact object. It has been demonstrated that dissipative effects, such as heat flow within the star can alter the adiabatic index. To this end a critical value for the adiabatic index ($\Gamma_{crit}$) was proposed which depends on two factors: (i) the compactness $(u = M/R^2)$ of the stellar model and (ii) a measure of the deviation from hydrostatic equilibrium (quantified as the amplitude of the Lagrangian displacement from equilibrium).  We cast the critical adiabatic parameter as: $\Gamma_{crit} = \frac{4}{3} + \frac{19}{21}u$ \citep{mou,maurcrit}.  In order to ensure stability against radial perturbations we must have $\Gamma > \Gamma_{crit}$ \citep{mou}. However, it has been pointed that for stable neutron stars including white dwarfs and supermassive compact objects, $\Gamma$ ranges between 2 and 4 while for matter obeying a polytropic EoS $\Gamma > \frac{4}{3}$ and is sensitive to the ratio of the central density to the central pressure \citep{ayan3}. 
    Figure \ref{fig5} shows that our models satisfy the Chandrasekhar stability criterion since the adiabatic index $(\Gamma)$ is increasing and more than $2$ throughout the strange star models. It is interesting to note that an increase in the EGB coupling constant tends to render the configuration less stable. This can be ascertained from the top and bottom left panels of Figure \ref{fig5}. In the $\theta^0_0=\hat{\epsilon}$ solution we observe that an increase in the decoupling constant has very little impact on $\Gamma$ (top right panel). In the bottom right panel (for $\theta^1_1=\hat{P}_r$ solution), an increase in $\beta$ while holding $\alpha$ constant tends to stabilise the configuration. \\
In addition to the above discussion, we further study the stability of the strange star model through Harrison-Zeldovich-Novikov (HZN) stability criterion. For this purpose, we discuss the stability for the model corresponding to  $\theta_0^0=\hat{\epsilon}$ solution only. In Fig.\ref{fig6}, we display the mass of the configuration as a function of the central density ($\epsilon_c$) for $\theta_0^0=\hat{\epsilon}$ model. According to the Harrison-Zeldovich-Novikov stability criterion \citep{harris,zeld}, we have the following constraints:
    \begin{eqnarray}
    \frac{dM}{d\rho_c} &>& 0 \hspace{0.5cm} \rightarrow \mbox{stable configuration}\\ \nonumber \\
  \frac{dM}{d\rho_c} &<& 0 \hspace{0.5cm} \rightarrow \mbox{unstable configuration}  
        \end{eqnarray}
   The left panel of Figure \ref{fig6} shows that we have stable configurations when the EGB coupling constant is varied while the decoupling constant is fixed. For densities in the range of 500 $MeV/fm^3$ to $1000 MeV/fm^3$, variation in $\alpha$ has no effect on $M(\rho_c)$. As the density increases, we observe that an increase in $\alpha$ results in a peeling away of $M(\rho_c)$ from the 5D EGB configuration. Furthermore, the gradient, $\frac{dM}{d\rho_c}$ increases with an increase in $\alpha$. The right panel of Figure 5 shows a similar trend for central densities less than $1000 MeV/fm^3$. Here we kept the EGB paramter constant and varied $\beta$. We also note that $\frac{dM}{d\rho_c}$ is smaller than their counterparts in the left panel of Figure \ref{fig6}. From this we can conclude that the EGB coupling constant has a greater influence in stabilising the fluid configuration compared to the decoupling constant. 
    \begin{figure*}
    \centering
    \includegraphics[width=8cm,height=6.3cm]{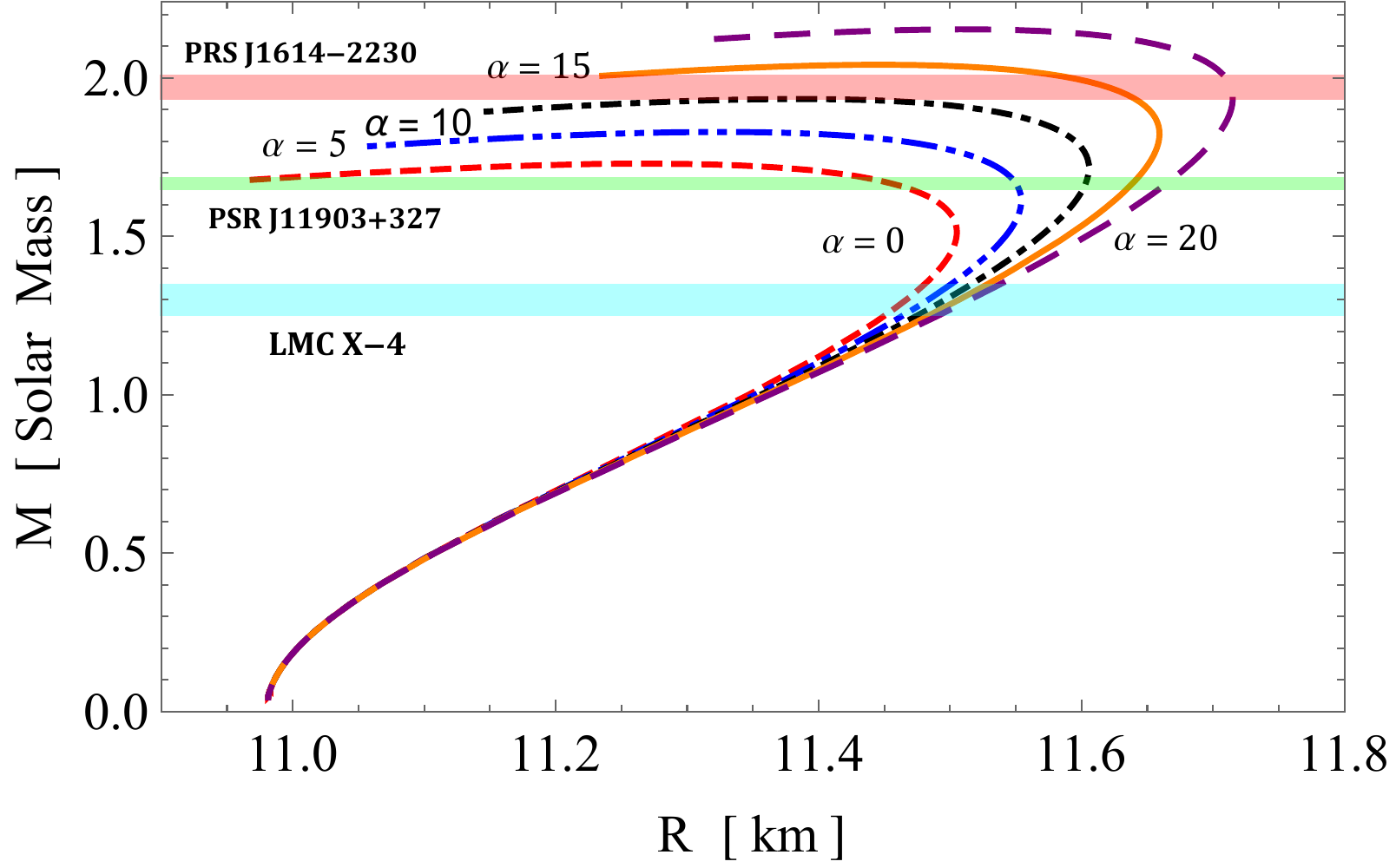}~~~ \includegraphics[width=8cm,height=6.3cm]{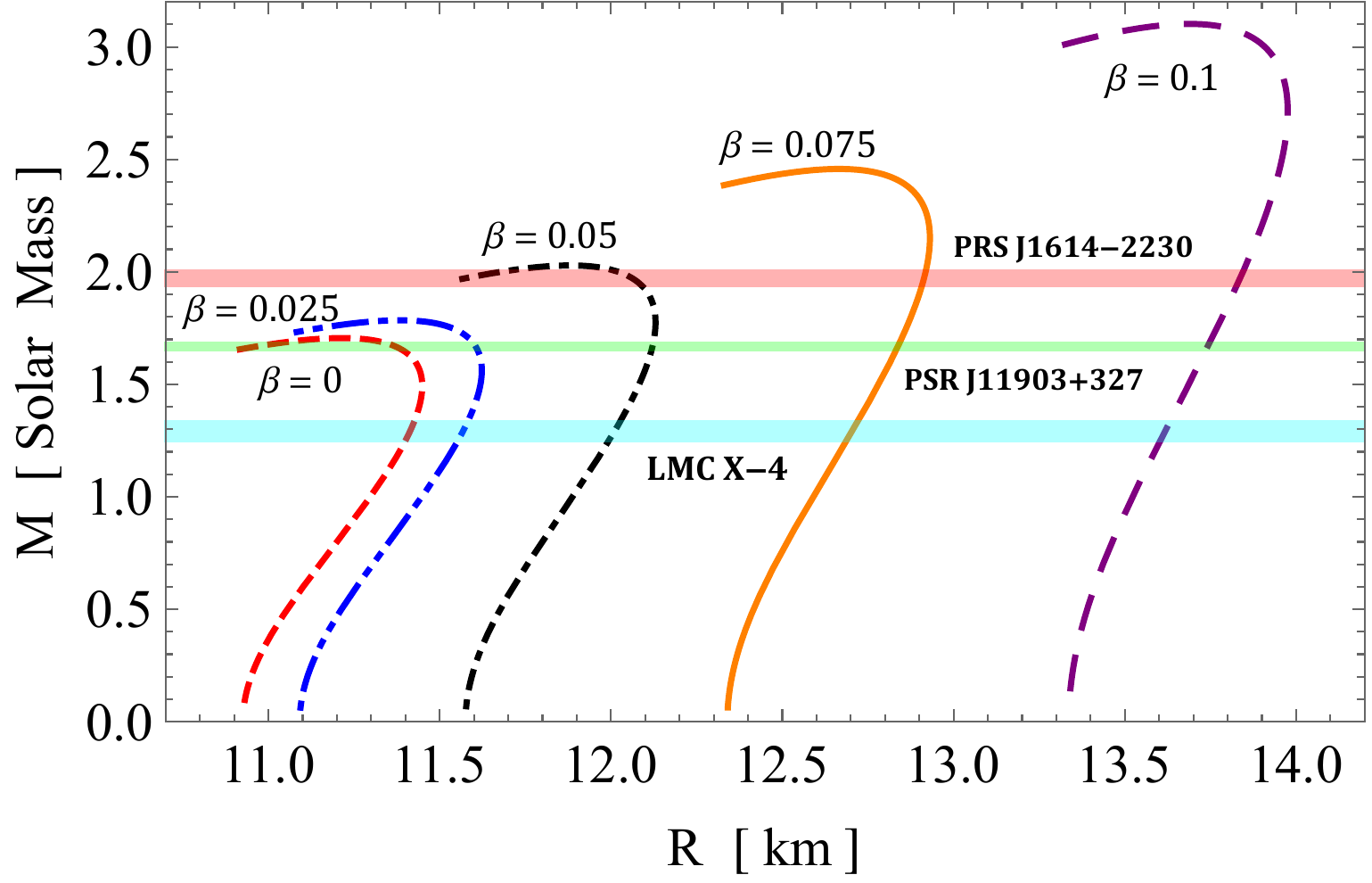}
    \caption{The left and right panels show the $M-R$ curves depending on different values $\alpha$ and  $\beta$, respectively when $\theta^0_0=\hat{\epsilon}$.}
    \label{fig7}
\end{figure*}  
\begin{figure*}
    \centering
    \includegraphics[width=8cm,height=6.3cm]{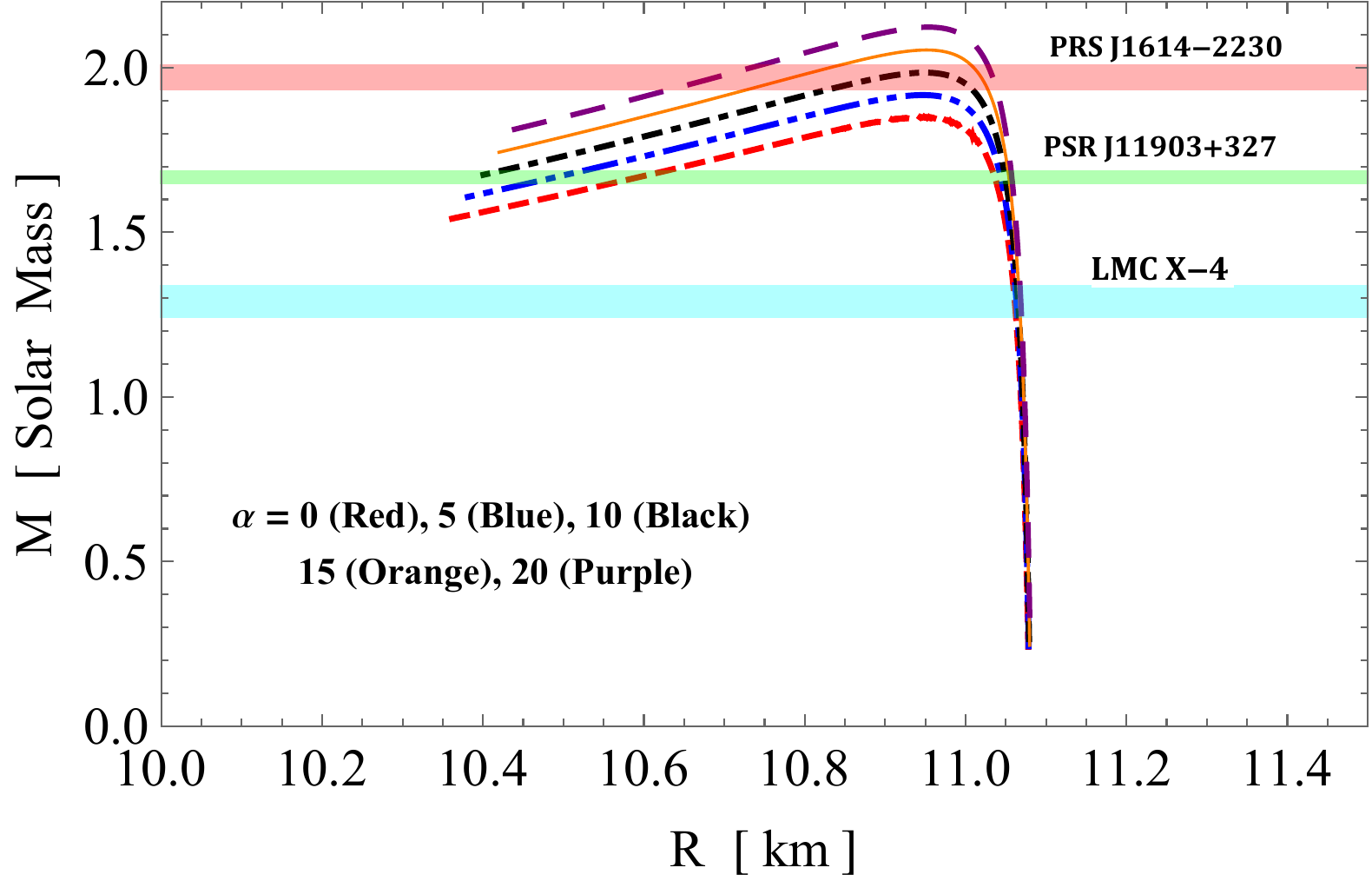}~~~ \includegraphics[width=8cm,height=6.3cm]{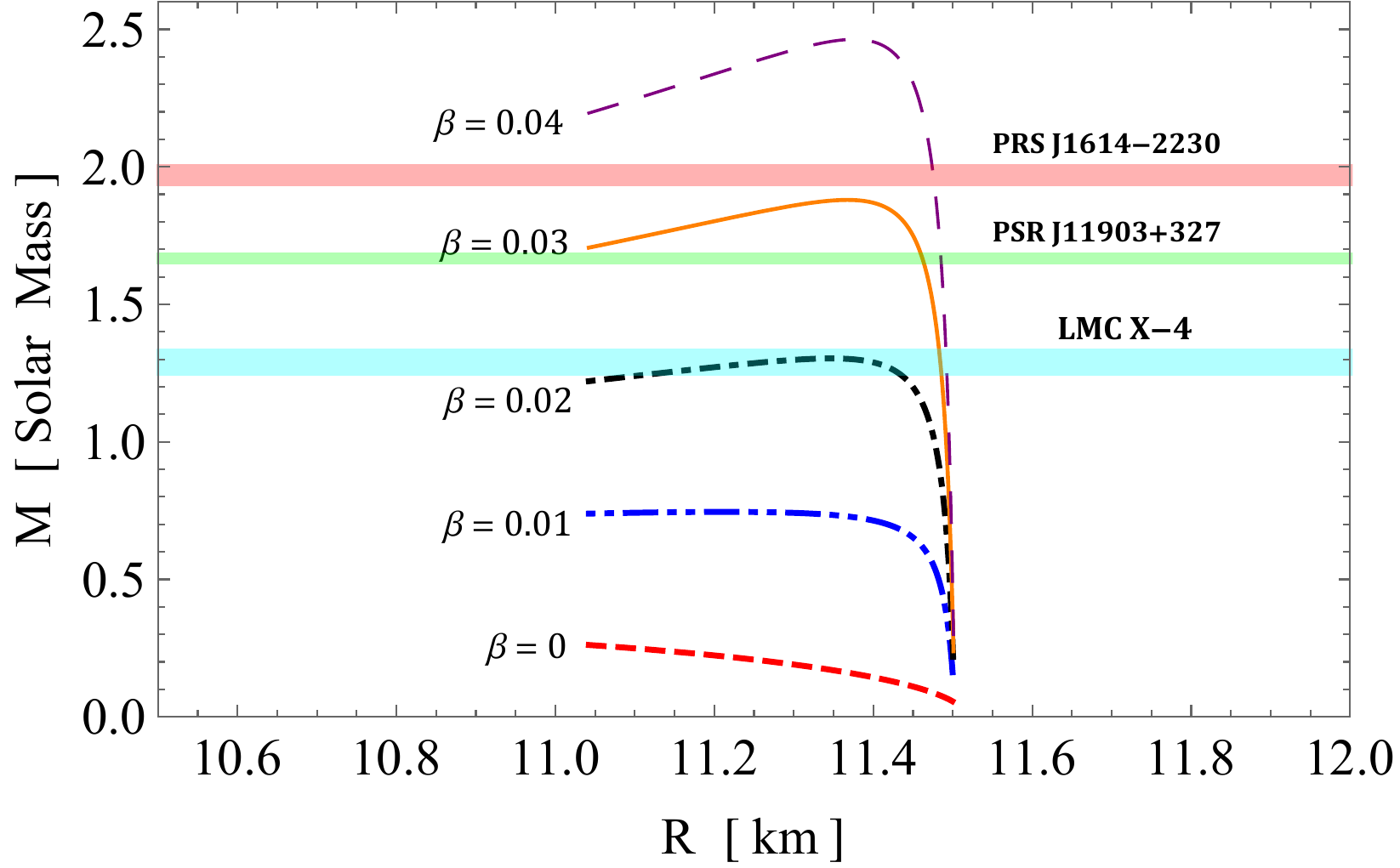}
    \caption{The left and right panels show the $M-R$ curves depending on different values of $\alpha$ and  $\beta$, respectively when $\theta^1_1=\hat{P}_r$. }
    \label{fig8}
\end{figure*}
\subsection{Mass and Radius Measurements of the Strange Stars via $M-R$ Curves} \label{sec5.3}
We now turn our attention to Figures \ref{fig7} and \ref{fig8} which depict the masses of three star candidates, viz., PSR J1614-2230, PSR J1903+317, and LMC X-4 as functions of stellar radius. Observations of pulsars such as PSR J1614-2230 and PSR J0348 +0432 have placed an upper bound on the mass of $M= 2 M_{\odot}$ \citep{Ayan1}. Looking at Figure \ref{fig7} in conjunction with Table \ref{table1}, we observe that for varying EGB parameter, with fixed decoupling constant (left panel) and a mass range of 1.29 $\pm$ 0.05 to 1.97 $\pm$ 0.04, the predicted radii fall within the range, $11.45^{+0.02}_{-0.01}~ \text{km}$  and 11.71$^{+0.005}_{-0.005} ~ \text{km}$. We note that the pure EGB ($\alpha = 0$) case rules out the existence of compact object PSR J1614-2230. Furthermore, the maximum mass allowable is around $M \simeq 1.5 M_{\odot}$ with radii being less than $11.5 ~ \text{km}$. As $\alpha$ increases, the predicted masses cover a wider range of observed values including configurations with masses greater than $M = 2 M_{\odot}$ and radii closer to $\sim$ $11.8 ~ \text{km}$. On the right panel of Figure \ref{fig7}, we observe the effect of varying the decoupling constant and $\alpha$ constant. An increase in $\beta$ is accompanied by an increase in mass and radii. It is possible to generate stellar structures with masses closer to $M \simeq 3 M_{\odot}$ and radii of approximately $\sim 14 ~ \text{km}$. These are outside the observed ranges for realistic compact objects such as pulsars and strange stars. 
\begin{figure*}
    \centering
    \includegraphics[scale=0.47]{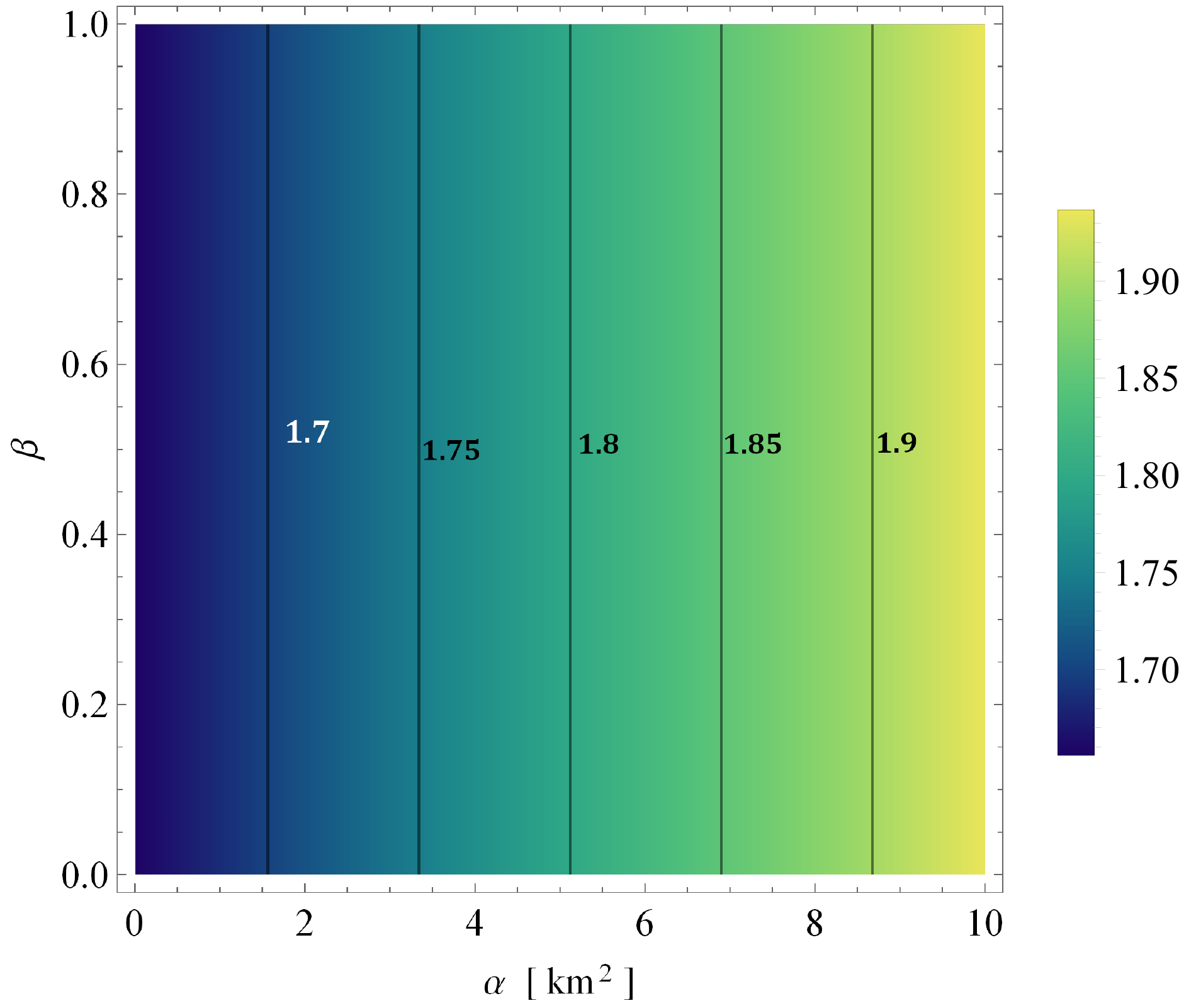}~~~~
     \includegraphics[scale=0.47]{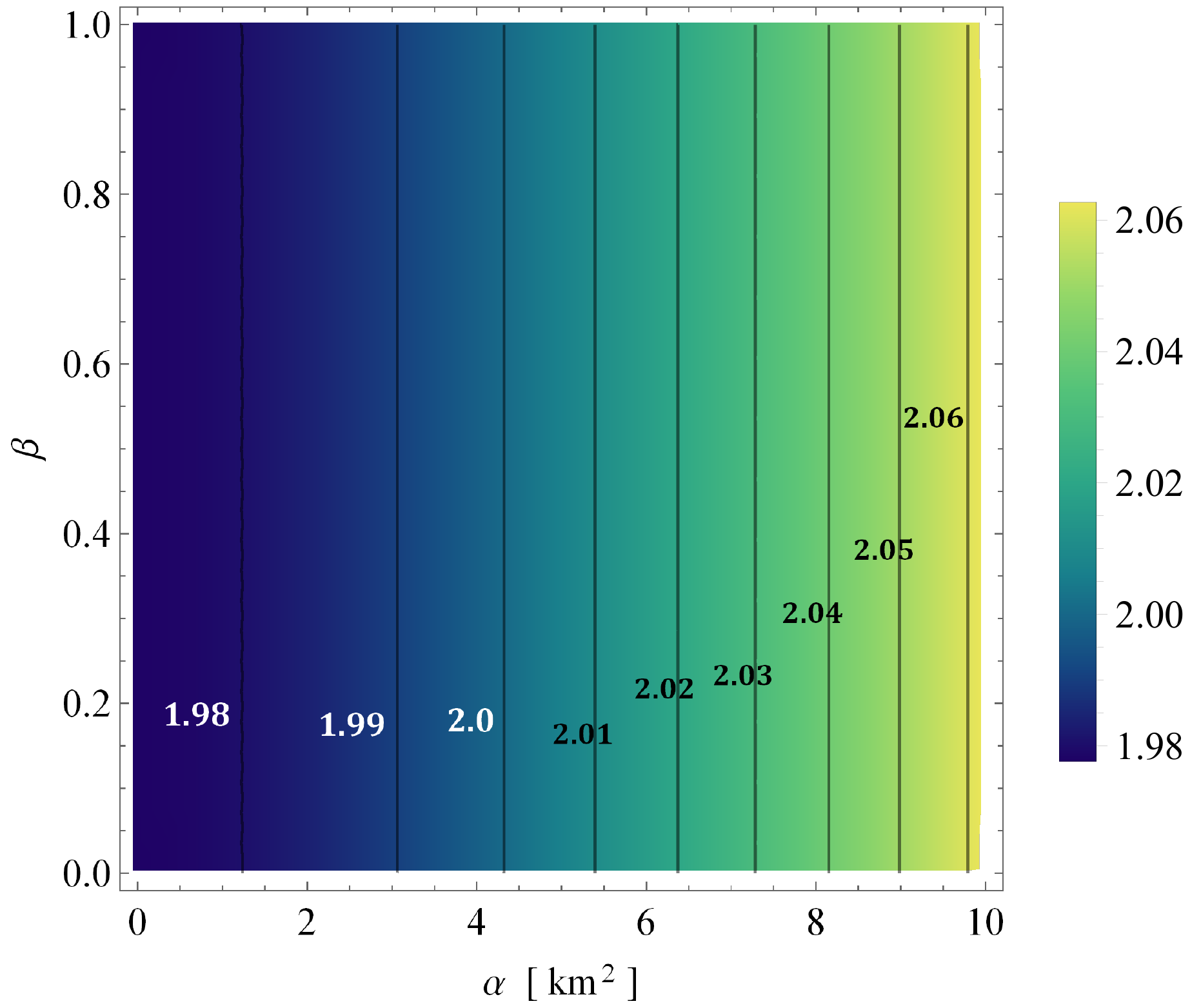}
\caption{\textit{Left panel:} $\alpha-\beta$ plane for equi-mass with $R = 11.3~km, ~N = 10^{-7}/km^4,~\mathcal{B}_g=58\,MeV/fm^3$ for the case $\theta^0_0=\hat{\epsilon}$. \textit{Right panel:} $\alpha-\beta$ plane for equi-mass with $R = 11.3~km, ~N = 10^{-7}/km^4,~\mathcal{B}_g=58\,MeV/fm^3$ for the case $\theta^1_1=\hat{P}_r$.}
    \label{fig9}
\end{figure*}
\begin{figure*}
    \centering
    \includegraphics[scale=0.47]{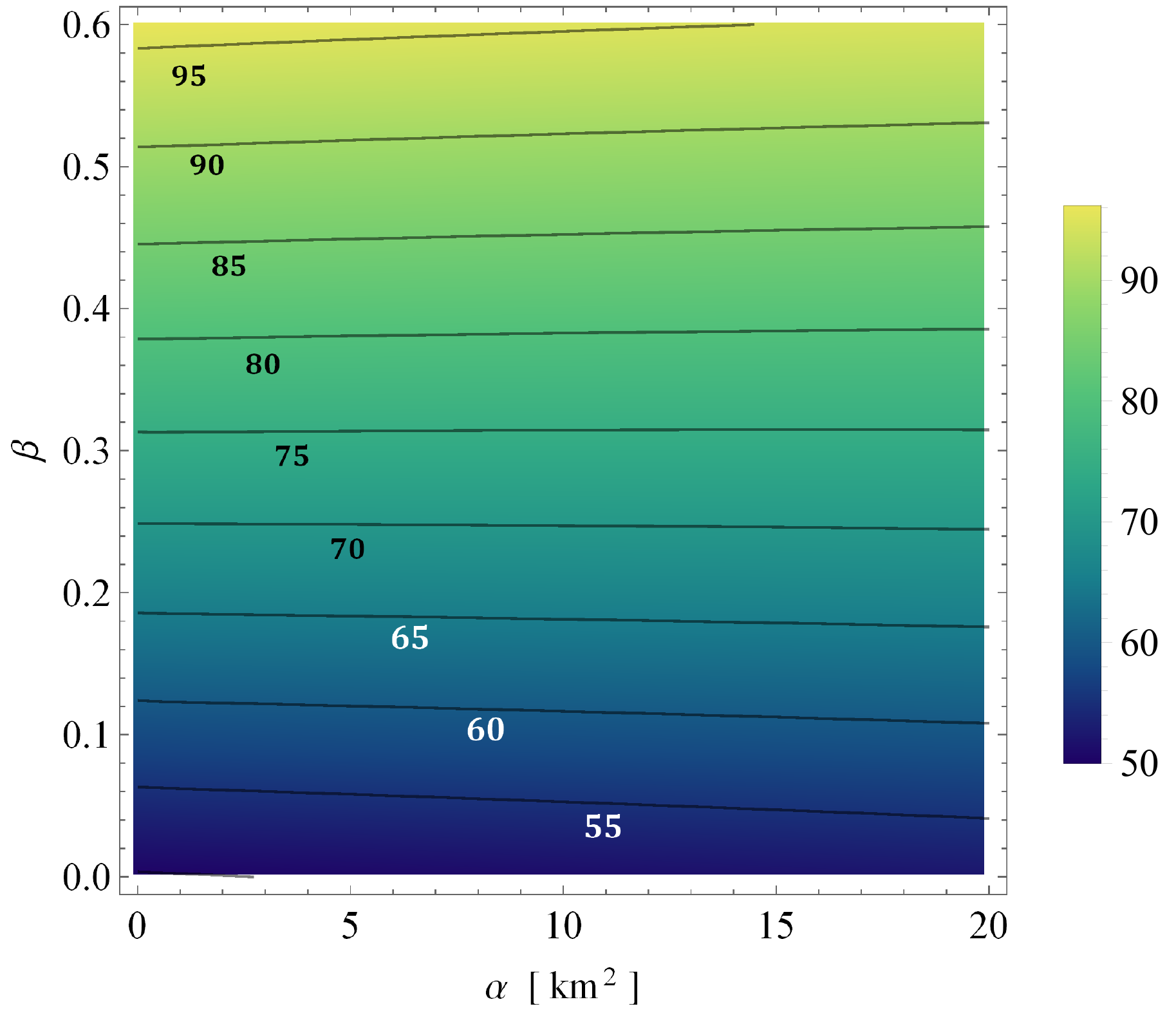}~~~~~~ \includegraphics[scale=0.47]{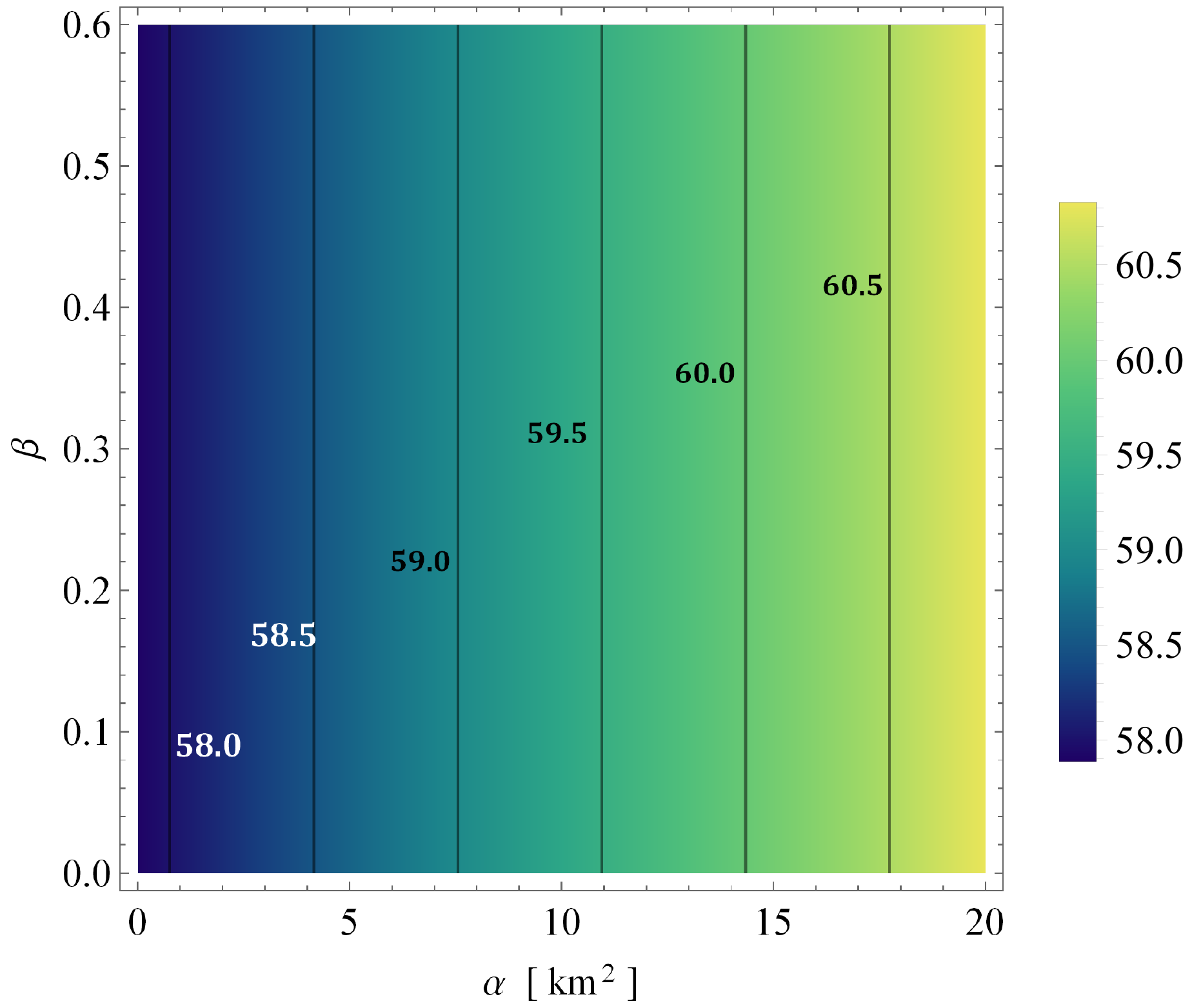} 
    \caption{\textit{Left panels}: $\alpha-\beta$ plane for equi-$\mathcal{B}_g$ with $R = 11.3~km, ~N = 10^{-7}/km^4,~L=0.001/km^3$ for the case $\theta^0_0=\hat{\epsilon}$.  \textit{Right panels}: $\alpha-\beta$ plane for equi-$\mathcal{B}_g$ with $R = 11.3~km, ~N = 10^{-7}/km^4,~L=0.0017/km^3$ for the case $\theta^1_1=\hat{P}_r$. } 
    \label{fig10}
\end{figure*}
In Figure \ref{fig8} ($\theta^1_1=\hat{P}_r$ solution) and Table \ref{table2},  for the same star candidates as in represented in Table 1., we note that the predicted radii lie in the range range of $10.98^{+0.03}_{-0.005} ~ \text{km} $ and $11.04^{+0.005}_{-0.01} ~ \text{km}$. Again, we note that in the 5D Einstein limit ($\alpha = 0$), the existence of star candidate  PSR J1614-2230 is ruled out. An increase in the EGB coupling constant tends to rule out the existence of  PSR J1903+317 and LMC X-4 while the there is an increase in the radius of PSR J1614-2230 as $\alpha$ increases.  Figure \ref{fig8} (right panel) reveals that in the case of fixed $\alpha$ and varying decoupling constant, all three stellar candidates considered here are ruled out. As the magnitude of the decoupling constant increases, these stellar structures do exist with radii typically in the range of $[11.38, 11.49] ~ \text{km}$  which are lower than the radii predicted for the $\theta^0_0=\hat{\epsilon}$ solution. These predictions fall well within observational results of quark stars as reported by \citep{Ayan1,Ayan2}.

\begin{figure*}
    \centering
   \includegraphics[scale=0.47]{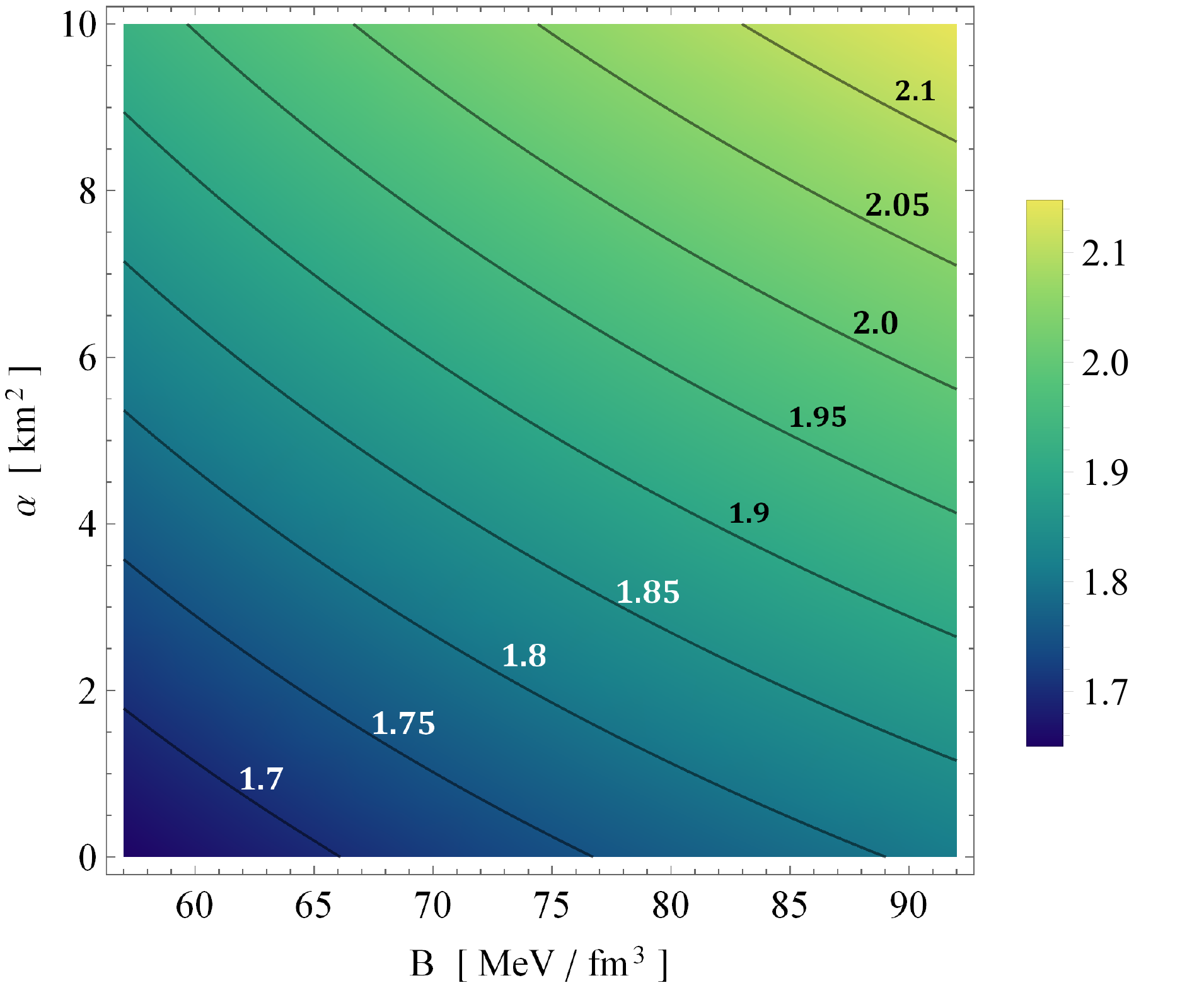}~~~~~~ \includegraphics[scale=0.47]{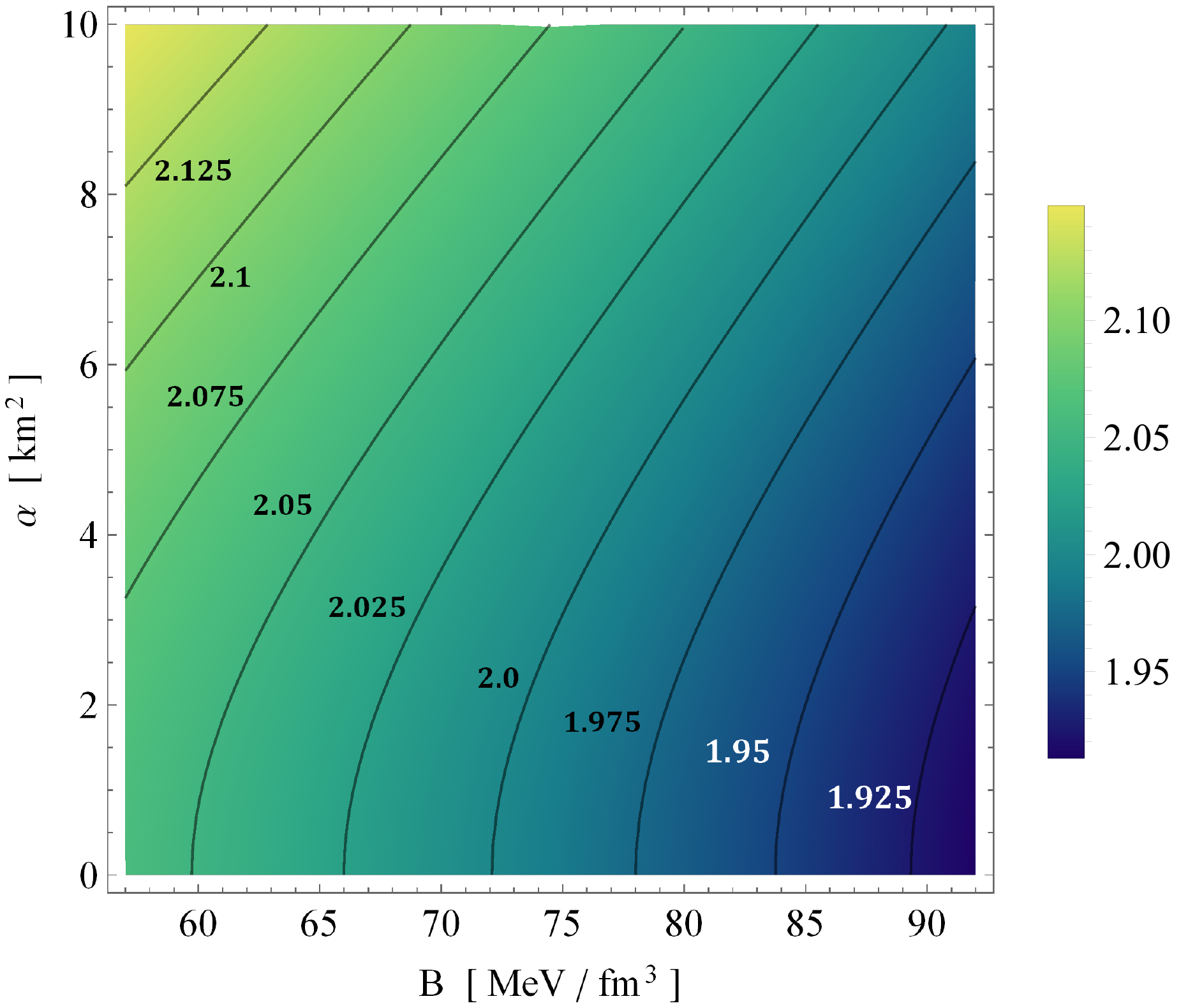}
\caption{\textit{Left panels}:  $\mathcal{B}_g-\alpha$ plane for equi-mass with $R = 11.3~km, ~N = 10^{-7}/km^4,~L=0.001/km^2$ for the case $\theta^0_0=\hat{\epsilon}$.   \textit{Right panels}: $\mathcal{B}_g-\alpha$ plane for equi-mass with $R = 11.3~km, ~N =10^{-7}/km^4,~L=0.00117/km^2$ for the case $\theta^1_1=\hat{P_r}$.} 
    \label{fig11}
\end{figure*}
\begin{figure*}
    \centering
    \includegraphics[scale=0.47]{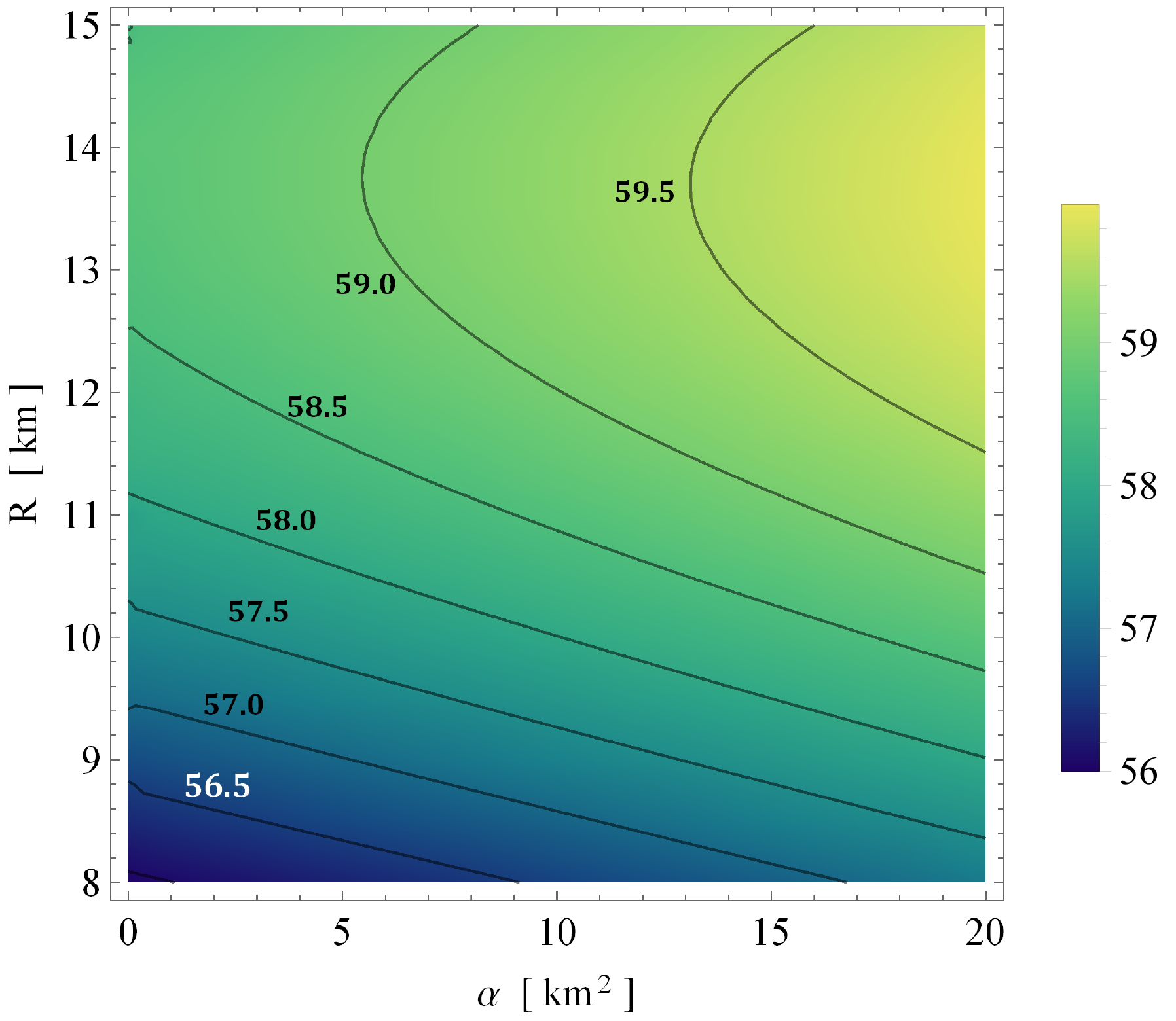}
   ~~~~~~\includegraphics[scale=0.47]{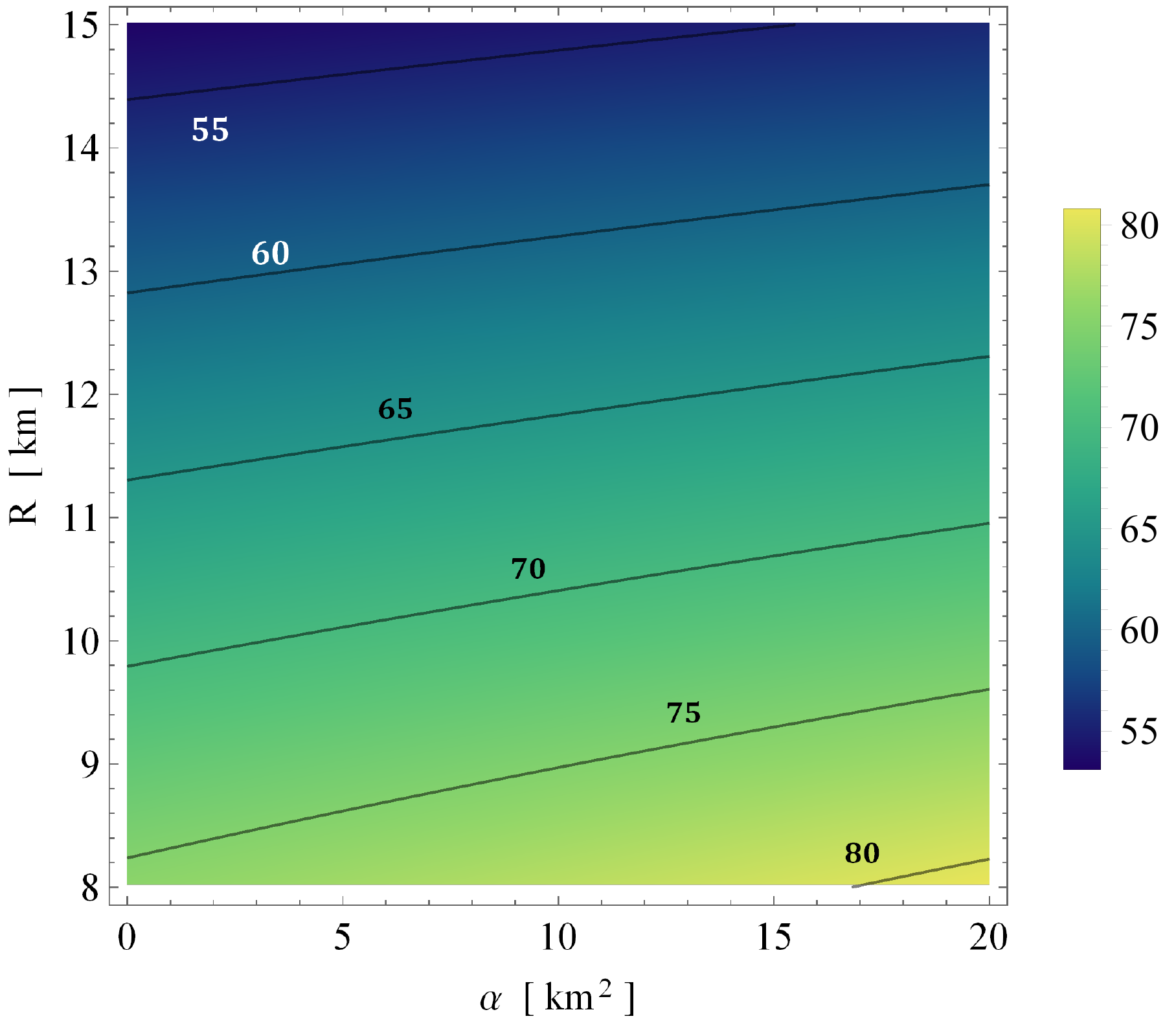}
    \caption{\textit{Left panels}:  $\alpha-R$ plane for equi-$\mathcal{B}_g$ with $\beta=0.1, ~N = 10^{-7}/km^4,~\beta=0.1$ for the case $\theta^0_0=\hat{\epsilon}$.  \textit{Right panels}: $\alpha-R$ plane for equi-$\mathcal{B}_g$ with $\beta=0.027, ~N = 10^{-7}/km^4,~L=0.002/km^2$ for the case $\theta^1_1=\hat{P}_r$.}
    \label{fig12}
\end{figure*}
\subsection{Mass and Bag constant Measurements of the Strange Stars via equi-plane diagrams} \label{sec5.4}
Further, the $\alpha-\beta$ planes in Fig. \ref{fig9} show the equi-mass contours for a fixed radius $11.3 ~ \text{km}$ and bag constant $\mathcal{B}=58~ MeV/fm^3$. Here the left panel of figure \ref{fig9} is for the the solution where $\theta_0^0$ mimicking density and on the right panel for $\theta_1^1$ mimicking pressure. In both the graphs one can see that the mass of the stellar system is unaffected by the change in $\beta-$parameter while strong dependence on EGB coupling constant, $\alpha$ is evident. Further, one can also observe that the mass for the same range of $\alpha$ is lower for $\theta_0^0=\hat{\epsilon}$ than $\theta_1^1=\hat{P}_r$ solution. Next, Figs. \ref{fig10} shows the $\alpha-\beta$ plane for equi-$\mathcal{B}$ contours. Here a very interesting phenomenon emerges where the bag constant is affected only by the MGD coupling $\beta$ and other only by EGB-coupling. In the solution I (when $\theta_0^0=\hat{\epsilon}$), the values of the bag constant is strongly affected by the MGD-$\beta$ coupling with the EGB coupling has very small role, however, in the solution II (when $\theta_1^1=\hat{P}_r$) the decoupling constant $\beta$ affects the bag constant negligibly while the $\alpha$ does significantly. In the first solution, as $\beta$ increases the bag constant in the range $50 ~MeV/fm^3$ to above $95~MeV/fm^3$. However, in the second solution while alpha increases from 0 to 20 $km^2$, the $\mathcal{B}$ lies in a narrow window between $\sim 58-61~MeV/fm^3$. To see the the effects of $\mathcal{B}$ and $\alpha$ for a fixed radius,  we have plotted contours of equi-mass in $\mathcal{B}-\alpha$ plane in both the solutions (Figs. \ref{fig11}). In the first solution where $\theta_0^0$ mimics energy density, for a fixed $\alpha$ and increasing the bag constant, the mass of the structure increases and decreases in the second solution with $\theta_1^1=\hat{P}_r$. However, fixing the bag constant and increasing the strength of the EGB coupling constant, the mass in both solutions increases. Hence, to obtain a high mass configuration (i) the bag constant and $\alpha$ must be high in solution I and (ii) low bag constant and high $\alpha$ in solution II (see Fig. \ref{fig11}). Further, one can also see that by fixing the mass and varying radius along with $\alpha$, we observe the variation in $\mathcal{B}$. Figure \ref{fig12} (left panel) shows that for a fixed $\alpha$ when radius increases the bag constant also increases. However, Fig. \ref{fig12} (right panel) shows the opposite trend i.e. at fixed $\alpha$ when $R$ increases the bag constant decreases. Summarising, we can conclude that to achieve high bag constant in solution I, the radius and EGB coupling must attain large magnitudes while in solution II, high $\alpha$ and low $R$ must be chosen. This implies that solution I ($\theta_0^0=\hat{\epsilon}$) is well suited to model comparatively larger radii configurations than solution II ($\theta_1^1=\hat{P}_r$). 
\begin{table*}
\footnotesize
\centering
\caption{The predicted Radii compact stars PSR J1614-2230, PSR J1903+317, and LMC X-4 for the case $\theta^0_0(r)=\hat{\epsilon}(r)$ }\label{table1}
 \scalebox{0.8}{\begin{tabular}{| *{12}{c|} }
\hline
{Objects} & {${\frac{M}{M_\odot}}$}   & \multicolumn{5}{c|}{{Predicted $R$ km}} & \multicolumn{5}{c|}{{Predicted $R$ km}} \\
\cline{3-12}
&& \multicolumn{5}{c|}{$\alpha$} & \multicolumn{5}{c|}{$\beta$} \\
\cline{3-12}
&  & 0 & 5 & 10 & 15 & 20 & 0 & 0.025 & 0.05 & 0.075 & 0.10  \\ \hline
PSR J1614-2230   &  1.97 $\pm$ 0.04  & -  &  -  &   -  &  11.62$^{+0.02}_{-0.04}$  &  11.71$^{+0.005}_{-0.005}$ & -&-& $12.07^{+0.03}_{-0.08}$ & $12.91^{+0.01}_{-0.01}$ & $13.84^{+0.01}_{-0.02}$  \\
\hline
PSR J1903+327& 1.667$\pm$0.021  & $11.45^{+0.02}_{-0.01}$  & $11.55^{+0.01}_{-0.01}$ & $11.60^{+0.01}_{-0.01}$ & $11.64^{+0.005}_{-0.01}$ & 11.67$^{+0.005}_{-0.01}$ & $11.37^{+0.02}_{-0.03}$ & $11.61^{+0.01}_{-0.02}$ & $12.12^{+0.01}_{-0.01}$ & $12.84^{+0.01}_{-0.01}$ & $13.74^{+0.01}_{-0.01}$ \\
\hline
LMC X-4 & 1.29 $\pm$ 0.05 & $11.46^{+0.02}_{-0.01}$ & $11.48^{+0.02}_{-0.01}$  & $11.49^{+0.03}_{-0.01}$ & $11.50^{+0.03}_{-0.02}$ & $11.51^{+0.03}_{-0.02}$ & $11.42^{+0.01}_{-0.02}$ & $11.57^{+0.02}_{-0.02}$ & $12.01^{+0.02}_{-0.02}$ & $12.70^{+0.02}_{-0.01}$ & $13.62^{+0.01}_{-0.02}$ \\
\hline
\end{tabular}}
\end{table*}

\begin{table*}
\centering
\caption{ The predicted Radii compact stars PSR J1614-2230, PSR J1903+317, and LMC X-4 for the case $\theta^1_1(r)=\hat{P}(r)$ }\label{table2}
 \scalebox{0.95}{\begin{tabular}{| *{12}{c|} }
\hline
{Objects} & {${\frac{M}{M_\odot}}$}   & \multicolumn{5}{c|}{{Predicted $R$ km}} & \multicolumn{5}{c|}{{Predicted $R$ km}} \\
\cline{3-12}
&& \multicolumn{5}{c|}{$\alpha$} & \multicolumn{5}{c|}{$\beta$} \\
\cline{3-12}
&  & 0 & 5 & 10 & 15 & 20 & 0 & 0.01 & 0.02 & 0.03 & 0.04 \\ \hline
PSR J1614-2230   &  1.97$\pm$ 0.04   & -  &  -  &  $10.98^{+0.03}_{-0.005}$   &  $11.02^{+0.01}_{-0.01}$  & $11.03^{+0.01}_{-0.005}$ &-&-&-&-&11.47   \\
\hline
PSR J1903+327& 1.667$\pm$0.021  & $11.04^{+0.005}_{-0.01}$  & 11.045& 11.05&11.055&  $11.06$ &-&-&-& 11.46 & 11.48 \\
\hline
LMC X-4 & 1.29$\pm$ 0.05  & 11.06 & -  &- &- & 11.07 &-&-& 11.38 & 11.48 & 11.49 \\
\hline
\end{tabular}}
\end{table*}
\section{Conclusions and Astrophysical Implications} \label{sec6}
We now provide an overview of our findings at the attempt to model a strange star utilising the gravitational decoupling via minimal geometric deformation (MGD) method within Einstein-Gauss-Bonnet gravity. We employed the MIT bag model equation of state and the Tolman ansatz for one of the gravitational potentials which formed the basis for constructing a bounded stellar configuration. The MGD formalism leads to two family of solutions, viz., one emanating from the $\theta_0^0=\hat{\epsilon}$ sector and the other from $\theta_1^1=\hat{P}_r$ sector. The EGB coupling constant, $\alpha$ and the decoupling constant, $\beta$ were shown to be the key protagonists in controlling the gravitational and thermodynamical behaviour of our models. In order to appreciate the impact of $\alpha$ and $\beta$ on the regularity and stability  of the stellar configurations we varied the EGB coupling constant and held the the decoupling parameter fixed and vice versa. Our models are regular throughout the interior of the stellar fluid. The energy density and pressure stresses are well-behaved and fit the profiles for a realistic star. For both solutions, the tangential pressure dominates the radial pressure which results in a repulsive anisotropic force which stabilises the stellar configuration. An interesting observation is that the decoupling constant suppresses the energy density and the anisotropic stresses while an increase in the EGB parameter enhances these quantities. A stability analysis via the anisotropic generalisation of the Chandrasekhar adiabatic index and the Harrison-Zeldovich-Novikov stability criterion revealed that our models are stable, with the stability being enhanced by the decoupling constant. The highlight of our work reveals that the magnitudes of $\alpha$ and $\beta$ have an intrinsic connection to the (non)-existence of strange star candidates; PSR J1614-2230, PSR J1903+317, and LMC X-4. A fine-tuning of these parameters is required to ensure that the observed masses and corresponding radii of these stellar candidates are attained. Furthermore, the MGD formalism within EGB gravity also predicted stable configurations above the observed mass $\sim 2 M_{\odot}$ for neutron stars. While the authors concede that there is no current evidence for higher dimensions or observational data on the magnitudes of $\alpha$ and $\beta$, we do believe that our work is novel and provides a theoretical basis for the existence of strange star models which fall within the range of observed masses and radii. In addition, we have provided a mechanism to generate stable configurations beyond the upper limit of mass $\sim 2 M_{\odot}$ without invoking exotic matter states such as dark energy and dark matter.

\section*{Acknowledgement}
The author SKM acknowledges that this work is carried out under TRC Project (Grant No. BFP/RGP/CBS-/19/099), the Sultanate of Oman. SKM is thankful for continuous support and encouragement from the administration of University of Nizwa. 

\section*{Appendix}
\begin{eqnarray}
&& \hspace{-0.2cm}  \hat{P}_{t1}(r)= 9 + 36 N r^4 + 8 L^3 r^6 + 35 N^2 r^8 + 8 N^3 r^{12} + 6 L^2 r^4  (5 + 4 N r^4) + L r^2 (31 + 65 N r^4 + 24 N^2 r^8) + 4 \alpha [14 L^2 r^2 \nonumber\\&& \hspace{0.9cm} + N r^2 (9 + 19 N r^4) + L (9 + 33 N r^4)],\nonumber\\
&& \hspace{-0.4cm} \hat{P}_{t2}(r)=- 9 (-16 L^6 r^{10} - L^5 r^6 (92 \alpha + 87 r^2 + 96 N r^6) - L^4 [-80 \alpha^2 r^2 + 12 \alpha r^4 (17 + 45 N r^4) +   5 r^6 (35 + 91 N r^4  + 48 N^2 r^8)] \nonumber\\&& \hspace{0.9cm} -  N r^2 [224 \alpha^2 N^2 r^4 (3-N r^4) + (1 + N r^4)^2  (45 +  72 N r^4 + 75 N^2 r^8 + 16 N^3 r^{12}) + 4 \alpha N r^2 (87 + 129 N r^4 + 85 N^2 r^8 \nonumber\\&& \hspace{0.9cm} + 43 N^3 r^{12})]+\hat{P}_{t3}(r)),\nonumber \\
&& \hspace{-0.4cm} \hat{P}_{t3}(r)=-  L [18 + 255 N r^4 + 715 N^2 r^8 + 897 N^3 r^{12} + 515 N^4 r^{16} + 96 N^5 r^{20} - 48 \alpha^2 N^2 r^4 (-29 + 13 N r^4) +  4 \alpha N r^2 (108  \nonumber\\&& \hspace{0.9cm} + 323 N r^4 + 322 N^2 r^8 + 195 N^3 r^{12})] -    L^3 [48 \alpha^2 (3 - 7 N r^4) +     4 \alpha r^2 (55 + 254 N r^4 + 310 N^2 r^8) +  r^4 (171 + 771 N r^4 \nonumber\\&& \hspace{0.9cm} + 950 N^2 r^8 + 320 N^3 r^{12})] -   L^2 [16 \alpha^2 N r^2 (54 - 41 N r^4)  +  4 \alpha (27 + 249 N r^4 + 440 N^2 r^8 + 350 N^3 r^{12}) +   r^2 (85 + 622 N r^4 \nonumber\\&& \hspace{0.9cm} + 1255 N^2 r^8 + 990 N^3 r^{12} +   240 N^4 r^{16})].\nonumber \\ 
&& \hspace{-0.4cm} \theta_{11}(r)=48 \alpha^2 (L^2 + 3 L N r^2 + 2 N^2 r^4) (1 - L r^2 - N r^4 + \beta \Phi (1  + L r^2 + N r^4)) + (1 + L r^2 + N r^4)^2 [4 \mathcal{B} (r + L r^3 + N r^5)^2 \nonumber\\&& \hspace{0.9cm} - 
    3 (3 + 4 L^2 r^4 + 9 N r^4 + 4 N^2 r^8 + 8 L (r^2 + N r^6))],\nonumber \\ 
&& \hspace{-0.4cm} \theta_{12}(r)=   
    -  4 \alpha (1 + L r^2 +  N r^4) \big[4 \mathcal{B} (1 + L r^2 + N r^4)^2 (1 - L r^2  - N r^4 +  \beta \Phi (1 + L r^2 + N r^4)) - 
    3 \big(4 (-1 + \beta \Phi) L^3 r^4  \nonumber\\&& \hspace{0.9cm} + 
       L^2 r^2 (-5 - 12 N r^4 + 3 \beta \Phi (3 + 4 N r^4)) + 
       N r^2 (3 - 7 N r^4 - 4 N^2 r^8 + 2 \beta \Phi (3 + 5 N r^4 + 2 N^2 r^8)) +  L (2\,(1 - 6 N r^4 \nonumber\\&& \hspace{0.9cm} - 6 N^2 r^8) + \beta \Phi (5 + 19 N r^4 + 12 N^2 r^8))\big)\big].\nonumber 
\end{eqnarray}

\section*{Data Availability}
The data underlying this article are available in the article and in its online supplementary material.



\begin{thebibliography}{99}

\bibitem[\protect\citeauthoryear{Abbas \& Tahir}{2018}]{abbas}  Abbas G.,  Tahir M.,  2018, {\em Advances in High Energy Physics}{\bf 2018},7420546.

\bibitem[\protect\citeauthoryear{Abell\'an et al.}{2020}]{Abell2020}  Abell\'an G.,  Rinc\'on A., Fuenmayor E.,  Contreras E., 2020, Eur. Phys. J. Plus \textbf{135}, 606.

\bibitem[\protect\citeauthoryear{Amendola et al.}{2007}]{amendola} Amendola L., Charmousis C., Davis S. C., 2007,  {\em JCAP 0710}, {  10} 004 .

\bibitem[\protect\citeauthoryear{Annala et al}{2020}]{annala}  Annala E.,  Gorda T.,  Kurkela A.,  Nättilä J.,  Vuorinen A.,  2020, {\em  Nature Physics} {\bf 16}, 907. 

\bibitem[\protect\citeauthoryear{Zubair \& Azmat}{2020b}]{Zubair2021}  Azmat H.,  Zubair M., 2021, Eur. Phys. J. Plus 136, 112

\bibitem[\protect\citeauthoryear{Banerjee et al}{2021}]{ayann1}  Banerjee A.,  Tangphat T.,  Samart D.,  Channuie P.,  2021, {\em Astrophys. J.} {\bf 906}, 114. 

\bibitem[\protect\citeauthoryear{Banerjee et al.}{2021}]{ayan3}  Banerjee A.,  Tangphati T.,  Channuie P.,  2021, {\em Astrophys. J.} {\bf 909}, 14.

\bibitem[\protect\citeauthoryear{Bhar \& Govender}{2019}]{bhar}  Bhar P.,  Govender M.,  2019, {\em  Astrophys Space Sci} {\bf  364}, 186.

\bibitem[\protect\citeauthoryear{Bodmer}{1971}]{bodmer}  Bodmer A. R.,  1971, {\em  Phys. Rev. D} {\bf 4}, 1601.

\bibitem[\protect\citeauthoryear{Boulware}{1985}]{boul}  Boulware D. G.,  Deser S.,  1985, {\em Phys. Rev. Lett.} {\bf 55},  2656.

\bibitem[\protect\citeauthoryear{Casadio et al.}{2015}]{casadio}  Casadio R.,  Ovalle J.,  da Rocha R.,  2015, {\em Class. Quant. Gravit.} {\bf  32}, 215020. 

\bibitem[\protect\citeauthoryear{Casadio et al.}{2015}]{casadio1}  Casadio R.,  Ovalle J.,  da Rocha R.,  2015, {\em Europhys. Lett.} {\bf 110}, 40003. 

\bibitem[\protect\citeauthoryear{Cavalcanti et al.}{2016}]{cavalcanti}  Cavalcanti R.T.,  Goncalves da Silva A.,  da Rocha R.,  2016, {\em Class. Quant. Grav.} {\bf  33} 215007.

\bibitem[\protect\citeauthoryear{Casadio \& da Rocha}{2016}]{casadio2}  Casadio R.,  da Rocha R., 2016, {\em Phys. Lett. B} {\bf 763}, 434.

\bibitem[\protect\citeauthoryear{Choudhury and Ghosh}{2016}]{choudhury} Choudhury D.,  Ghosh K.,  2016, {\em  Physics Letters B} {\bf 763}, 155.

\bibitem[\protect\citeauthoryear{Chandrasekhar}{1964a}]{chandra1} Chandrasekhar S. 1964a, {\em Astrophys. J.}{, 140}, 417.

\bibitem[\protect\citeauthoryear{Chandrasekhar}{1964b}]{chandra2}   Chandrasekhar S. 1964b, {\em Phys. Rev. Lett.}{, 12}, 114.

\bibitem[\protect\citeauthoryear{Chodos et al.}{1974}]{Chodos:1974} Chodos A., Jaffe R. L., Johnson K., Thorn C. B., Weisskopf V. F., 1974, Phys. Rev. D 9, 3471 

\bibitem[\protect\citeauthoryear{Chodos et al}{1974a}]{chodos1}   Chodos A.,  Jaffe R. L.,  Johnson K.,  Thorn C.B.,  1974, {\em  Phys. Rev. D} {\bf 10}, 2599.

\bibitem[\protect\citeauthoryear{Chodos et al}{1974b}]{chodos2}    Chodos A.,  Jaffe R. L.,  Johnson K.,  Thorn C.B., Weisskopf V.,  1974, {\em  Phys. Rev. D} {\bf 9}, 3471. 

\bibitem[\protect\citeauthoryear{Chilambwe}{2015}]{hans6}  Chilambwe B.,  Hansraj S.,  Maharaj S. D.,  2015, {\em Int. J. Mod. Phys. D} {\bf 24}, 1550051.

\bibitem[\protect\citeauthoryear{Contreras \& Bargue\~no}{2018}]{Contreras2018}  Contreras E., Bargue P.,\~no,  2018, Eur. Phys. J. C \textbf{78}, 558.

\bibitem[\protect\citeauthoryear{Contreras \& Bargue\~no}{2019}]{Contreras2019} Contreras E., Bargue P.,\~no,  2019, Class. Quantum Grav. \textbf{36}, 215009.


\bibitem[\protect\citeauthoryear{Contreras et al.}{2020}]{Contreras2020} Contreras E., et al. 2020, Classical and Quantum Gravity 37 , 155002.

\bibitem[\protect\citeauthoryear{Contreras et al.}{2021}]{Ovalleprd3} Contreras E. , Ovalle J., Casadio R., 2021, Phys. Rev. D 103, 044020.

\bibitem[\protect\citeauthoryear{Darabi et al}{2018}]{darabi}  Darabi F.,  Moradpour H.,  Licata I.,  Heydarzade Y., Corda C.,  2018, {\em  Eur. Phys. J. C} {\bf  78},  25. 

\bibitem[\protect\citeauthoryear{Darmois}{1927}] {darmois}  Darmois G. J., 1927, {\em Memorial des Sciences Mathematiques}, 25, 1.

\bibitem[\protect\citeauthoryear{Davis}{2003}]{davis} Davis S. C.,  2003, {\em Phys. Rev. D} {\bf 67}, 024030.

\bibitem[\protect\citeauthoryear{da Rocha}{2017a}]{rocha1} da Rocha R.,  2017a, {\em  Phys. Rev. D} {\bf 95}, 124017. 

\bibitem[\protect\citeauthoryear{da Rocha}{2017b}]{rocha2} da Rocha R.,  2017b, {\em Eur. Phys. J. C} {\bf 77}, 355.


\bibitem[\protect\citeauthoryear{Dehghani}{2004}]{dehghani}  Dehghani M. H., 2004, {\em Phys. Rev. D}, {  70}, 064009 .

\bibitem[\protect\citeauthoryear{Doneva \& Yazadjiev}{2021}]{doneva} Doneva D. D., Yazadjiev S. S., 2021, {\em JCAP} {\bf 2021}, 024 .


\bibitem[\protect\citeauthoryear{Drake et al}{2002}]{drake} Drake J.J.  et al.  2002, {\em  The Astrophysical Journal} {\bf  572}, 996.

\bibitem[\protect\citeauthoryear{Ellis et al.}{2011}]{ellis1}  Ellis G. F. R.,  van Elst H.,  Murugan J.,  Uzan J.P., 2011, {\em  Class. Quantum Grav.} {\bf 28} 225007 

\bibitem[\protect\citeauthoryear{Ellis}{2014}]{ellis2}  Ellis G. F. R., 2014, {\em Gen. Relativ. Gravit.}  {\bf 46} 1619

\bibitem[\protect\citeauthoryear{Freire et al.}{2011}]{Ayan2}  Freire P. C. C.,  Bassa C. G.,  Wex N., et al., 2011, MNRAS {\bf 412}, 2763 

\bibitem[\protect\citeauthoryear{Farhi \& Jaffe}{1984}]{Farhi:1984} Farhi E., Jaffe R. L., 1984, Phys. Rev. D 30, 2379 

\bibitem[\protect\citeauthoryear{Gross}{1999}]{gross}  Gross D., 1999, {\em Nucl. Phys. Proc. Suppl.} {\bf 74},  426.

\bibitem[\protect\citeauthoryear{Glavan \& Lin}{2020}]{glavan}   Glavan D.,  Lin C., 2020, {\em Phys. Rev. Lett.} {\bf 124}, 081301 .

\bibitem[\protect\citeauthoryear{Gurses et al.}{2020a}]{gurses1}   Gurses M.,  Sisman T. C.,  Tekin B., 2020a, {\em Eur. Phys. J. C} {\bf 80}, 647 

\bibitem[\protect\citeauthoryear{Gurses et al.}{2020b}]{gurses2}  Gurses M.,  Sisman T. C.,  Tekin B., 2020b, {\em Phys. Rev. Lett.} {\bf 125}, 149001 .

\bibitem[\protect\citeauthoryear{Gravanis \& Willison }{2003}]{gravanis}	Gravanis E., Willison S., 2003,
{\em Phys. Lett. B} {\bf 562}, 118 .

\bibitem[\protect\citeauthoryear{Hansraj \& Banerjee}{2020}]{hans1}  Hansraj S.,  Banerjee A.,  2020, {\em Mod.Phys.Lett.A} {\bf 35}, 2050105.

\bibitem[\protect\citeauthoryear{Harko et al.}{2011}]{Harko}  Harko T.,  Lobo F.S.N.,  Nojiri S.,  Odintsov S.D., 2011, {\em Phys. Rev. D}, {\bf 84}, 024020.

\bibitem[\protect\citeauthoryear{Hansraj}{2018}]{hans2}  Hansraj S., 2018, {\em  Eur. Phys. J. C} {\bf 78},,  700 

\bibitem[\protect\citeauthoryear{Hansraj et al.}{2017}]{hans3}  Hansraj S.,  Goswami R.,  Ellis G.,  Mkhize N.,  2017, {\em  Phys.Rev.D} {\bf  96}, 044016.

\bibitem[\protect\citeauthoryear{Hansraj et al.}{2015}]{hans5}  Hansraj S.,  Chilambwe B.,  Maharaj S. D.,  2015, {\em Eur. Phys. J. C } {\bf 27}, 277.

\bibitem[\protect\citeauthoryear{Hansraj \& Mkhize}{2020}]{hans7}  Hansraj S.,  Mkhize N.,  2020, {\em Physical Review D} {\bf  8},  084028.

\bibitem[\protect\citeauthoryear{Hansraj et al.}{2019}]{hans8}    Hansraj S., Maharaj S. D.,  Chilambwe B.,  2019, {\em Physical Review D} {\bf 12}, 124029.

\bibitem[\protect\citeauthoryear{Heintzmann \& Hillebrandt}{1975}]{hill}  Heintzmann H.,  Hillebrandt W., 1975, {\em Astron. Astrophys.} {\bf 38}, 51 

\bibitem[\protect\citeauthoryear{Harrison}{1965}]{harris}  Harrison B. K., 1965, Gravitational Theory and Gravitational Collapse, University of Chicago Press, Chicago.

\bibitem[\protect\citeauthoryear{Israel}{1966}]{israel} Israel W., 1966, {\em Nuovo Cim. B}, { 44}, 1.

\bibitem[\protect\citeauthoryear{Kaluza}{1921}]{kaluza}   Kaluza T.,  1921, {\em  Sitz. Ber. Preuss. Akad. Wiss.},  966.

\bibitem[\protect\citeauthoryear{Klein}{1926}]{klein}   Klein O.,  1926, {\em  Zeit. f. Physik} \textbf{37},  895.

\bibitem[\protect\citeauthoryear{Kang}{2012}]{kang}  Kang Z.,  Zhan-Ying Y.,  De-Cheng Z.,  Rui-Hong Y.,  2012, {Chinese Physics B}, {\bf 21} 020401.

\bibitem[\protect\citeauthoryear{Lovelock}{1971}]{love1}  Lovelock D., 1971, {\em J. Math. Phys.} {\bf 12}, 498.

\bibitem[\protect\citeauthoryear{Lovelock}{1972}]{love2}  Lovelock D.,  1972, {\em J. Math. Phys.} {\bf 13}, 874.


\bibitem[\protect\citeauthoryear{Maartens \& Koyama}{2010}]{maartens}  Maartens R., Koyama K.,  2010, {Living Reviews in Relativ.}, {\bf 13}, 10.

\bibitem[\protect\citeauthoryear{Maharaj et al.}{2015}]{hans4}  Maharaj S. D.,  Chilambwe B.,  Hansraj S.,  2015, {\em Phys. Rev. D} {\bf 91},  084049.

\bibitem[\protect\citeauthoryear{Maurya}{2019}]{Maurya2019} Maurya S.K., 2019, Eur. Phys. J. C 79, 958.

\bibitem[\protect\citeauthoryear{Maurya}{2020}]{Maurya2020} Maurya S.K., 2020, Eur. Phys. J. C 80, 429.

\bibitem[\protect\citeauthoryear{Maurya et al.}{2020}]{Maurya11} Maurya S.K. et al., 2020, Eur. Phys. J. C 80, 918. 

\bibitem[\protect\citeauthoryear{Maurya et al.}{2021}]{Maurya12} Maurya S.K. et al., 2021, Eur. Phys. J. C 81, 701.

\bibitem[\protect\citeauthoryear{Moustakidis}{2020}]{mou}  Moustakidis C. C., 2020, {\em Gen. Relativ. Grav.} {\bf 49}, 68

\bibitem[\protect\citeauthoryear{Merafina \& Ruffini}{1989}]{Merafina} Merafina M., \& Ruffini R. 1989, A \& A, 221, 4 

\bibitem[\protect\citeauthoryear{Maurya \& Ortiz}{2020a}]{Maurya2020a} Maurya S.K., Tello-Ortiz F.  2020, Physics of the Dark Universe 27, 100442

\bibitem[\protect\citeauthoryear{Maurya \& Ortiz}{2020b}]{Maurya2020b} Maurya S.K., Tello-Ortiz F.,  2020, Physics of the Dark Universe 29, 100577

\bibitem[\protect\citeauthoryear{Maurya \& Al-Farsi}{2021}]{Maurya2021} Maurya S.K., Al-Farsi L.S.S.,  2021, Eur. Phys. J. Plus 136, 317.

\bibitem[\protect\citeauthoryear{Ovalle}{2016}]{ovalle1}   Ovalle J., 2016, {\em Int. J. Mod. Phys. Conf. Ser.} {\bf 41}, 1660132.

\bibitem[\protect\citeauthoryear{Ovalle et al.}{2015}]{ovalle}  Ovalle J., Gergely L.A., Casadio R.,  2015, {\em  Class. Quant. Gravit.} {\bf 32}, 045015.

\bibitem[\protect\citeauthoryear{Ovalle}{2019}]{ovalle3}  Ovalle J.,  2019, {\em Phys. Lett. B} {\bf  788},213. 

\bibitem[\protect\citeauthoryear{Ovalle et al.}{2018}]{Ovalle2018}  Ovalle J.,  Casadio R., da Rocha R.,  Sotomayor A.,  2018, Eur. Phys. J. C {\bf78}, 122.

\bibitem[\protect\citeauthoryear{Ovalle \& Linares}{2013}]{Ovalle2013} Ovalle J., Linares F.,  2013, Phys. Rev. D 88, 104026

\bibitem[\protect\citeauthoryear{Ovalle et al.}{2021}]{Ovalleprd2}  Ovalle J. et al., 2021, Phys. Dark Univ. 31, 100744.

\bibitem[\protect\citeauthoryear{Leon \& Cruz}{2000}]{Ponce}  Ponce de Leon J.,  Cruz N., 2000, {\em Gen. Relativ. Gravit.} {\bf 32}, 1207 

\bibitem[\protect\citeauthoryear{Peshier et al}{2000}]{peshier}  Peshier A.,  Kampfer B., Soff G.,  2000, {\em  Phys. Rev. C} {\bf 61}, 045203. 

\bibitem[\protect\citeauthoryear{Pani et al. }{2011}]{pani}  Pani P., Berti E.,  Cardoso V., Read J.,  2011, {\em Phys. Rev. D} { \bf{84}}, 104035.

\bibitem[\protect\citeauthoryear{Panotopoulos \& Rinc$\acute{\text{o}}$n}{2019}]{Rincon12}  Panotopoulos G.,  Rinc$\acute{\text{o}}$n $\acute{\text{A}}$., 2019, Eur.Phys.J.Plus 134, 472

\bibitem[\protect\citeauthoryear{Rastall}{1972}]{rastall1}   Rastall P.,  1972, {\em Phys. Rev. D}, {\bf 6}, 3357.

\bibitem[\protect\citeauthoryear{Rastall}{1976}] {rastall2}   Rastall P.,  1976, { \em Can. J. Phys.}, {\bf 54}, 66.

\bibitem[\protect\citeauthoryear{Randall \& Sundrum }{1999a}]{randall1}  Randall L.,  Sundrum R.,  1999a, {\em  Phys. Rev. Lett.} {\bf 83}, 3370.

\bibitem[\protect\citeauthoryear{Randall \& Sundrum}{1999b}]{randall2}  Randall L.,  Sundrum R.,  1999b, {\em Phys. Rev. Lett.} {\bf 83}, 4690. 

\bibitem[\protect\citeauthoryear{Rinc$\acute{\text{o}}$n et al.}{2020}]{Rincon2020}  Rinc$\acute{\text{o}}$n $\acute{\text{A}}$. et al., 2020, Eur. Phys. J. C \textbf{80}, 490.

\bibitem[\protect\citeauthoryear{Rinc$\acute{\text{o}}$n et al.}{2019}]{Rincon11}   Rinc$\acute{\text{o}}$n $\acute{\text{A}}$. et al., 2019, Eur.Phys.J.C 79, 873 

\bibitem[\protect\citeauthoryear{Rocha}{2020}]{Ovalleprd1}  Rocha R. D., 2020, Phys.Rev. D 102, 024011.

\bibitem[\protect\citeauthoryear{Starobinsky}{1980}]{staro}  Starobinsky A. A., 1980, {\em Phys. Lett. B.} {\bf 91}  99 

\bibitem[\protect\citeauthoryear{Sharif \& Ramzan}{2020}]{sharif}  Sharif M., Ramzan A.,  2020, {\em  Phys.Dark Univ.} {\bf  30},  100737.

\bibitem[\protect\citeauthoryear{Maurya \& Ortiz}{2020}]{Sharif2020}  Sharif M.,  Majid A., 2020, Astrophys. Space Sci.365, 42. 

\bibitem[\protect\citeauthoryear{Sharif and Ama-Tul-Mughani}{2020}]{sharif1}  Sharif M., Ama-Tul-Mughani Q.,  2020, {\em  Annals Phys.} {\bf  415}, 168122.

\bibitem[\protect\citeauthoryear{Sharif and Majid}{2020}]{sharif2}  Sharif M.,  Majid A.,  2020, {\em  Phys. Dark Univ.} {\bf  30}, 100610,.

\bibitem[\protect\citeauthoryear{Tangphati et al.}{2021}]{tangphati}  Tangphati T.,  Pradhan A., Errehymy A., Banerjee A.,  2021, {\em Physics Letters B}, {\bf 819},  136423.

\bibitem[\protect\citeauthoryear{Ortiz et al.}{2020}]{maurcrit}  Tello-Ortiz F.,  Maurya S. K.,  Gomez-Leyton Y., 2020, {\em Eur. Phys. J. C} {\bf 80}, 324 .

\bibitem[\protect\citeauthoryear{Ortiz et al.}{2020}]{Ortiz2020} Tello-Ortiz F., et al., 2020, Eur. Phys. J. C 80, 324 

\bibitem[\protect\citeauthoryear{Tomozawa}{2012}]{tomozawa}   Tomozawa Y., 2012, arXiv:1107.1424 [gr-qc].

\bibitem[\protect\citeauthoryear{Tangphati et al.}{2021}]{Ayan1}  Tangphati T.,  Pradhan A.,  Errehymy A.,  Banerjee A., {\em 2021, Phys. Lett. B} {\bf 819}, 136423 

\bibitem[\protect\citeauthoryear{Visser}{2018}]{visser}  Visser M.,  2018, {\em Phys. Lett. B.} {\bf  782},  83.

\bibitem[\protect\citeauthoryear{Witten}{1984}]{witten}  Witten E.,  1984, {\em  Phys. Rev. D} {\bf 30}, 272. 

\bibitem[\protect\citeauthoryear{Wiltshire}{1988}]{wilt}  Wiltshire D. L.,  1988, {\em Phys. Rev. D} {38}, 2445.

\bibitem[\protect\citeauthoryear{Wright}{2016}]{Wright}  Wright M., 2016, Gen. Relativ. Gravit.  48:93

\bibitem[\protect\citeauthoryear{Zeldovich \& Novikov}{1971}]{zeld}  Zeldovich Y. B., Novikov I. D., 1971, Relativistic Astrophysics, Vol. I: Stars and Relativity, University of Chicago Press, Chicago .

\bibitem[\protect\citeauthoryear{Zubair \& Azmat}{2020a}]{Zubair2020} Zubair M.,  Azmat H., 2020, Ann of Phys. 420, 168248 

\end{thebibliography}
\end{document}